\pdfoutput=1
\documentclass[a4paper,11pt]{article}
\usepackage{jcappub}

\usepackage[english]{babel}
\usepackage[T1]{fontenc}
\usepackage{lmodern}
\usepackage{pifont}
\usepackage{fontawesome}
\usepackage{microtype,xspace,soul}
\usepackage{textcomp}
\usepackage{footmisc}
\usepackage{amsmath}
\usepackage{siunitx}
\DeclareSIUnit\astrounit{au}
\usepackage[version=4]{mhchem}
\usepackage{mathtools,relsize}

\usepackage[dvipsnames]{xcolor}
\usepackage{graphicx}
\usepackage{booktabs,tabularx,enumitem,threeparttable,makecell}
\usepackage[sort&compress,numbers]{natbib}

\def\araa{ARA\&A}%
\def\apj{ApJ}%
\def\apjl{ApJ}%
\def\apjs{ApJS}%
\def\apss{Ap\&SS}%
\def\jcap{JCAP}%
\def\mnras{MNRAS}%
\def\prd{Phys.\ Rev.\ D}%
\def\prl{Phys.\ Rev.\ Lett.}%
\def\physrep{Phys.\ Rep.}%
\definecolor{Blue}{RGB}{0,0,122}
\definecolor{Red}{RGB}{173,42,26}
\hypersetup{colorlinks=true, linktoc=all, linkcolor=Blue, citecolor=Red, urlcolor=Blue}

\newcommand*{\tabindent}{\hspace{0.5em}}
\newcommand{\cmark}{{\color{OliveGreen}\ding{51}}}
\newcommand{\cmarkvar}{{\color{OliveGreen}\textbf{(}\ding{51}\textbf{)}}}
\newcommand{\xmark}{{\color{red}\ding{55}}}
\newcommand{\rplus}{\protect\hspace{-.1em}\protect\raisebox{.35ex}{\smaller{\smaller\textbf{+}}}}
\newcommand{\cpp}{\mbox{C\rplus\rplus}\xspace}

\newcommand{\reffig}[1]{figure~\ref{#1}}
\newcommand{\citeref}[1]{ref.~\cite{#1}}
\newcommand{\citerefs}[1]{refs.~\cite{#1}}

\newcommand{\alphaEM}{\alpha_{\scriptscriptstyle\text{EM}}}

\newcommand{\ax}{a}
\newcommand{\ma}{m_\ax}

\newcommand{\gagg}{g_{\ax\gamma\gamma}}

\newcommand{\gaee}{g_{{\ax}ee}}

\newcommand{\sol}{\ensuremath{\odot}}
\newcommand{\Rsol}{\mathrm{R}_\sol}
\newcommand{\dE}{d_\text{E}}
\newcommand{\rCZ}{r_\text{\tiny CZ}}
\newcommand{\rtach}{\rCZ}
\newcommand{\rupper}{r_\text{upper}}
\newcommand{\dtach}{d_\text{tach}}
\newcommand{\dupper}{d_\text{upper}}

\newcommand{\ABC}{ABC\xspace}
\newcommand{\Phiae}{\Phi_{\ax}^{\scriptscriptstyle\text{ABC}}}
\newcommand{\Gae}{\Gamma_{\ax}^{\scriptscriptstyle\text{ABC}}}
\newcommand{\Phiag}{\Phi_{\ax}^\text{P}}
\newcommand{\GPrim}{\Gamma_{\ax}^\text{P}}
\newcommand{\GLP}{\Gamma_{\ax}^\text{LP}}
\newcommand{\GTP}{\Gamma_{\ax}^\text{TP}}
\newcommand{\Gee}{\Gamma_{\ax}^{ee}}

\newcommand{\Gff}{\Gamma_{\ax}^\text{ff}}
\newcommand{\Gffz}{\Gamma_{\ax,z}^\text{ff}}
\newcommand{\Gabs}{\Gamma_{\gamma,\,\text{abs}}}
\newcommand{\Gprod}{\Gamma_{\gamma,\,\text{prod}}}
\newcommand{\GCompt}{\Gamma_{\ax}^\text{C}}
\newcommand{\ks}{\kappa_\text{s}}
\newcommand{\ompl}{\omega_\text{pl}}
\newcommand{\GammaL}{\Gamma_\text{L}}
\newcommand{\GammaT}{\Gamma_\text{T}}

\newcommand*\diff{\mathop{}\!\mathrm{d}}
\newcommand{\ee}{\mathrm{e}}
\newcommand{\ii}{\mathrm{i}}
\newcommand{\amu}{m_\text{u}}
\newcommand{\dd}{\mathrm{d}}
\newcommand{\vc}[1]{\mathbf{#1}}
\def\lsim{\mathrel{\rlap{\lower4pt\hbox{$\sim$}}\raise1pt\hbox{$<$}}}
\def\gsim{\mathrel{\rlap{\lower4pt\hbox{$\sim$}}\raise1pt\hbox{$>$}}}

\newcommand{\ztz}{z_0}

\newcommand{\ver}{\texttt{v0.8b}\xspace}

\newcommand{\updated}[1]{#1}

\title{Quantifying uncertainties in the solar axion flux and their impact on determining axion model parameters}

\author[a,\dagger]{Sebastian Hoof,\note[$\dagger$]{Contact authors}}
\author[b]{Joerg Jaeckel}
\author[b,\dagger]{and Lennert J.\ Thormaehlen}

\affiliation[a]{Institut f\"{u}r Astrophysik, Georg-August-Universit\"{a}t {G\"{o}ttingen},\\Friedrich-Hund-Platz~1, 37077\ {G\"{o}ttingen}, Germany}
\affiliation[b]{Institut f\"{u}r theoretische Physik, Universit\"{a}t Heidelberg,\\Philosophenweg 16, 69120 Heidelberg, Germany}
\emailAdd{hoof@uni-goettingen.de}
\emailAdd{jjaeckel@thphys.uni-heidelberg.de}
\emailAdd{l.thormaehlen@thphys.uni-heidelberg.de}

\abstract{We review the calculation of the solar axion flux from axion-photon and axion-electron interactions and discuss the size of various effects neglected in current calculations. For the Primakoff flux we then explicitly include the partial degeneracy of electrons. We survey the available solar models and opacity codes and develop a publicly available C++/Python code to quantify the associated systematic differences and statistical uncertainties. The number of axions emitted in helioseismological solar models is systematically larger by about 5\% compared to photospheric models, while the overall statistical uncertainties in solar models are typically at the percent~level in both helioseismological and photospheric models. However, for specific energies, the statistical fluctuations can reach up to about~5\% as well. Taking these uncertainties into account, we investigate the ability of the upcoming helioscope IAXO to discriminate KSVZ axion models. Such a discrimination is possible for a number of models, and a discovery of KSVZ axions with high $E/N$ ratios could potentially help to solve the solar abundance problem. We discuss limitations of the axion emission calculations and identify potential improvements, which would help to determine axion model parameters more accurately. \href{https://github.com/sebhoof/SolarAxionFlux}{\faGithub}}

\begin{document}
\maketitle
\flushbottom

\section{Introduction}
QCD~axions~\cite{1977_pq_axion1,1977_pq_axion2,1978_weinberg_axion,1978_wilczek_axion} and axion-like particles~(ALPs)~\cite{Kim:1986ax,1002.0329} are intriguing hypothetical particles, which could play an important role as a solution for the Strong~CP problem~\cite{1977_pq_axion1,1977_pq_axion2}, as dark matter~(DM) candidates~\cite{Preskill:1982cy,Abbott:1982af,Dine:1982ah,Turner:1983he,Turner:1985si,1201.5902}, or in explaining anomalous observations in a variety of astrophysical objects~\cite[e.g.][]{1512.08108,1708.02111}. The additional energy loss of stars through the emission of ALPs has also long been used to place limits on their properties~(see e.g. refs~\cite{Vysotsky:1978dc,Dicus:1978fp,PhysRevD.22.839,PhysRevLett.48.1522,PhysRevD.26.1840,PhysRevLett.56.26,Raffelt:1985nk} for early works on this subject). The production of ALPs inside the Sun has recently received renewed attention due to an increased activity in studying previously neglected interactions of solar ALPs~\cite{2005.00078,2006.10415,2006.12431,2010.06601} as well as in the context of the observation of an excess of electronic recoil events in the XENON1T experiment~\cite{2006.09721}.

This interest is also driven by the upcoming helioscope~\cite{1983_sikivie,1985_sikivie,1989_vanbibber} experiment IAXO~\cite{1401.3233,1904.09155}, which will improve on the sensitivity of previous experiments by more than an order of magnitude. Helioscopes use magnetic fields inside a light-tight experimental setup to convert axions originating from the Sun into detectable photons. Previous helioscopes~\cite{Lazarus:1992ry,hep-ex/9805026,astro-ph/0204388,hep-ex/0411033,hep-ex/0702006,0806.2230,0906.4488,1106.3919,1302.6283,1307.1985,1503.00610} placed some of the strongest limits on axion-photon interactions for sub-meV axion masses -- currently $\gagg < \SI{0.66e-10}{\GeV^{-1}}$~(at 95\% CL) from the CAST experiment~\cite{1705.02290}. IAXO even has potential to measure the axion's mass and couplings~\cite{1811.09278,1811.09290}. It could also open the door for studying solar properties such as metal abundances~\cite{1908.10878} and magnetic fields inside the Sun~\cite{2006.10415}.

While this prospect is very exciting, both the precise determination of axion properties as well as the study of solar physics using axions depend on our ability to accurately predict the expected solar axion flux. For a possible ``axion precision age'', it is therefore unavoidable to quantify the systematic and statistical uncertainties related to axion generation inside the Sun.

To this end, we developed a \cpp code~(with a Python frontend) that provides a state of the art calculation of the solar axion flux.\footnote{The code has been made available as open source software under the BSD-3-Clause licence at \url{https://github.com/sebhoof/SolarAxionFlux}. We used version \ver for this work, which can be obtained from the version tagged with the corresponding label. Details on the Python frontend can be found in \hyperref[appendix:solaxlib]{appendix~\ref{appendix:solaxlib}}.} Our implementation of the equations is based on the calculations of refs.~\cite{Raffelt:1985nk,Raffelt:1987np,1310.0823,2005.00078,2006.10415,2010.06601} as well as code written for our previous works~\cite{1810.07192,1908.10878}. We revisit all currently available solar models, assessing their suitability for axion studies and quantify the systematic differences between them. Using the results of the Monte~Carlo simulations for representative helioseismological and photospheric solar models developed in refs.~\cite{astro-ph/0511337,Serenelli:2009yc,1611.09867}, we respectively estimate the statistical uncertainties on the solar axion flux and investigate the experimental consequences.

The main aim of our work is to quantify important sources of uncertainties in the solar axion flux. To demonstrate how this can affect uncertainties in the determination of the axion properties (cf.\ refs.~\cite{1811.09278,1811.09290}) from experimental helioscope measurements, we show that the flux calculations are precise enough such that KSVZ axion models within the reach of IAXO can be distinguished from each other.

This work is structured as follows; in section~\ref{sec:equations} we review the calculation of the solar axion flux and highlight changes and improvements in our treatment compared to previous work. In section~\ref{sec:solarmodelscodes} we discuss the publicly available solar models and opacity codes. Sections~\ref{sec:uncertainties} and~\ref{sec:axionmodels} are dedicated to our results regarding the statistical uncertainties and axion models. Finally, in section~\ref{sec:conclusions} we provide conclusions from this study and provide an outlook with respect to upcoming helioscope searches.

\section{Calculation of the solar axion flux}\label{sec:equations}
Let us review the calculation of the different contributions to the solar axion flux. For simplicity we only consider ALPs that are completely defined by their mass~($m_a$) and couplings to photons~($\gagg$) and electrons~($\gaee$),
\begin{equation}
\label{eq:ALPlagrangian}
    \mathcal{L}_\mathrm{ALP} = \frac{1}{2}(\partial_\mu a)^2 -\frac{1}{2} m_a^2 a^2 +  \frac{\gaee}{2m_e} \, (\partial_\mu a)\, \bar{e}\gamma^\mu \gamma^5  e - \frac{\gagg}{4}a\, F_{\mu \nu} \widetilde{F}^{\mu\nu} \, .
\end{equation}
Note that, in QCD~axion models, both $\gaee$ and $\gagg$ are proportional to the inverse of the axion decay constant i.e.\ $\gagg$,~$\gaee \propto 1/f_a$ or, equivalently, proportional to~$m_a$. We return to this connection in section~\ref{sec:axionmodels}, but ignore it until then to provide a general description. Further note that we write the axion-electron interactions with the derivative term instead of a pseudoscalar Yukawa term, viz.\ $-i \gaee \,a\, \bar{e}\gamma^5e$. Even though the two descriptions are related through the equation of motion, the Yukawa term can give a  sizable contribution to $\gagg$ through a triangle  diagram with an electron in the loop, while the same diagram with derivative couplings vanishes in the limit of a small momentum transfer compared to the electron mass~\cite{Quevillon:2019zrd}. This limit can safely be taken for all interactions inside the Sun that are considered in this work. By choosing this description, we thus avoid having to calculate additional axion-photon interactions induced by triangle diagrams.

\begin{figure}
    \centering
    \includegraphics[width=5.99in]{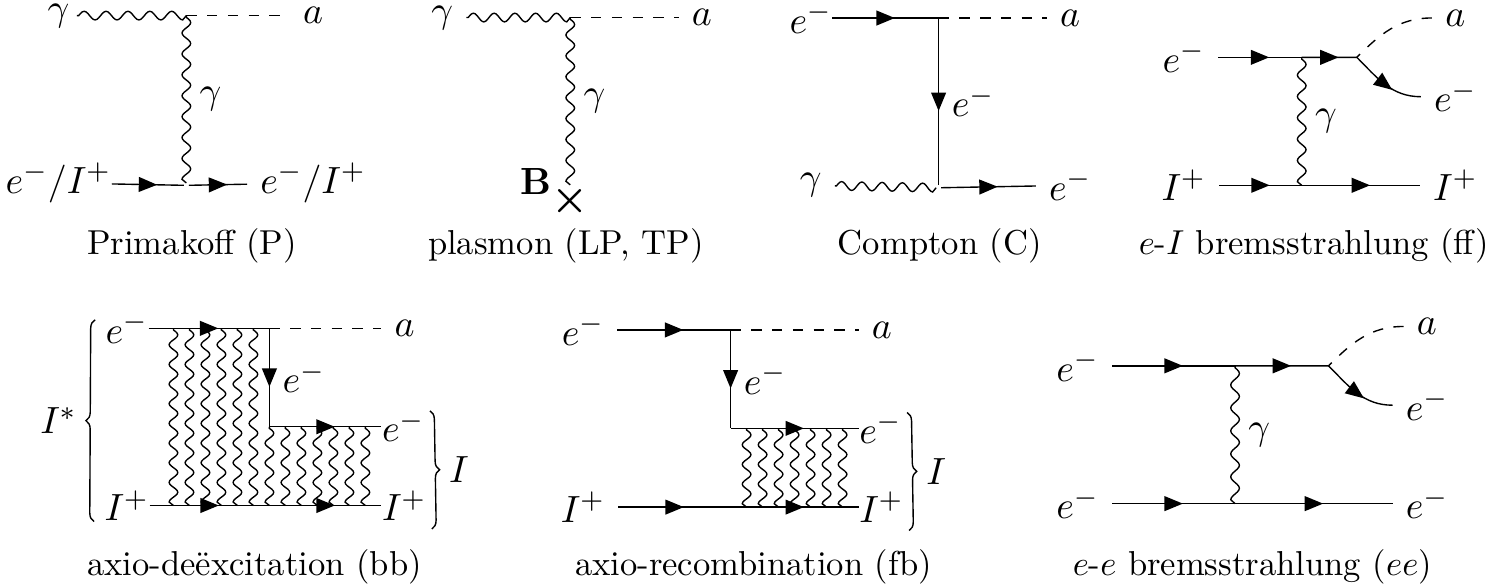}
    \caption{Diagrams for the axion interactions included in our code~(cf.\ refs.~\cite{Raffelt:1985nk,1310.0823}). The letter $\vc{B}$ denotes the macroscopic magnetic field inside the Sun. The letter~$I$ denotes an arbitrary chemical element, which may either be charged~(as ion $I^{+}$) or be in an excited state~($I^{\ast}$). Bound ion states in the (bb) and (fb) processes are indicated by many photon~lines ``binding'' the charged constituents. For simplicity, we show all ions as simply-charged but all processes will also involve multiply-charged ions~(i.e.\ $I^{2+}$, $I^{3+}$, \dots).
    \label{fig:feynman_diagrams}}
\end{figure}
All axion production processes induced by eq.~\eqref{eq:ALPlagrangian} can therefore be split into two categories: the ones that involve a coupling to the electromagnetic field via $\gagg$ and ones that involve the axion-electron coupling. Figure~\ref{fig:feynman_diagrams} presents an overview of the interactions considered in this work. We do not include axion-nucleon interactions in eq.~\eqref{eq:ALPlagrangian} since the resulting axion fluxes are subdominant~(due to other astrophysical constraints) and monochromatic~\cite{1982PhLB..119..323R,0906.4488,1209.2800,Moriyama:1995bz,Krcmar:1998xn,Gavrilyuk:2014mch}. We further assume that the axions produced in the Sun are ultrarelativistic i.e.\ $\ma \ll \omega \sim \mathcal{O}(\si{\keV})$. 

\begin{table}
	\caption{Corrections and uncertainties relative to the respective solar axion flux. Where possible, we provide a parametric estimate and the averaged and maximal effect between 1~and \SI{10}{\keV}. See section~\ref{sec:uncertainties} for details on the various corrections or sources of uncertainty. The intrinsic plasmon conversion flux uncertainties are completely dominated by the uncertainty of the solar magnetic field strength (varies up to a factor of 225), and we hence do not discuss smaller corrections to these contributions.\label{tab:corrections}}
	\renewcommand{\arraystretch}{1.109}
	\centering
	\begin{threeparttable}
	    {\small
		\begin{tabular}{llcc}
			\toprule
			\textbf{Correction/Uncertainty} & \textbf{Order} & \textbf{Averaged} & \textbf{Maximal}  \\
			\midrule
			\textit{Primakoff flux} & & &  \\
			\tabindent{}Solar model uncertainty (systematic) & & $\sim 5.1\%$ & $\sim 11\%$ \\
			\tabindent{}Solar model uncertainty (statistical) & & $\sim 1\%$ & $\sim 2.5\%$ \\
			\addlinespace[4pt]
			\tabindent{}Atomic transition (ff, bf, \& bb) emitting $\gamma + a$ & & $< 0.2\%$ & $\sim 4\%$ \\
			\tabindent{}Higher-order QED effects & $\alphaEM$& $< 0.7\%$ & $< 0.7\%$\\
			\tabindent{}Electro-Primakoff effect  &  & $< \num{4e-5}$ & $<0.4\%$ \\
			\tabindent{}Non-vanishing axion mass ($\ma \sim \mathrm{eV}$) & $\ma/\omega $ & $< 0.1\%$ & $< 0.1\%$ \\
			\tabindent{}Inelastic Primakoff &  & $\lsim 0.1\%$ & $\lsim 0.1\%$\\
			\tabindent{}\updated{Form factor for non-static charges} &  & \updated{$\lsim 0.02\%$} & \updated{$\lsim 0.02\%$}\\
			\tabindent{}\updated{Full relativistic dispersion relation} &  & \updated{$\lsim \num{5e-6} $} & \updated{$\lsim \num{e-6}$}\\

			 \addlinespace[4pt]
			 \tabindent{}Non-resonant transverse plasmon conversion\tnote{\textdagger} & & 0.01\%--5\%  &  0.02\%--50\% \\
			 \tabindent{}Resonant longitudinal plasmon conversion\tnote{\textdagger} & & $< \num{2e-6}$ & $< \num{0.7e-4}$ \\
			\midrule
			\textit{Resonant longitudinal plasmon conversion}  & & &  \\
			\tabindent{}Solar magnetic field strength  & & factor 225 & factor 225   \\
			\midrule
			\textit{Non-resonant transverse plasmon conversion} & & &   \\
			\tabindent{}Solar magnetic field strength  & & factor 225 & factor 225   \\
			\tabindent{}\makecell[l]{Total scattering rate $\GammaT$\\ \tabindent\tabindent{}(Rosseland mean vs. monochromatic)}
			 &  & factor 2.2 & factor 15.2  \\
			\midrule
			\textit{\ABC flux} & & &  \\
			\tabindent{}Solar model uncertainty (systematic) & & $\sim 5.4\%$ & $\sim 19\%$    \\
			\tabindent{}Solar model uncertainty (statistical) & & $\sim 1.5\%$ & $\sim 5\%$ \\
			\addlinespace[4pt]
			\tabindent{}Opacity uncertainty (systematic) & & 1\%--3\% & $<440\%$ \\
			\tabindent{}OP opacity uncertainty (statistical) & & $\lsim 1\%$ & $\sim 17\%$ \\
			\addlinespace[4pt]
			\tabindent{}Born approximation in eqs.~\eqref{eq:ffRate} and \eqref{eq:eeRate} &  & $\updated{\lsim 10\%}$ & $\updated{\lsim 10\%}$ \\
			\tabindent{}Electron degeneracy &  & $\lsim 7\%$ & $\lsim 8 \%$ \\
			\tabindent{}\makecell[l]{Higher order in multipole expansion in eq.~\eqref{eq:axion_photon_ratio}} & $\omega/(m_e Z\alphaEM)$ & $\lsim 2\%$ & $\lsim 7\%$\\
			\tabindent{}Effects of relativistic electron & & $\lsim 3\%$ & $\lsim 6\%$ \\
			\tabindent{}Higher-order QED effects & $\alphaEM$ & $< 0.7\%$ & $< 0.7\%$ \\
 			\tabindent{}Spatial and spin wave function non-separable & $E_{n}\,(Z\alphaEM)^2/n$& $\lsim 0.5\%$ & $\lsim 0.5\%$\tnote{*}\\
			\tabindent{}Non-vanishing axion mass ($\ma \sim \mathrm{eV}$) & $\ma/\omega $ & $< 0.1\%$ & $< 0.1\%$\\
			\bottomrule
		\end{tabular}
		\begin{tablenotes}\footnotesize
		    \item[\textdagger] While plasmon fluxes also arise from~$\gagg$, they are distinct and experimentally distinguishable from the Primakoff flux. We include them here to give an idea about their potential relevance. Strictly speaking, resonant longitudinal plasmon conversions are vanishingly small compared to the Primakoff flux in the energy range of 1--\SI{10}{\keV}. This is because the plasma frequency is always smaller than~\SI{1}{\keV}. The values quoted here correspond to the off-resonance conversions taken into account by using eq.~\eqref{eq:LPrate}. 
		    \item[*] The assumption of separability of spatial and spin wave functions only works if we do not resolve the atomic fine structure. In practice this limits the resolution to about $\sim \SI{25}{\eV}$. We estimate this from the lines listed in the NIST database~\cite{NIST_ASD}, which is in line with the parametric estimate given in the table. The uncertainty given here corresponds to a spectrum smoothed out at this level of resolution. At better resolution, the local uncertainty can be much larger. See section~\ref{sec:limitations} for details.
		\end{tablenotes}
		}
	\end{threeparttable}
\end{table}
Before going into the details of the calculations and describing the various uncertainties that affect them, we present an overview of all effects and the associated uncertainties considered in this work in table~\ref{tab:corrections}.

\subsection{Primakoff effect without degeneracy}\label{sec:primakoff}
The dominant production mechanism due to axion-photon interactions is the Primakoff effect. It can be viewed as excitations of the electromagnetic field inside the plasma -- called plasmons -- converting into an axion in the electromagnetic field of electrons or ions~(charged nuclei). 

The Primakoff process has always been the focus of solar axion searches, for which there are two main reasons: one is that nearly all axion models feature a non-vanishing~$\gagg$. Second, and more practically, \updated{helioscope} detection relies on the photon coupling and hence cannot discover axions with $\gagg = 0$.
Therefore, in helioscope searches, a contribution from the Primakoff production is essentially unavoidable -- even though other productions mechanisms, such as the ones discussed below for non-vanishing electron coupling, may be dominant.

The Primakoff production rate ($\GPrim$) is crucially affected by charge screening in the plasma, first included by Raffelt~\cite{Raffelt:1985nk,Raffelt:1987np}. \updated{His screening prescription assumes a static potential during each scattering event. First steps to go beyond this approximation, and a calculation of the relative size of associated corrections, are included in appendix~\ref{appendix:formfactor}.} Raffelt's result can be expressed in terms of the Debye screening scale~$\ks$, the plasma temperature~$T$, the axion energy~$\omega$, the number density of free electrons~$n_e$ and the axion-photon coupling~$\gagg$, such that
\begin{align}
	\GPrim(\omega) &= \gagg^2 \alphaEM \, \frac{n_e + \bar{n}}{8} \left[\left(1+\frac{\ks^2}{4\omega^2}\right)\ln\left(1+\frac{4\omega^2}{\ks^2}\right)-1\right] \frac{2}{\ee^{\omega/T}-1}\, , \label{eq:PrimaRate}
\end{align}
where the quantity $\bar{n}$ is defined as the sum of number densities of each ion~($n_z$), weighted by the square of its electric charge~($Q_z^2$) in units of the elementary charge squared i.e.\
\begin{equation}
	\bar{n} \equiv \sum_{z}Q_z^2 \, n_z \, . \label{eq:nbar_definition}
\end{equation} 
$z$ is a label for all types of ions in the plasma. Note that, even close to the centre of the Sun, $Q_z$ is in general not equal to the atomic number since we cannot assume full ionisation of all atoms~(see section~\ref{sec:electron-densities}). Moreover, $\alphaEM \approx 1/137.036$ is the fine-structure constant. 
In the non-degenerate limit, the Debye screening scale $\ks$ is given by~\cite{Raffelt:1985nk}
\begin{equation}
	\ks^2 = \frac{4\pi \alphaEM}{T} (n_e + \bar{n}) \, .
	\label{eq:screening_nondeg}
\end{equation}

Equation~\eqref{eq:PrimaRate} does not take into account that photons inside a plasma have a non-trivial dispersion relation. \updated{In the non-relativistic and non-degenerate limit, the dispersion relation reads~\cite{Raffelt:1996wa}
\begin{equation}
    \omega^2 = \ompl^2 + k_\gamma^2 \, . \label{eq:dispersion}
\end{equation}
Higher order corrections to this relation are suppressed by powers of $T/m_e$. We discuss these in section~\ref{sec:limitations}.}
The plasma frequency $\ompl$ is given by
\begin{equation}
    \ompl^2  = \frac{4\pi \alphaEM}{m_e} \, n_e \, ,
    \label{eq:plasmafreq_nondeg}
\end{equation}
where $m_e \approx \SI{511.0}{\keV}$ is the electron mass.

We stress that eq.~\eqref{eq:PrimaRate} is only valid in the limit of $\omega \gg \ompl$. The rate vanishes for $\omega < \ompl$, and it is significantly suppressed when $\omega \gsim \ompl$.
The full result, which includes the plasma frequency, and is valid for all $\omega > \ompl$, is given by~\cite{Jaeckel:2006xm,2006.10415}
\begin{align}
    &\GPrim(\omega) = \gagg^2\alphaEM \, \frac{n_e + \bar{n}}{8} \left[\frac{1}{2} \int_{-1}^{1} \! \dd x \; \frac{1-x^2}{(\xi_1-x)(\xi_2-x)}\right] \frac{2}{\ee^{\omega/T}-1}\frac{\diff k_\gamma}{\diff \omega} \label{eq:PrimaRate_fullintegral} \\
    &\text{with} \quad \xi_1 \equiv \frac{k_a^2+k_\gamma^2}{2k_ak_\gamma}   \quad \text{and} \quad \xi_2 \equiv \xi_1 + \frac{\ks^2}{2k_ak_\gamma} \, .
\end{align}
We determine the analytical result of this integral to be
\begin{equation}
    \GPrim(\omega) = \gagg^2\alphaEM \, \frac{n_e+\bar{n}}{8} \left[\frac{(\xi_1^2-1)\ln\left(\frac{\xi_1+1}{\xi_1-1}\right) - (\xi_2^2-1)\ln\left(\frac{\xi_2+1}{\xi_2-1}\right)}{2\, (\xi_1-\xi_2)} - 1\right] \frac{2}{\ee^{\omega/T}-1}\frac{\omega}{k_\gamma} \, , \label{eq:PrimaRate_full}
\end{equation}
where $k_a$ and $k_\gamma$ denote the axion and photon momenta, respectively. Since we work in the limit of ultrarelativistic axions, we have $k_a \simeq \omega$, while $k_\gamma$ can be computed from eq.~\eqref{eq:dispersion}.

In summary, there are only three solar quantities that enter eq.~\eqref{eq:PrimaRate_full}, namely $T$, $n_e$, and~$\bar{n}$. Typically only $T$~is tabulated in solar models. We discuss the calculation of $n_e$ and $\bar{n}$ in section~\ref{sec:integrate}.

\subsection{Primakoff effect with degeneracy}\label{sec:primakoff_deg}
Since electrons in the solar plasma are partially degenerate, treating them as an ideal gas in the previous section is only an approximation.\footnote{We are indebted to Georg Raffelt for pointing out that this can have an appreciable effect on our results.}
Going beyond this approximation, we note that the screening scale, the plasma frequency, and total scattering rates are affected by the degeneracy of electrons at the percent level.

To include the electron degeneracy, we have to evaluate the phase space integrals with the full Fermi-Dirac distribution,
\begin{equation}
    f(p) = \frac{1}{\ee^{{(E(p)-\mu)}/T}+1} \, , \label{eq:fd_distr}
\end{equation}
with energy~$E$, momentum $p$ and chemical potential~$\mu$. We can calculate the latter at every position inside the Sun by numerically inverting the relation for the electron number density in terms of the chemical potential. In the non-relativistic limit and assuming an ideal Fermi gas, this is given by~\cite{FermiGas}
\begin{equation}
  n_e = 2 \left(\frac{m_e T}{2\pi}\right)^{3/2} \, \mathcal{F}^\text{\tiny FD}_{1/2}(z) \, ,
\end{equation}
where $z = (\mu-m_e)/T$ and the Fermi-Dirac integral is defined as
\begin{equation}
    \mathcal{F}^\text{\tiny FD}_j(z) \equiv \frac{1}{\Gamma(j+1)} \, \int_0^\infty \dd t \; \frac{t^j}{\ee^{t-z}+1} = -\sum_{k=1}^\infty (-1)^{k} \, \frac{\exp(z)^k}{k^{j+1}} \, , \label{eq:F_FD_j}
\end{equation}
and $\Gamma(\cdot)$ is the usual gamma~function.

\begin{figure}
    \centering
    \includegraphics[width=3.75in]{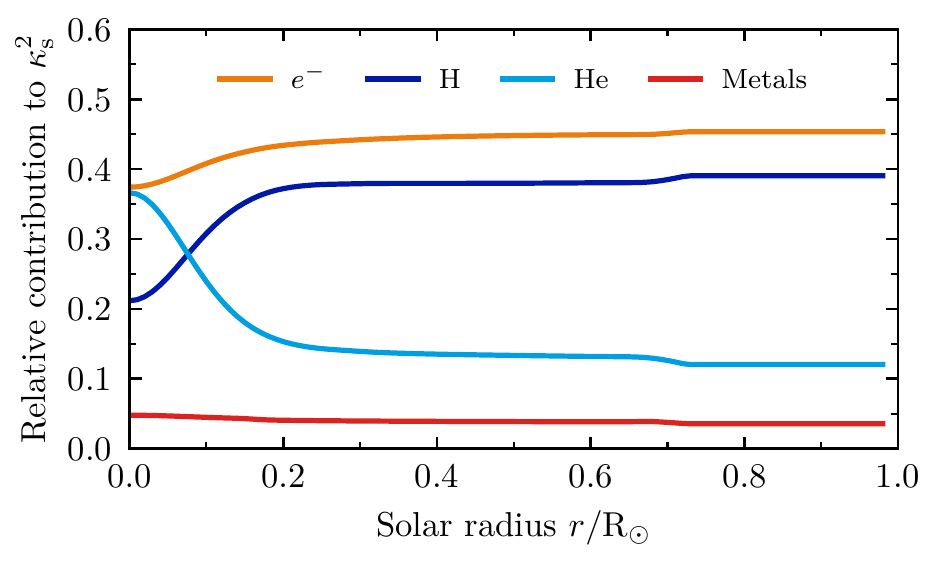}
    \caption{\updated{Relative contributions of electrons, \ce{H}, \ce{He}, and heavier elements~(metals) to $\ks^2$ for the AGSS09 model.}}
    \label{fig:el_contrib_to_ks}
\end{figure}
\updated{After solving eq.~\eqref{eq:F_FD_j}} for $\mu$, we can evaluate the full expression for the screening scale, including the degeneracy of electrons, which is~\cite{Raffelt:1996wa}
\begin{equation}
    	\ks^2 = \frac{4\pi \alphaEM}{T}  \bar{n} + \frac{4\alphaEM}{\pi}\int_0^\infty \dd p \; f(p) \, p \, \left(v + v^{-1}\right) \, , \label{eq:degen_ks}
\end{equation}
where $v = p/E = p/\sqrt{m_e^2 + p^2}$ is the electron's velocity. It is straightforward to check that in the non-degenerate limit this reproduces eq.~\eqref{eq:screening_nondeg}. Depending on the location inside the Sun, the effect of degeneracy on the screening scale is at most a reduction of~1.2\%. \updated{In fig.~\ref{fig:el_contrib_to_ks}, we show the different contributions to~$\ks^2$ as a function of the solar radius, according to eq.~\eqref{eq:degen_ks}.}

Similarly, the full expression of the plasma frequency is~\cite{Raffelt:1996wa}
\begin{equation}
    \ompl^2 = \frac{4\alphaEM}{\pi}\int_0^\infty \dd p \; f(p) \, p \, \left(v - \frac{1}{3}v^3\right) \, . \label{eq:degen_wpl}
\end{equation}
In contrast to the expression~\eqref{eq:degen_ks} for the screening scale, only positive powers of the velocity appear. As the degeneracy is strongest at small velocities, the plasma frequency is less affected by the electron degeneracy than the screening scale. While the difference to the non-degenerate result in eq.~\eqref{eq:plasmafreq_nondeg} is at most of order $10^{-3}$, we always include the full expression in our calculations.

The occupation numbers can reach values of up to $\sim0.2$ in the solar core, which can lead to Pauli blocking of scattering processes with small momenta of final state electrons. To include the corresponding suppression, we follow a strategy similar to the one used in ref.~\cite{Raffelt:1996wa} to include degeneracy effects in Coulomb scattering. 
We define a suppression factor $F_\mathrm{deg}$ that affects (only) the part of the production rate in eq.~\eqref{eq:PrimaRate_full} related to scattering of electrons,\footnote{Since the mass of the ions in the plasma is at least a factor of~2000 larger than $m_e$, they are highly non-degenerate.}
\begin{align}
    n_e \mapsto F_\text{deg} \, n_e \, , \label{eq:degen_correction}
\end{align}
with the suppression factor
\begin{equation}
    F_\text{deg} = \frac{\int_0^\infty \dd p_1 \int_{-1}^{1}\dd y \int_{-1}^{1}\dd x \; f(p_1) \, \left(1-f(p_2)\right) \frac{\dd\sigma}{\dd\Omega}}{\int_0^\infty \dd p_1 \int_{-1}^{1} \dd y \int_{-1}^{1}\dd x \; f(p_1) \, \frac{\dd\sigma}{\dd\Omega}} \, .
    \label{eq:degen_factor}
\end{equation}
Here, $p_1$ and $p_2$ are the incoming and outgoing electron momenta, respectively, and we average over the relative angle $y \equiv \cos{\theta_{12}}$ between the incoming momentum and the momentum transfer, as well as over the scattering angle $x \equiv \cos{\theta}$. The integrals have to be performed for the appropriate Fermi-Dirac distributions\footnote{In the non-degenerate case and taking the limit of stationary electron targets everything can be expressed in terms of the electron number density and the statistical distribution of the electron energy drops out. To normalise appropriately we can therefore use the Fermi-Dirac distribution in the denominator.}~$f(p)$ and the differential Primakoff cross~section~$\frac{\diff\sigma}{\diff\Omega}$ from the integrand of eq.~\eqref{eq:PrimaRate_fullintegral}, 
\begin{equation}
    \frac{\dd\sigma}{\dd\Omega} = \frac{\alphaEM}{32\pi} \, \gagg^2 \, \frac{1-x^2}{(\xi_1-x)(\xi_2-x)}.
\end{equation}
The final electron momentum $p_2$ depends on the transferred momentum $q$ and the relative angle $\theta_{12}$,
 \begin{align}
  p_2^2 &= p_1^2 + 2 \, p_1 q \cos(\theta_{12}) + q^2 \, .
\end{align}
while the momentum transfer $q$ can be computed as 
\begin{align}
  \left(\frac{q}{\omega}\right)^2 &= 2 - \left(\frac{\ompl}{\omega}\right)^2 - 2 \, \sqrt{1 - \left(\frac{\ompl}{\omega}\right)^2} \cos(\theta) \, .
  \end{align}

\begin{figure}
    \centering
    {\hfill
    \includegraphics[width=2.75in]{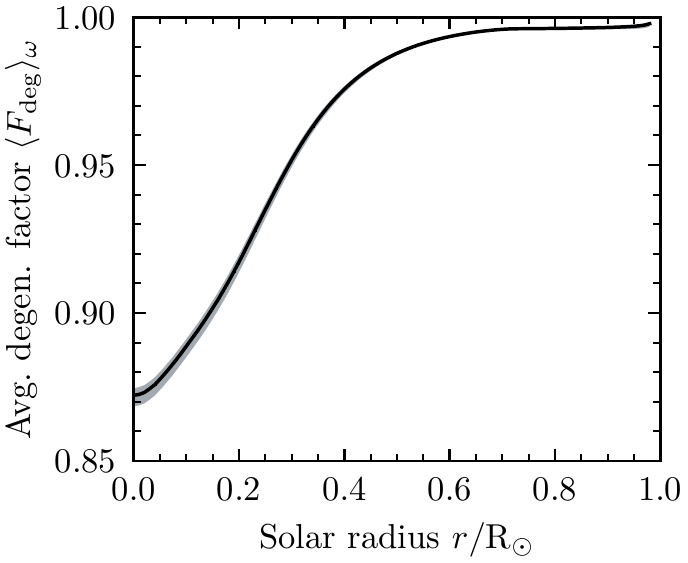}
    \hfill
    \includegraphics[width=2.75in]{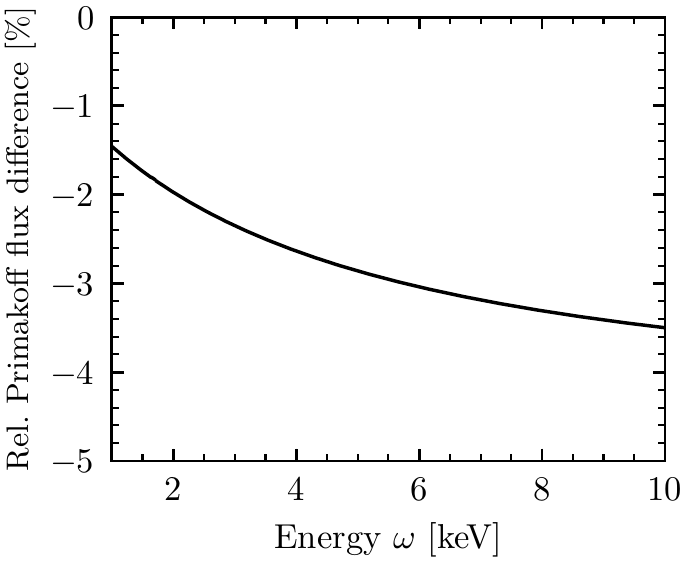} \hfill}
    \caption{Electron degeneracy effects. \textit{Left:} Energy-averaged electron degeneracy factor for the solar model AGSS09~(black line). The shaded grey region indicates the range of values of $F_\text{deg}$ between \SIrange[range-phrase=--]{1}{10}{\keV}. \textit{Right:} Primakoff flux reduction (in percent) due to \emph{all} degeneracy effects in the energy range of interest and for the AGSS09~model. \label{fig:electron_degen_factor}}
\end{figure}
Evaluating the degeneracy factor numerically,\footnote{Note that the integrals over $x = \cos(\theta)$ in both the numerator and denominator, as well as the integral over $y = \cos(\theta_{12})$ in the denominator, can be performed analytically. The latter can be calculated with the help of $\int_{-1}^{1} \dd y \; f(p_2) = \left. -T^2 [z \ln(1+\ee^{\ztz-z}) + \mathcal{F}^\text{\tiny FD}_1(\ztz-z)]/(p_1 q) \right|_{z_-}^{z_+}$, where $\ztz \equiv \mu/T$ and where $z$ needs to be replaced by $z_\pm \equiv \sqrt{m_e^2 + (p_1 \pm q)^2}/T$ for the upper/lower limit of~$y = \pm1$.} we find that it is around $F_\text{deg} \sim 0.87$ at the centre of the Sun, approaching unity as $r \rightarrow 1$ with only weak dependence on the energy of the emitted axion $\omega$ in the range of interest (1-10)~keV. To further simplify the calculation, we may therefore compute~$\langle F_\text{deg}\rangle_\omega$, the energy-averaged values between \SIrange{1}{10}{\keV}, by adding an integration over energy for both the numerator and denominator in eq.~\eqref{eq:degen_factor}. The result for the example of AGSS09~model is shown in \reffig{fig:electron_degen_factor}. As a result of the combined electron degeneracy effects, the integrated Primakoff flux in the relevant energy range is reduced by about~2.6\%.

\subsection{Plasmon conversion in the solar magnetic field}\label{sec:axionplasmon_interactions}
Apart from the electromagnetic fields of ions, which vary on small length scales, the Sun is also permeated by a large-scale magnetic field. This macroscopic magnetic field plays a similar role to the microscopic electromagnetic fields for the Primakoff effect, and plasmons can convert into axions through axion-photon interactions. This effect has recently gained attention because the conversion rates of both longitudinal and transverse plasmons feature resonances, which enhance this contribution at the resonance energy scale. As a consequence, the plasmon contributions are potentially detectable~\cite{2005.00078,2006.10415} or relevant for stellar energy loss arguments~\cite{2005.00078,2010.06601}. 

We adopt the magnetic field model used in ref.~\cite{2006.10415}, where the field is toroidal, $\vc{B}(\vc{r}) = -3 \, B(r) \cos(\theta)\sin(\theta) \; \hat{\vc{e}}_\phi$, $\theta$~is the azimuthal angle inside the Sun and
\begin{equation}
 B(r) =
  \begin{cases}
    B_\text{rad}(1+\lambda)(1+\frac{1}{\lambda})^{\lambda}\left(\frac{r}{\rCZ}\right)^{2}\left[1-\left(\frac{r}{\rCZ}\right)^{2}\right]^{\lambda} & \text{for } r < \rCZ - \dtach\\
    B_\text{tach}\left[1-\left(\frac{r-\rtach}{\dtach}\right)^{2}\right] & \text{for } |r-\rtach|< \dtach \\
    B_\text{outer}\left[1-\left(\frac{r-\rupper}{\dupper}\right)^{2}\right] & \text{for } |r - \rupper| < \dupper\\
    0 & \text{otherwise}
  \end{cases} \, . \label{eq:solar_b_fields}
\end{equation}
Here, $\rCZ \approx 0.712\,\Rsol$ is the approximate radius of the radiative zone, $\lambda\equiv 10\,\rCZ/\Rsol + 1$, $\rupper \approx 0.96\,\Rsol$ is the nominal beginning of the outer layers of the Sun, and $\dtach \approx 0.02\,\Rsol$ and $\dupper \approx 0.035\,\Rsol$ are the shell thickness of the tachocline and the outer layers of the Sun, respectively. We choose the same ranges for the $B$-field normalisations as considered in \updated{Fig.~3 of }ref.~\cite{2006.10415}, viz.\ $B_\text{rad} \in [\SI{200}{\tesla},\,\SI{3000}{\tesla}]$, $B_\text{tach} \in [\SI{4}{\tesla},\,\SI{50}{\tesla}]$, and $B_\text{outer} \in [\SI{3}{\tesla},\,\SI{4}{\tesla}]$.

The conversion rate of longitudinal plasmons into axions is given by~\cite{2005.00078,2006.10415}
\begin{equation}
    \GLP(\omega)= \frac{\gagg^2B_\parallel^2}{\ee^{\omega/T}-1}\frac{\omega^2 \GammaL }{(\omega^2 -\ompl^2)^2+(\omega \GammaL)^2} \label{eq:LPrate} \, ,
\end{equation}
where $B_\parallel$ is the magnetic field strength projected onto the propagation direction of the plasmon, and $\GammaL$ is the collision rate of longitudinal plasmons.
$\GLP$ is dominated by the resonance at $\omega = \ompl$, and $\GammaL$ only sets the (narrow) width of this peak.\footnote{\updated{In fact, the resonance is so narrow that we had to set $\GammaL = \mathrm{max}(\GammaT,\SI{0.1}{\eV})$ to improve convergence for general integration intervals.}}
This is why the rate $\GLP$ is commonly approximated by a delta function~\cite{2005.00078,2006.10415,2010.06601}.
Since the plasma frequency inside the Sun is limited to values $\ompl \lsim \SI{0.3}{\keV}$, the axion flux from resonant longitudinal photon conversion only contributes at small energies with three identifiable peaks corresponding to the plasma frequencies in the radiative zone, the tachocline and the upper layers of the Sun~(see also Fig.~2 in ref.~\cite{2006.10415}).
\updated{The non-resonant conversion of longitudinal plasmons is suppressed due to its peculiar dispersion relation. In the limit of non-relativistic, non-degenerate electrons, and to leading order in $T/m_e$, it is~\cite{Raffelt:1996wa}
\begin{equation}
    \omega^2 = \ompl^2 \left(1 + 3\, \frac{k^2}{\omega^2} \, \frac{T}{m_e} \right) \, .
\end{equation}
For momenta of the order of the energy, this implies $\omega \approx \ompl$~\cite{Raffelt:1996wa,2010.06601}, which is precisely the resonance condition that can be implemented by the delta function. At momenta much larger than $\omega$, the modes are not on resonance but -- because their four-momentum is space-like ($k>\omega$) -- they are affected by \updated{Landau damping~\cite{Landau:1946jc,Raffelt:1996wa}. Raffelt} concludes that in this case \emph{longitudinal modes no longer exist}, which is why we only integrate over a small region around the resonance peak of $\GLP$. This is effectively the same as using a delta function as done in refs.~\cite{2005.00078,2006.10415,2010.06601}. A closer investigation of how the off-resonance contributions are suppressed may be interesting, but we expect the ensuing corrections to be very small.}

\begin{figure}
    \centering
    \includegraphics[width=4in]{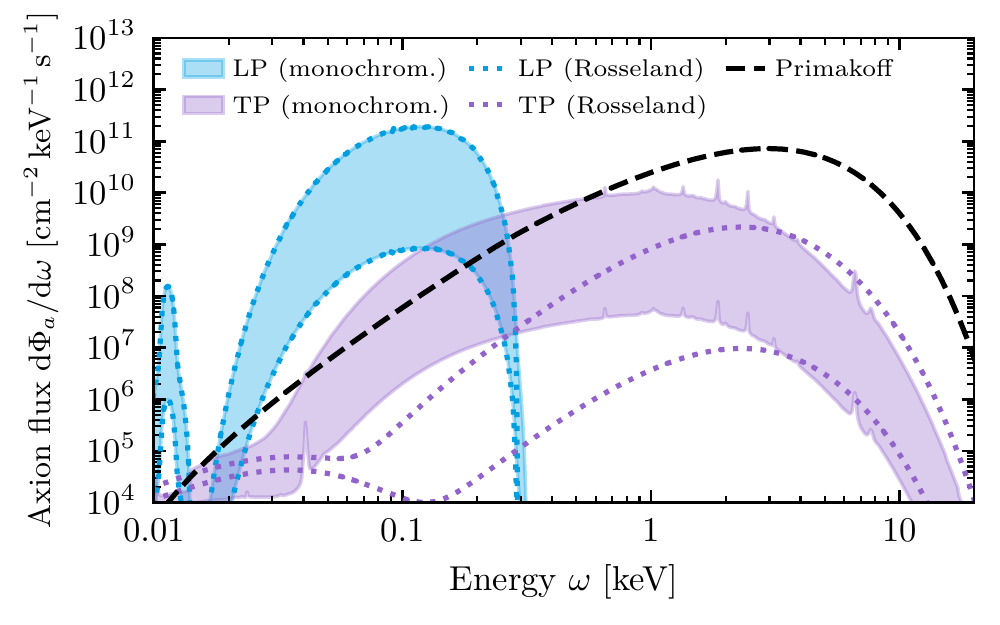}
    \caption{Axion flux from longitudinal~(LP) and transverse~(TP) plasmon interactions for $\gagg = \SI{e-10}{\GeV^{-1}}$. The blue and purple shaded contours show the LP and TP flux between our minimal and maximal $B$-field reference values, respectively, \updated{using monochromatic opacities~$\kappa(\omega,r)$. Dotted lines show the LP and TP fluxes when the Rosseland mean opacities are used. In both cases, the opacities are calculated using OP~data.} The dashed black line shows the Primakoff flux for comparison.\label{fig:plasmons}}
\end{figure}
Transverse plasmons, on the other hand, obey the dispersion relation given in eq.~\eqref{eq:dispersion}. As a consequence, resonant conversion is only possible when the axion is massive with $\ma = \ompl$. The corresponding rate is given by~\cite{2006.10415}
\begin{equation}
    \GTP(\omega)= \frac{2 \, \gagg^2B_\perp^2}{\ee^{\omega/T}-1}\frac{\omega^2 \GammaT}{(\ompl^2 -\ma^2)^2+(\omega \GammaT)^2} \; \label{eq:TPrate},
\end{equation}
with the average square of the magnetic field projected onto the polarisation vector $B_\perp^2$ defined as in ref.~\cite{2006.10415}. 
$\GammaT$ is the collision rate of transverse plasmons and the prefactor of two accounts for the two linearly independent polarisation states.

As mentioned previously, we only consider ultrarelativistic axions with negligible mass compared to the plasma frequency such that we are only interested in the limit $\ma \rightarrow 0$ of eq.~\eqref{eq:TPrate}. Far from the resonance, the axion production rate becomes proportional to the collision rate $\GammaT$.
This introduces a complication since $\GammaT$ is frequency-dependent. We can, however, deduce $\GammaT$ from the monochromatic opacity $\kappa(\omega)$, which is defined as the absorption coefficient $k$ (inverse of the mean free path) per plasma density $\rho$:
\begin{equation}
\label{eq:MonoOpacity}
 \kappa(\omega) = \frac{k(\omega)}{\rho} \, .
\end{equation}
The total collision rate is then given by \cite{2010.06601},
\begin{equation}
    \GammaT = k(\omega) \, \left(1-\ee^{-\omega/T}\right) = \kappa(\omega) \, \rho \, \left(1 - \ee^{-\omega/T}\right) \, .
    \label{eq:GammaTransverse}
\end{equation}
Details about different opacity codes are given in sections~\ref{sec:axionelectron_interactions} and~\ref{sec:solarmodelscodes}.

\updated{Previous studies of the non-resonant transverse axion flux have used the frequency-independent Rosseland mean opacity~$\kappa_\text{R}$ for the evaluation of $\GammaT$~\cite{2010.06601}}, which is defined by~\cite{Krief:2016znd}
\begin{equation}
\label{eq:opacity}
    \frac{1}{\kappa_\text{R}} = \int_0^\infty \frac{\mathcal{R}({\omega}/{T})}{T \, \kappa(\omega)} \diff \omega
\end{equation}
with the Rosseland weight function
\begin{equation}
    \mathcal{R}(u) \equiv \frac{15}{4\pi^4}\frac{u^4 \, \ee^u}{(\ee^u-1)^2} \, .
\end{equation}
\updated{The Rosseland mean opacity is therefore \emph{derived} from the more fundamental monochromatic opacity in the sense that it is an effective description that depends on a weight function. Similar to the Planck mean opacity, and other possible definition of mean opacities, it provides a useful description in certain contexts but may not be meaningful in others~(see e.g.\ ref.~\cite{2014_opacity}). Based on this, we expect that the more fundamental monochromatic opacities provide a more accurate description in situations where the results obtained from monochromatic and mean opacities disagree.

In figure~\ref{fig:plasmons}, we can clearly see that the choice of opacity function makes a rather large difference for the TP flux -- especially for smaller energies where the non-resonant conversion of longitudinal plasmons can potentially become comparable, or even larger, than the Primakoff flux. Furthermore, unlike after Rosseland averaging, the peaks from the monochromatic opacity are still present in the flux, which results in distinct, potentially interesting features at these frequencies.

In summary, since the monochromatic opacity is -- as explained above -- the more fundamental quantity, we propose that using it to estimate the mean free path provides a more accurate description of the axion spectrum from transverse plasmons than the Rosseland mean opacity.}

We show the spectral axion fluxes from longitudinal and transverse plasmon conversions in figure~\ref{fig:plasmons} along with the Primakoff flux for comparison.\footnote{\updated{Note that we do not include a geometrical factor of 1.8, which is required to calculate the time averaged flux on Earth from $\GLP$ as discussed in Sec.~III~D of ref.~\cite{2006.10415}. This is because, in general, this factor depends on the data taking times throughout the year and therefore is -- at least slightly -- different for each experiment. Instead, we show the flux averaged over all emission directions from the Sun, for which the geometric factor was found to be $1/3$ in ref.~\cite{2010.06601}. To recover the result in ref.~\cite{2006.10415}, one has to multiply the spectrum labelled LP in fig.~\ref{fig:plasmons} by $3\times 1.8$.}}
\updated{The large possible range of plasmon flux values for a given energy} is due to the fact that the large-scale solar magnetic field is poorly constrained: the normalisations of its inner components differ by factors larger than an order of magnitude.

Although plasmon conversions inside large-scale magnetic fields are due to~$\gagg$, they are not simply corrections or sources of uncertainty of the Primakoff prediction. This is because the peaks in the axion spectrum corresponding to longitudinal plasmon conversion are well separated from the bulk of the Primakoff contribution, and they can therefore be detected independently. As was pointed out in ref.~\cite{2006.10415}, this would enable a measurement of the magnetic fields in the deep solar interior. Such a measurement would also allow for a more accurate prediction of the non-resonant conversion rate of transverse plasmons. In addition, the angular distribution of the flux from plasma conversions depends on the geometry of the solar magnetic field. This results in an anisotropic axion emission and an annually oscillating flux on earth, giving us another handle to distinguish the two flux components~\cite{2006.10415}.

In what follows, we consider the Primakoff flux as the only process that arises from~$\gagg$. We do this since it is currently not possible to quantify the uncertainties on the plasmon flux normalisations. As mentioned before, it is possible to distinguish the plasmon and Primakoff components of the flux from their different spectral behaviour and angular distributions.

\subsection{Axion-electron interactions}\label{sec:axionelectron_interactions}
There are numerous processes by which axions can be produced through their coupling to electrons. On the level of Feynman diagrams, an outgoing photon of any ``ordinary'' tree level scattering process can usually be replaced by an axion, even though the corresponding interaction terms differ in their spin structure. This is why electron scattering processes, which are constantly occurring at a high rate in the solar plasma, have a (highly-suppressed) axion equivalent. Examples for these axion production mechanisms are given by the diagrams in figure~\ref{fig:feynman_diagrams}. 

The interaction rates of some of these processes, such as electrons scattering off nuclei~(ff), electron bremsstrahlung~($ee$) and via the Compton effect~(C) can be computed analytically similar to the Primakoff rate. Again, charge screening in the stellar plasma has to be taken into account. Here, we only cite the results in their compact form following ref.~\cite{1310.0823}:
\begin{align}
    \Gffz (\omega) &= \gaee^2 \, \alphaEM^2 \, \frac{8\pi}{3 \sqrt{2\pi}} \frac{Q_z^2 n_z n_e}{\sqrt{T}m_e^{7/2}\omega} \, \ee^{-\omega/T} \, \mathcal{I}(\omega/T,y) \, , \\
    \Gff (\omega) &=  \sum_{z}\Gffz\, , \label{eq:ffRate}\\
    \Gee (\omega) &= \gaee^2 \, \alphaEM^2 \, \frac{4 \sqrt{\pi}}{3} \frac{n_e^2}{\sqrt{T}m_e^{7/2}\omega} \, \ee^{-\omega/T} \, \mathcal{I}(\omega/T,\sqrt{2}y) \, , \label{eq:eeRate}\\
    \GCompt (\omega) &= \gaee^2 \,  \alphaEM\hphantom{^2} \, \frac{n_e}{3m_e^4} \, \frac{\omega^2}{\ee^{\omega/T}-1} \, , \label{eq:CRate}
\end{align}
where we have used the function\footnote{Note that the second integral as a function of~$t$ has an analytical solution, which reads $ \ln(t^2 + y^2)/2$.}
 \begin{equation}
     \mathcal{I}(u,y) = \int_0^\infty \! \dd x \, x \,  \ee^{-x^2}\,\int_{\sqrt{x^2+u}-x}^{\sqrt{x^2+u}+x} \dd t \; \frac{t}{(t^2+y^2)} \, , \label{eq:aux_fun_F}
 \end{equation}
and where $y$ is defined as
\begin{equation}
    y \equiv \frac{\ks}{\sqrt{2 m_e T}} \, .
\end{equation}
Recall that $\alphaEM$ denotes the fine-structure constant and that $\bar{n}$ is defined as in eq.~\eqref{eq:nbar_definition}. In contrast to ref.~\cite{1310.0823}, we use a different screening description, which results in a change of the integrand in the $\mathcal{I}$~function in eq.~\eqref{eq:aux_fun_F}. Following the arguments given in refs.~\cite{Raffelt:1985nk,Raffelt:1996wa,Vitagliano:2017odj}, we concluded that charge screening for ion- and electron-bremsstrahlung is best described by an effective form factor $q^2/(q^2+\ks^2)$ in the transition amplitude, where $q$~is the transferred momentum. Reference~\cite{1310.0823} used the square of this form factor. More fundamentally, the two choices for the form factor arise from either taking the average over charge distributions at the level of the matrix element squared or the matrix element itself~\cite{Raffelt:1996wa}. The former description is better if the time it takes for the scattering electron to cross the potential is smaller than the typical timescale of screening cloud formation.
While this is expected to be a reasonable approximation for the screening contribution of the ions, it is not so straightforward for the part arising from the electrons. 
A more systematic investigation of this effect is beyond the scope of this work but would be highly desirable.

Comparing the fluxes arising from both descriptions (for simplicity we apply the different prescriptions for the whole screening), we find that our integrated ABC flux between \SIrange{1}{10}{\keV} is about 5.7\% \emph{larger} compared to the form factor used in ref.~\cite{1310.0823} while the difference in the spectral flux within the same energy range is at most~26\%.

The production rates in eqs.~\eqref{eq:ffRate}--\eqref{eq:CRate} are affected by the partial degeneracy of electrons in two ways. First, the degeneracy suppresses the screening scale compared to the non-degenerate case. We account for this by using eq.~\eqref{eq:degen_ks}. Second, the rates can be suppressed by Pauli blocking since all three processes contain at least one electron in the final state~(see also refs.~\cite{Raffelt:1996wa,Carenza:2021osu}). We do not correct for this non-negligible effect, and we discuss its size in more detail in section~\ref{sec:limitations}.

Bound-bound~(bb) and free-bound~(fb) transitions in the atomic shell require additional attention because atomic structures of the involved elements enter the calculation. The first thorough calculation involving all of theses processes was done by Redondo~\cite{1310.0823}. In two steps, he relates the production of axions to the absorption of photons. This is commonly given via the monochromatic opacity $\kappa$ defined in eq.~\eqref{eq:MonoOpacity}. 

The opacity is a crucial ingredient for solar modelling since it determines the amount of energy transport by radiative diffusion. Accordingly, large efforts have been made to compute opacities of stellar plasmas (see refs.~\cite{1965ApJS...11...22C,Rogers:1992ud,astro-ph/0410744,1995_LEDCOP,2012_OPAS,Krief:2016znd} for an overview of works in this direction). These calculations include the aforementioned atomic structures as well as plasma effects. Following Redondo we can therefore avoid a detailed modelling of atomic transitions by relating these results to the solar axion emission.

In a first step, the production and absorption of photons from interactions with ions can be related to one another by detailed balance under the assumption of a local thermal equilibrium~\cite{1310.0823},
\begin{equation}
    \Gabs^{i} \, \ee^{-\omega/T}= \Gprod^{i} \, . \label{eq:detailed_balance}
\end{equation}
Here, $\Gabs^{i}$ and $\Gprod^{i}$ are the absorption and production rate per photon phase space for process~$i$~(with $i = \text{bb},\text{fb},\text{ff}$), respectively. The factor $\ee^{-\omega/T}$ is required in order to have the same overall production and absorption rates per solar volume, taking into account stimulated emission as well as the thermal occupation number. As discussed in ref.~\cite{1310.0823}, detailed balance implies that eq.~\eqref{eq:detailed_balance} is satisfied for all processes individually. All the following rates will be production interaction rates, which is why we drop the respective subscripts.

In a second step, the ratio of axion relative to spin averaged photon production for each of the three processes can be calculated via~\cite{1310.0823}
\begin{equation}
    \frac{\Gamma^{i}_a}{\Gamma^{i}_\gamma} = \frac{1}{8\pi}\frac{\gaee^2\omega^2}{\alphaEM m_e^2} \, .
    \label{eq:axion_photon_ratio}
\end{equation}
Combining these relations, Redondo was able to express the total axion emission rate $\Gae$ in terms of the monochromatic absorption coefficient $k(\omega)$ as well as the Compton and $ee$~bremsstrahlung rates~\cite{1310.0823},
\begin{align}
    \Gae  &\equiv  \Gamma^{\text{ff}}_a +\Gamma^{\text{fb}}_a + \Gamma^{\text{bb}}_a +\Gamma^{\text{C}}_a +  \Gamma^{\text{$ee$}}_a \label{eq:abc_rate} \\
    &= \frac{1}{8\pi} \frac{\gaee^2 \omega^2}{\alphaEM m_e^2} \frac{k(\omega)}{\ee^{\omega/T}-1} + \frac{1}{2}\frac{\ee^{\omega/T}-2}{\ee^{\omega/T}-1}\Gamma^{\text{C}}_a + \Gamma^{\text{$ee$}}_a \label{eq:totalrate} \, .
\end{align}

We will use this final expressions together with eqs.~\eqref{eq:CRate} and~\eqref{eq:eeRate}, whenever we work with the total monochromatic opacity of the solar plasma. However, when monochromatic opacities are available for each element individually, we follow the suggestion of Redondo to calculate the ff~processes involving hydrogen and helium directly with eq.~\eqref{eq:ffRate} and only include the opacities from elements heavier than helium in the first term of eq.~\eqref{eq:totalrate}. In this case, also  the full Compton  rate has to be included.

\subsection{Electron densities and ionisation states}
\label{sec:electron-densities}
We saw in sections~\ref{sec:primakoff} to~\ref{sec:axionelectron_interactions} that the overall axion production rate can be computed from the monochromatic opacity and a small set of plasma parameters. These parameters are the temperature~$T$, the number density of free electrons~$n_e$, and the number density of ions weighted by their charge squared~$\bar{n}$. Publicly available solar models typically only contain the values of $T$ at every radius, as well as other solar properties and element abundances. The values of $n_e$, $\bar{n}$, $\ks$, and $\kappa$ are usually not tabulated, and hence need to be computed.

\begin{figure}
	\centering
	\includegraphics[width=6in]{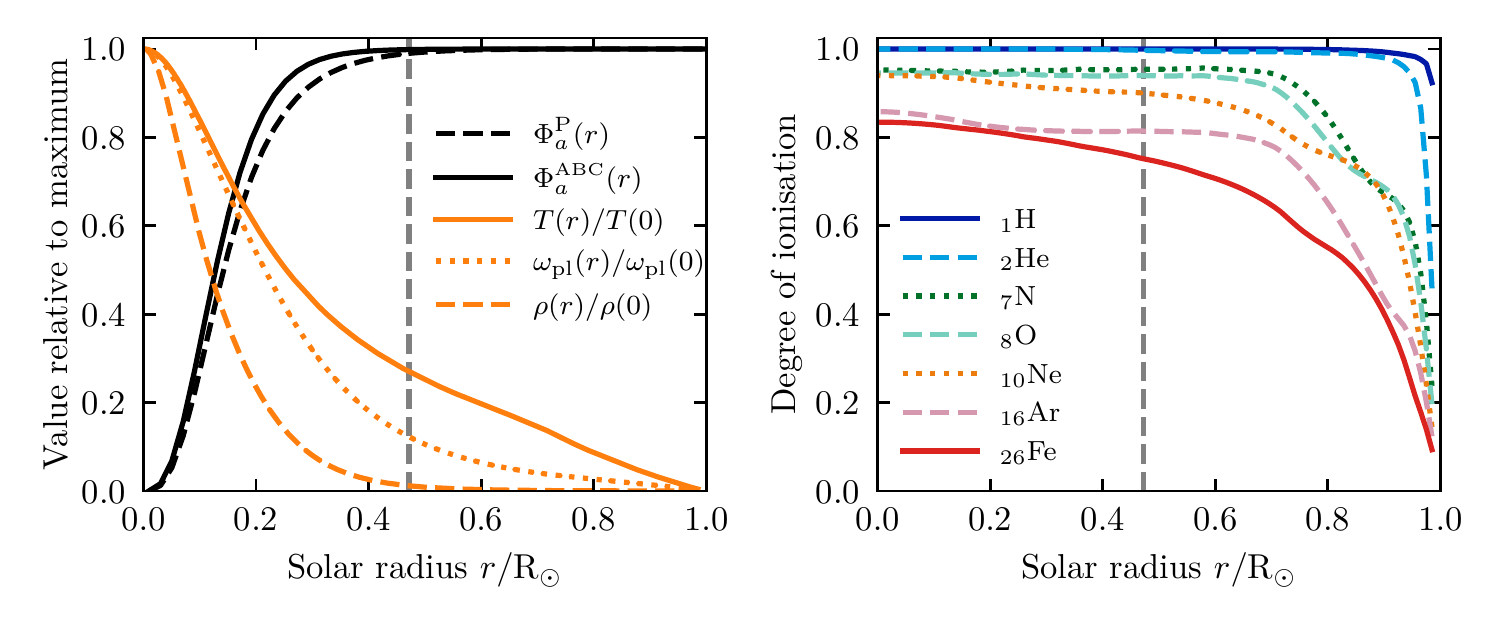}
	\caption{Overview of various properties as a function of solar radius for the B16-AGSS09 solar model~\cite{1611.09867}. \textit{Left:} The integrated Primakoff~($\Phiag$) and \ABC~fluxes~($\Phiae$) in addition to the solar temperature~($T$), density~($\rho$), and plasma frequency~($\ompl$) w.r.t.\ their maximum values. \textit{Right:} The degree of ionisation for various elements, including those with the largest mass fractions at the core (full ionisation corresponds to a value of~1; extracted from the OP code~\cite{astro-ph/0410744,astro-ph/0411010}). The vertical, dashed grey lines in both panels correspond to the radius by which 99\% of the total Primakoff and \ABC fluxes have been emitted.\label{fig:solar_radius_dependence}}
\end{figure}
An overview of solar properties as a function of radius and the integrated solar axion fluxes is shown in \reffig{fig:solar_radius_dependence}. Note already that most axions from Primakoff and \ABC fluxes are generated inside the solar core ($r/\Rsol < 0.25$) and almost the entire flux is emitted until about $r/\Rsol = 0.5$. The harder Primakoff spectrum is produced even more closely to the centre than the \ABC~flux. This will be relevant for the discussion of ionisation stages in this section.

We consider two options for the calculation of the electron density. First, Raffelt~\cite{Raffelt:1985nk} approximated $n_e$ by calculating the number of nucleons and assuming that there is one free electron per hydrogen core and one free electron per every two nucleons of all other elements i.e.
\begin{equation}
    n_e \approx \frac{X+1}{2} \frac{\rho}{\amu} \, , \label{eq:neRaff}
\end{equation}
where $\amu \approx 1820\,m_e \approx \SI{0.93}{\GeV}$ denotes the atomic mass unit and $X$ denotes the hydrogen mass fraction.\footnote{We adopt the commonly used notation of denoting the mass fraction of all hydrogen isotopes as~$X$ and the mass fraction of all helium isotopes as~$Y$, which in turn defines the \textit{metallicity} $Z \equiv 1-X-Y$.} This approximation works extremely well since it is correct for all of the (fully ionised) hydrogen and helium in the Sun which is already more than 98\% of its total density. Heavier elements also contribute almost one electron per two nucleons when they are fully ionised.

Second, since our solar models give the mass fractions of all relevant heavy elements, we can directly assume full ionisation and sum over the atomic number~$z$,
\begin{equation}
    n_e \approx\sum_z z \, n_z = \sum_z z \, X_z \, \frac{\rho}{A_z \amu} \, , \label{eq:neFullion}
\end{equation}
were $X_z$ is the mass fraction of each element and $A_z$ is its standard atomic weight.\footnote{We use the exact weight where isotopes are individually tracked in the solar abundance and average (solar or terrestrial) weight where elements are only tracked in bulk~(cf.\ ref.~\cite{Asplund:2009fu,1810.07192}).} This agrees with Raffelt's approximation at the sub-percent level everywhere inside the Sun.\footnote{Note that there is a third option for approximating $n_e$ by deducing it from the pressure of the plasma i.e.\ $n_e \approx (p - p_\text{rad})/T - {\scriptscriptstyle \sum}_z^{} n_z$ with the radiation pressure $p_\text{rad} = \pi^2 T^4/45$. The problem with this calculation is that it assumes the ideal gas equation for electrons in the plasma, which only becomes reasonable close to the core but is always far from exact.}

Since the mass fractions in our solar models add up to exactly one, we know that both the full ionisation result~\eqref{eq:neFullion} and Raffelt's approximation~\eqref{eq:neRaff} are larger than the true value of~$n_e$. We hence use the smaller of the two, i.e.\ the full ionisation result in eq.~\eqref{eq:neFullion}, as our default option.

Full ionisation is a particularly good approximation for $n_e$ close to the core~(for high temperatures) where most of the axion flux is generated.\footnote{We can estimate the systematic uncertainty of this calculation by doing a {na\"ive} comparison of ionisation energies with the thermal energy in the core of the Sun. For instance, the last ionisation stage from \ce{Fe^{25+}} to \ce{Fe^{26+}} requires around \SI{9.3}{\keV}~\cite{NIST_ASD} compared to $T \approx \SI{1.3}{\keV}$ at the centre of the Sun, see also right panel of figure~\ref{fig:solar_radius_dependence}.} Because metals contribute some fraction smaller than the metallicity $Z$ to the total number of free electrons and their ionisation fraction close to the core is of the order of 90\%, we can conservatively estimate that the relative error of the full ionisation assumption should be smaller than 0.1\%. This renders it negligible compared to the intrinsic uncertainties of the solar models themselves. 

The next parameter of interest, $\bar{n}$, is most significant for the Primakoff flux which we expressed in terms of the Debye screening scale in eq.~\eqref{eq:PrimaRate}, but it also enters the \ABC~flux through the function $\mathcal{I}$ in eq.~\eqref{eq:aux_fun_F}. To calculate $\bar{n}$, Raffelt used a similar approximation as the one for $n_e$ in equation~\eqref{eq:neRaff}~\cite{Raffelt:1985nk},
\begin{equation}
	\bar{n} = \sum_z Q_z^2 \, n_z  = \sum_z Q_z^2 \, X_z \, \frac{\rho}{A_z \, \amu} \approx \frac{\rho}{\amu} \, . \label{eq:nbar_raffelt}
\end{equation}

This approximation is, however, significantly less accurate than the previous one since the densities of heavy nuclei enter weighted by the square of their charges. Consequently, the metallic components become more significant and treating all metals as if they were helium-like underestimates the overall result. As already discussed above, the energy required to fully ionise heavy elements far exceeds the temperature at the solar core. This means that the assumption of full ionisation, which worked well for $n_e$, would be problematic for calculating $\bar{n}$. To make matters worse, the metallicity and the abundances of each metal are poorly known parameters.

To resolve this issue, \updated{we follow ref.~\cite{1310.0823} and} make use of data from the Opacity Project~(OP)~\cite{astro-ph/0411010}. Besides calculating opacities for arbitrary chemical compositions of Sun-like plasmas, the OP also provides tables of ionisation states for heavy ions in such plasmas. These tables can be interpolated to calculate $Q_z^2$ as a function of $T$ and $n_e$ (cf.\ right panel of figure~\ref{fig:solar_radius_dependence}). Due to the dependence on $n_e$, we could not use these tables for calculating $n_e$ itself. We can however make a consistency check by calculating $n_e$ using OP ionisation states with $n_e$ from full ionisation as input and compare it with the full ionisation result. We find that the relative difference does not exceed 0.2\% in the axion production region which is small enough for our purposes.\footnote{The OP ionisation tables agree well with a similar calculation performed in the context of another opacity code named STAR~\cite{Krief:2016znd}.}

Some solar models (BP and BS models in table~\ref{tab:solar_models_overview}) do not track the mass fractions $X_z$ for all relevant elements. In these cases we resort to calculating $\bar{n}$ under the assumption of full ionisation. 
As we have discussed above, this is a rough approximation. The use of these models should therefore be disfavoured for the prediction of the solar axion flux.

\subsection{Integration over the solar model}\label{sec:integrate}
Helioscopes can track the solar disc throughout the day, and the year, as the Sun moves across the sky. Since the (variable)~Sun-Earth distance~$\dE$ is much larger than the solar radius, $\Rsol/\dE \approx 0.005$, a reasonable first approximation is to treat the Sun as a point source~(an assumption we will later relax).

Using this approximation, the total spectral axion flux ${\diff \Phi_\ax/\diff \omega}$ can be written as an integral of the production rate $\Gamma$ over the solar volume~$V$,
\begin{equation}
    \frac{\diff \Phi_\ax}{\diff \omega} = \frac{1}{4\pi \dE^2} \, \int \! \diff V \, \frac{4\pi \omega^2}{(2\pi)^3} \, \Gamma(\omega,r) \, , \label{eq:integration_master}
\end{equation}
where $\dE$ is the (averaged) distance between the Sun and the Earth during the observation and $\Gamma$ can be any~(or the sum of all) of the production rates discussed earlier in this section. All axion fluxes in this work use $\dE = \SI{1}{\astrounit} \approx \SI{1.496e11}{\m}$~\cite{NASA}.

Note that eq.~\eqref{eq:integration_master} assumes that ALPs are effectively massless as far as the solar plasma is concerned i.e.\ ALPs with masses $\ma \ll \mathcal{O}(\si{\keV})$. Otherwise, eq.~\eqref{eq:integration_master} would have to be modified to include the axion dispersion relation between $k$ and $\omega$ (see e.g.\ equation~(7) of ref.~\cite{hep-ex/0702006}).

Since all quantities only depend on the position inside the Sun, the integration reduces to a one-dimensional integral,\footnote{The macroscopic magnetic field in eq.~\eqref{eq:solar_b_fields} is not spherically symmetric. Hence, the projected magnetic fields $B_\perp$ and $B_\parallel$ have to be averaged over plasmon directions and the spherical shell inside the Sun, as was done in refs.~\cite{2005.00078,2006.10415,2010.06601}. \updated{The integration of the plasmon fluxes for specific geometries is currently not fully supported in our code -- e.g.\ over parts of the solar disc -- and the results cannot be simply rescaled for more complex geometries.}}
\begin{equation}
\frac{\diff \Phi_\ax}{\diff \omega} = \frac{\Rsol^3}{4\pi \dE^2} \, \frac{4\pi \omega^2}{(2\pi)^3} \, \int_{0}^{2\pi} \! \diff \phi \, \int_{0}^{\pi} \! \diff \theta \, \sin(\theta) \, \int_{0}^{1} \! \diff r \, r^2 \, \Gamma(\omega,r) = \frac{\Rsol^3 \omega^2}{2\pi^2 d_E^2}\int_{0}^{1} \! \diff r \, r^2 \, \Gamma(\omega,r) \, ,
\end{equation}
where we express the radius~$r$ in units of the solar radius~$\Rsol$.

By employing good optics and observing a sufficiently large number of photon events, helioscopes may also resolve different regions of the solar disc. Therefore, another interesting geometry for the solar integration is to consider only parts of the solar disc e.g.\ the central part of the solar disc. Neglecting parallax~\cite{hep-ex/0702006},\footnote{We may neglect parallax because $\Rsol/d_\text{E} \approx 0.005$ and also because the outer parts of the Sun, where parallax is more pronounced, are less relevant for the calculation of the solar axion flux, cf.\ Fig~\ref{fig:solar_radius_dependence}.} the line-of-sight~(LOS) integration corresponds to a cylindrical field-of-view. This means that we have to perform an additional integration as we break one rotational degree of freedom. If~$\rho$ is the radius on the solar disc~(from its centre) and~$r$ the position inside the Sun, we have to integrate along the LOS coordinate~$z$ with $r^2 = \rho^2 + z^2$ and $|z| < z_\text{max} = \sqrt{1 - \rho^2}$. If we integrate between $\rho \in [0,\rho_1]$, we find
\begin{align}
	\frac{\diff \Phi_\ax}{\diff \omega} &= \frac{\Rsol^3 \omega^2}{(2\pi)^3\dE^2} \, \int_{0}^{2\pi} \! \diff \phi \, \int_{0}^{\rho_1} \! \diff \rho \, \rho \, \int_{-z_\text{max}(\rho)}^{z_\text{max}(\rho)} \! \diff z(\rho) \; \Gamma(\omega,r)\\
	&= \frac{\Rsol^3 \omega^2}{(2\pi)^2\dE^2} \, \int_{0}^{\rho_1} \! \diff \rho \, \rho \, \left( 2\int_{\rho}^{1} \! \diff r \, \left|\frac{\diff z}{\diff r}\right| \, \Gamma(\omega,r) \right) \\
	&= \frac{\Rsol^3 \omega^2}{2\pi^2\dE^2} \, \int_{0}^{\rho_1} \! \diff \rho \, \rho \, \int_{\rho}^{1} \! \diff r \, \frac{r}{\sqrt{r^2 - \rho^2}} \, \Gamma(\omega,r) \, ,
\end{align}
where the factor of~two is due to the $\mathbb{Z}_2$~symmetry of the~$z$ direction inside the Sun.

\section{Solar models and opacity codes}\label{sec:solarmodelscodes}
\begin{table}
	\caption{Overview of solar models, ordered by their publication dates. Green ticks and red crosses denote, respectively, whether or not the models agree with helioseismological~(``S''; high $Z$) or photospheric measurements~(``Ph''; low $Z$). In BP and BS models only a few metal abundances are tabulated. These can therefore only agree with the photospheric measurements of these particular metals~(tick in parentheses). $Z$~denotes the total mass fractions of metals, which is larger than the surface metallicity (see refs.~\cite{Serenelli:2009yc,0910.3690}) due to diffusion and gravitational settling~\cite{Asplund:2009fu}.\label{tab:solar_models_overview}}
	\renewcommand{\arraystretch}{1.15}
	\centering
	\begin{threeparttable}
	    {\small
		\begin{tabular}{llcccl}
			\toprule
			\textbf{Solar model} & \textbf{Comments} & \textbf{S} & \textbf{Ph} & \textbf{$\boldsymbol{Z}$} & \textbf{References} \\
			\midrule
			BP98 & no heavier metals tracked & \cmark & \xmark & 0.0201 & \citeref{astro-ph/9805135}\tnote{a} \\
			\midrule
			BP00 & no heavier metals tracked & \cmark & \xmark & 0.0188 & \citeref{astro-ph/0010346}\tnote{a} \\
			\midrule
			BP04 & no heavier metals tracked & \cmark & \xmark & 0.0188 & \citeref{astro-ph/0402114}\tnote{a} \\
			\midrule
			BS05-OP & no heavier metals tracked & \cmark & \xmark & 0.0190 & \citeref{astro-ph/0412440}\tnote{a} \\
			BS05-AGSOP & no heavier metals tracked & \xmark & \cmarkvar & 0.0141 & \\
			\midrule
			AGS05 & &  \xmark & \cmark & 0.0140 & \citeref{Serenelli:2009yc}\\
			\midrule
			GS98 & & \cmark & \xmark & 0.0188 &  \citerefs{Serenelli:2009yc,0910.3690}\tnote{b} \\
			AGSS09 & same as AGSS09met & \xmark & \cmark & 0.0150 & \\
			AGSS09ph & & \xmark & \cmark & 0.0152 & \\
			\midrule
			B16-GS98 & updated version of GS98 & \cmark & \xmark & 0.0188 & \citeref{1611.09867}\tnote{c}\\
			B16-AGSS09 & updated version of AGSS09(met) & \xmark & \cmark & 0.0150 & \\
			\bottomrule
		\end{tabular}
		}
		\begin{tablenotes}\footnotesize
			\item[a]Model files available at \href{http://www.sns.ias.edu/~jnb/SNdata/solarmodels.html}{this URL}
			\item[b]Model files available at \href{https://wwwmpa.mpa-garching.mpg.de/~aldos/solar\_main.html}{this URL}
			\item[c] Model files available at \href{http://www.ice.csic.es/personal/aldos/Solar\_Data.html}{this URL}
		\end{tablenotes}
	\end{threeparttable}
\end{table}
Let us first discuss the publicly available solar models considered in this work, which we list in table~\ref{tab:solar_models_overview}. Importantly, note that some solar models provide less information than others on the abundances of heavier solar metals, e.g.\ \ce{Fe}, which is relevant for axion-electron interactions. Another observation is that all of the available solar models fail to consistently explain both seismological observations and photospheric measurements. This is known as the solar metallicity or abundance problem~\cite{Bahcall:2004yr,Antia:2005mg,PenaGaray:2008qe,Asplund:2009fu,Serenelli:2009yc,Mendoza:2018iyx} in the literature, which has not yet been resolved.

\begin{table}
	\caption{Overview of opacity codes, ordered by their publication dates. Green ticks and red crosses denote whether or not it is possible to calculate opacities for arbitrary solar composition~(``Arb.\ comp.''). Note that with the LEDCOP and ATOMIC codes, this is only possible via an online interface~(tick in brackets).\label{tab:opacity_codes_overview}}
	\renewcommand{\arraystretch}{1.15}
	\centering
	\begin{threeparttable}
	    {\small
		\begin{tabular}{llcl}
			\toprule
			\textbf{Opacity codes} & \textbf{Comments} & \textbf{Arb.\ comp.} & \textbf{References} \\
			\midrule
			OP & & \cmark & \citerefs{astro-ph/0410744,astro-ph/0411010} \\
			\midrule
			LEDCOP & & \cmarkvar &  \citeref{1995_LEDCOP}\tnote{a} \\
			ATOMIC & & \cmarkvar &  \citeref{1601.01005}\tnote{a}\\
			\midrule
			OPAS & Only for AGSS09 comp. & \xmark & \citerefs{2012_OPAS,2015_OPAS}\tnote{b} \\
			\bottomrule
		\end{tabular}
		}
		\begin{tablenotes}\footnotesize
			\item[a] See \href{https://aphysics2.lanl.gov/apps/}{https://aphysics2.lanl.gov/apps/}
			\item[b] 
			The monochromatic opacities for the AGSS09 composition used in this work are not publicly available. We thank C.~Blancard and the OPAS collaboration for kindly making the additional data available to us.
		\end{tablenotes}
	\end{threeparttable}
\end{table}
Table~\ref{tab:opacity_codes_overview} shows an overview of all opacity codes that we have considered. Note that we could only use the Opacity Project~(OP) code to easily accommodate a solar model with arbitrary metal abundances. The LEDCOP and ATOMIC codes also work for opacities of arbitrary mixtures but they can only be generated via an online interface which is unfeasible for the very large number of different models which we considered. The OPAS collaboration only provided opacities for the abundances in the AGSS09~model.

\begin{figure}
	\centering
	\includegraphics[width=5in]{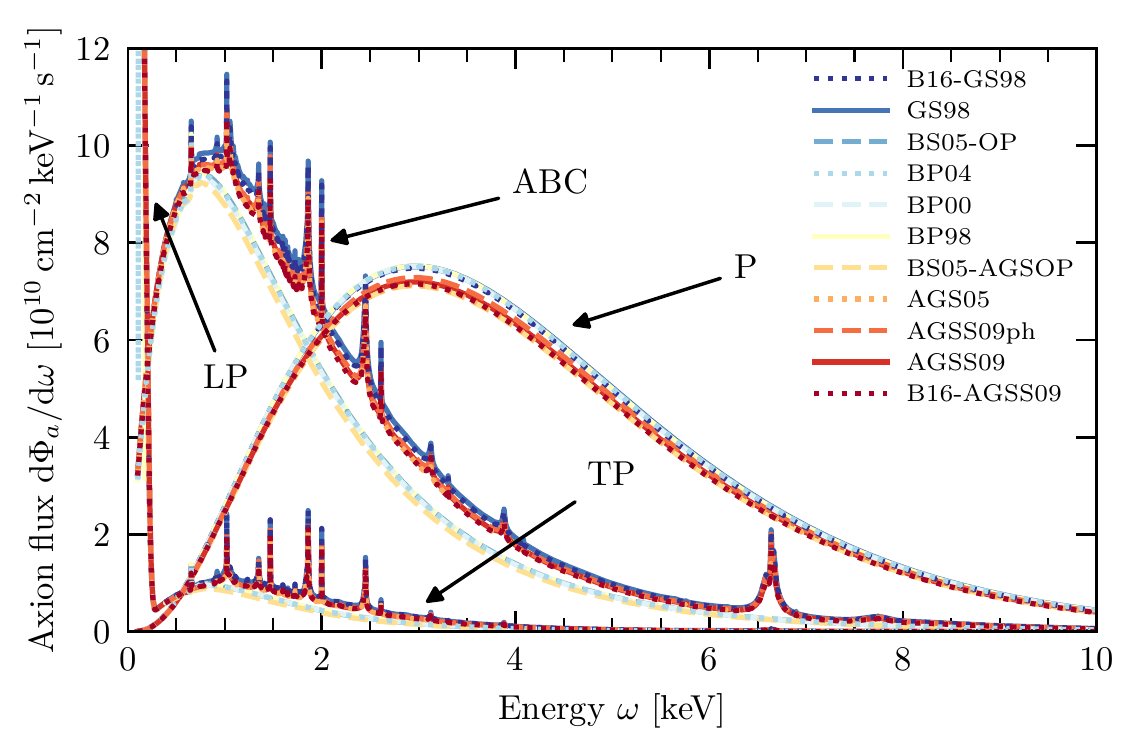}
	\caption{Overview of solar axion fluxes for different solar models between \num{0.1}~and \SI{10}{\keV}, using opacities from the OP~code and a stepsize of $\Delta \omega = \SI{1}{\eV}$. We show the \ABC fluxes for $\gaee = \num{e-12}$ and the Primakoff~(P) as well as combined LP and TP fluxes for $\gagg = \SI{e-10}{\GeV^{-1}}$. \label{fig:solar_models_overview}}
\end{figure}
To illustrate how differences in the solar models of table~\ref{tab:solar_models_overview} manifest themselves with regard to axion production inside the Sun, we show the corresponding axion fluxes in figure~\ref{fig:solar_models_overview}. Since the BP~and BS~models are missing information on the abundance of heavier metals, the flux from axion-electron interactions~($\Phiae$) is missing contributions from these high-$Z$ constituents. As a consequence, the smooth part of the spectrum is \emph{smaller} than the B16-AGSS09 model by about 20--40\% for most parts of the spectrum. Moreover, the peaks from these heavier metals are absent, and these distinct features could therefore not be compared to observation.

Clearly, the lack of information on the heavier metals makes these models unsuitable for investigating the axion-electron interactions.

For the Primakoff~spectrum~($\Phiag$) the deviations between the BP/BS~models and all other models enter through differences in the Debye screening~scale. The spectral flux can be up to about 12.5\%~\emph{larger} than the Primakoff flux from e.g.\ the B16-AGSS09 model. While the differences are not as striking as in the $\gaee$~case, the lack of information on the heavier metals still causes a relatively large systematic difference.
\begin{figure}
	\centering
	{
		\hfill
		\includegraphics[width=3in]{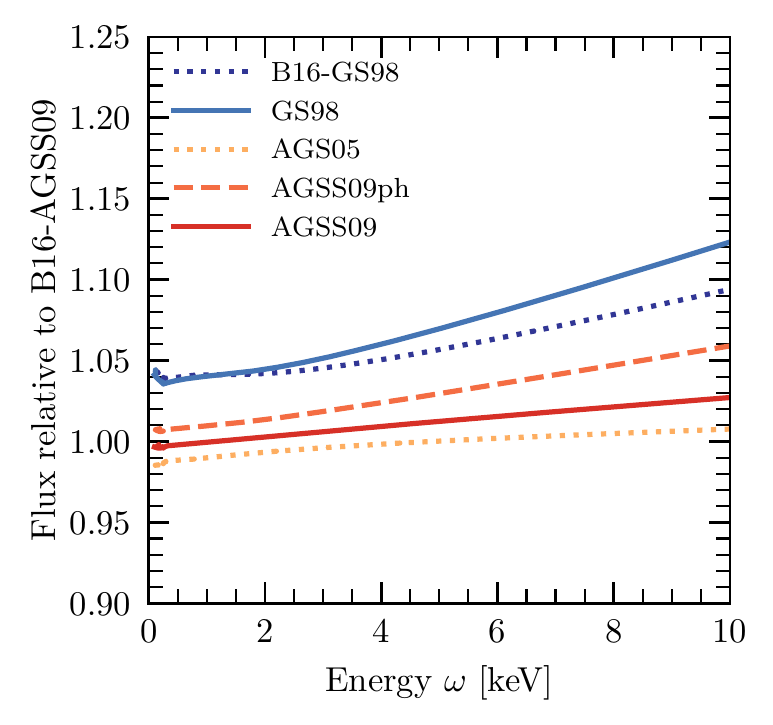}
		\hfill
		\includegraphics[width=3in]{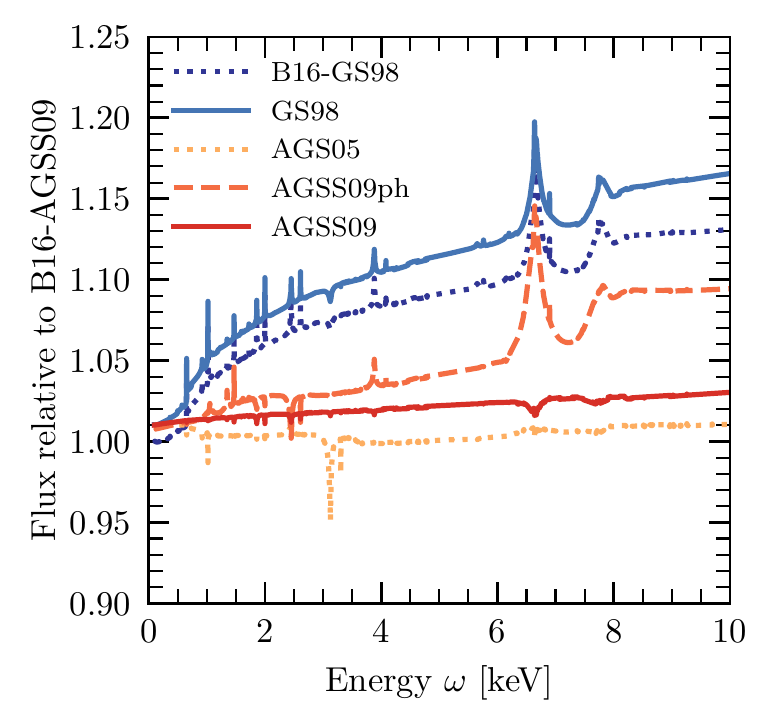}
		\hfill
	}
	\caption{Relative deviations in the axion flux from different solar models~(compared to the B16-AGSS09 model) for axion-photon interactions~(left panel) and axion-electron interactions~(right panel).\label{fig:solar_models_deviations}}
\end{figure}
Due to these shortcomings, from now on we will not consider the BP~and BS~models investigating solar models in the context of axion physics. In figure~\ref{fig:solar_models_deviations} we only show the deviations for the models that provide the most complete information on individual isotope abundances in order to get a better idea of what the systematic differences between competing solar models is. For the Primakoff flux, we see that deviations are typically a few, but no more than about~10\%. The differences become more pronounced at higher energies. For $\gaee$~interactions, we observe a similar trend with deviations of up to about~15\%.

\begin{figure}
	\centering
	{
		\hfill
		\includegraphics[width=3in]{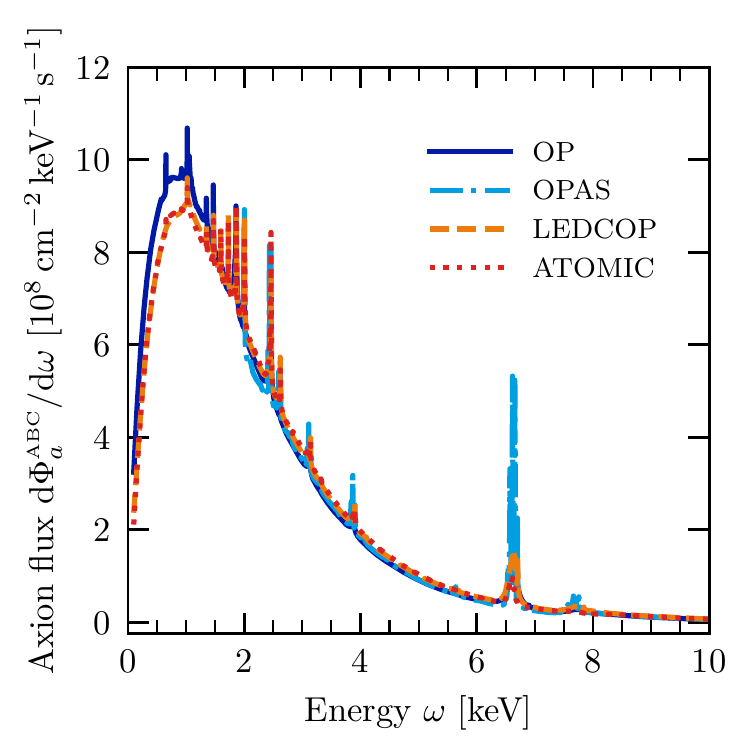}
		\hfill
		\includegraphics[width=3in]{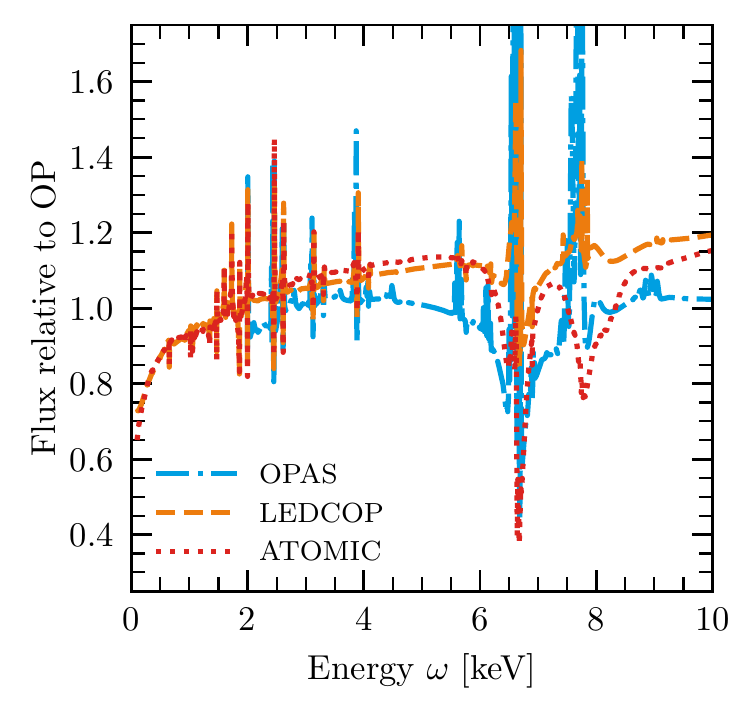}
		\hfill
	}
	\caption{\ABC flux (left panel) and relative deviations compared to the OP code (right panel) for the AGSS09~model and different opacity codes. Note that, in the right panel, for OPAS we do not show the huge deviations~(up to 440\%) for the \ce{Fe}~peak around~\SI{6.5}{\keV} and also not the low-energy region, for which the this code may not be fully accurate~\cite{1310.0823}.\label{fig:opacity_codes_deviations}}
\end{figure}
Analogously, figure~\ref{fig:opacity_codes_deviations} summaries how different opacity codes affect the solar axion flux, using the AGSS09~model as a reference solar model.\footnote{Recall that the capacities from the OPAS code are based on the AGSS09 composition, which is hence the only possible choice for a consistent comparison.} The outcome of this comparison is rather diffuse, with differences of the 10--20\% level being typical across the whole energy range and huge deviations around some of the peaks in the $\gaee$~spectrum. The latter are particularly important for studying the metal content of the Sun with suitable energy resolution~\cite{1908.10878}, but become less of an issue for integrated flux measurements i.e.\ experiments with an energy resolution somewhat larger than the width of the peaks. Due to the flexibility offered by the OP~code, it is our preferred choice for calculating the opacities. However, it should be kept in mind that a different choice of opacity code can significantly affect the outcome of a solar axion study. A~very detailed and thorough investigation of the different opacity codes in the context of axions has already been provided by ref.~\cite{1310.0823}.

\section{Intrinsic uncertainties in solar models and opacity codes}\label{sec:uncertainties}
While it is interesting to compare how the solar axion flux differs for various solar models and opacity codes, we want to arrive at a more complete representation of the total uncertainty. To do this, we need to investigate the intrinsic, statistical uncertainties associated with a given combination of solar model and opacity code. Since only the OP~code allows for full flexibility regarding different solar compositions, it is the natural choice as a benchmark for the opacity calculation.

In this section we focus on the uncertainties arising from the modelling of the solar plasma properties. 
Our baseline is the Primakoff flux~\eqref{eq:PrimaRate_fullintegral} including the electron degeneracy factor~\eqref{eq:degen_correction}, and the full \ABC~flux (without Pauli blocking)~\eqref{eq:abc_rate}.
For simplicity we do not include the flux from plasmon conversions in the magnetic field which are much more strongly affected by the uncertainties in the modelling of the solar magnetic fields and which is anyway expected to be subdominant in the 1--\SI{10}{\keV} range, cf.\ figure~\ref{fig:plasmons}. 

\subsection{Solar model uncertainties from Monte Carlo simulations}
To explore how uncertainties of solar models translate into uncertainties of the solar axion flux, we use results produced by the authors of ref.~\cite{astro-ph/0511337}, who generate a sample of about 10,000 solar models from a Monte Carlo~(MC) simulation.\footnote{We thank Aldo Serenelli for providing us with new MC samples for 9978~(9971) converged AGSS09~(GS98) solar models.} 

As mentioned earlier, the solar abundance problem~\cite{Bahcall:2004yr,Antia:2005mg,PenaGaray:2008qe,Asplund:2009fu,Serenelli:2009yc,Mendoza:2018iyx} currently still persists, with models from the high-$Z$ as well as the low-$Z$ category being favoured by helioseimology and photospheric observations, respectively. To cover both extremes, we pick representative models from both categories.

The authors of ref.~\cite{astro-ph/0511337} generate MC samples in the following manner: the Garching stellar evolution code~\cite{Weiss2008} is employed to evolve an initially homogeneous star over the full lifetime of the Sun. By varying three free parameters (initial helium abundance, initial metallicity, and mixing length parameter), one can ensure that the model converges to the desired present solar properties (luminosity, radius, and chemical composition at the surface) within a relative accuracy of~\num{e-4}, ending up with a realistic model for the current state of the Sun.

Apart from the three free parameters, which are varied in order to fulfill the criteria of convergence, this procedure depends on a total of 21~input parameters, which can have a strong effect on the final solar model. These include rates of nuclear fusion reactions, solar properties (e.g.\ luminosity, age, and diffusion coefficient), chemical composition as well as opacities and equations of state. All of these inputs come with their own uncertainty, which can be expressed in the form of a set of independent probability distributions, one for each input parameter. For the 10,000 solar models in~\cite{astro-ph/0511337} the input parameters are drawn from these probability distributions, before applying the procedure summarised above. This procedure ensures the generation of a statistically representative sample of solar models.

The authors of ref.~\cite{astro-ph/0511337} use this sample to quantify uncertainties of the solar neutrino fluxes, helioseismological quantities, and other characteristic values such as the core temperature. This demonstrates the flexibility of this approach and we can readily apply it to solar axion fluxes.

For our study, we use an updated set of around 10,000 realisations of each the AGSS09 and GS98 solar model. By calculating all of the corresponding solar axion spectra as described in section~\ref{sec:equations}, we find the statistical uncertainty of the axion flux directly from the statistical uncertainties of the input parameters used to build a solar model. The MC samples for solar models automatically take into account all possible dependencies between the solar parameters that enter our axion flux calculation.

\subsection{Opacity code uncertainties}
As discussed, the \ABC flux requires external information in form of monochromatic opacities, which depend on~$T$ and $n_e$. We illustrated in figure~\ref{fig:opacity_codes_deviations} that using different opacity codes can in general result in significantly different powers of the flux. Even though some of these deviations can be discarded as unphysical~\cite{1310.0823}, the question remains how we can estimate the uncertainty of the monochromatic opacity. 

Recall that we choose the OP code for our uncertainty estimation since it allows us to compute opacities for arbitrary chemical compositions, and therefore to use eq.~\eqref{eq:ffRate} for all ff~processes involving \ce{H} and \ce{He}. It also allows us to calculate opacities for each mixture of the $2 \times 10,000$ solar models individually. Furthermore, the OP does not make assumptions or include effects that are fundamentally not applicable to axion production, which is also why Redondo picked it for his ``best~guess'' result~\cite{1310.0823}. The OP does not supply its own uncertainty estimate for monochromatic opacities. However, recent efforts in solar modelling have carefully analysed uncertainties of their opacity input~\cite{1611.09867}. They found that opacity uncertainties are much larger at the bottom of the convective zone~(CZ) compared to the solar centre~(i.e.\ at $r=0$). This led them to the idea of a temperature-dependent relative opacity uncertainty $\delta\kappa(T)$, which we too adopt for our MC simulation. The authors of ref.~\cite{1611.09867} use a logarithmic interpolation between the uncertainty level at the core~(2\%) and at the bottom of the convection zone~(7\%). This can be achieved by modifying the opacity $\kappa_\text{OP}$ as calculated from the OP to 
\begin{equation}
    \kappa (\omega;\, T,\, n_e) = \kappa_\text{OP}(\omega;\, T,\, n_e) \, \left(1 + \delta \kappa (T)\right) \, ,
\end{equation}
where 
\begin{equation}
    \delta \kappa (T) = a + b \, \frac{\log_{10}\left(T_0/T\right)}{\log_{10}\left(T_0/T_\text{CZ}\right)} \, , \label{eq:opacity_variation}
\end{equation}
with $\log_{10}(T_0/T_\text{CZ}) \simeq 0.9$ and where we take $T_0 \equiv T(r=0)$ directly from the solar model. With these definitions, we can include opacity uncertainties into our MC simulation by drawing the parameters $a$ and $b$ from normal distributions with $\sigma_a = 0.020$ and $\sigma_b = 0.067$~\cite{1611.09867}.

A significant difference between our application and solar modelling, for which this parameterisation of the opacity uncertainty was developed, is that we are working with monochromatic opacities. In solar modelling, one is mostly interested in radiative energy transport and hence only frequency averaged opacities such as the Rosseland mean opacity are required~(as e.g.\ in ref.~\cite{1611.09867}). One may therefore worry if simply taking the uncertainty of the mean as the uncertainty at every energy $\omega$ is valid. One reason why we think it should be reasonable is that we indeed vary the spectral shape of the monochromatic opacity but merely through its dependence on the chemical composition, which is different for each of the 10,000 solar models.

Another reason is that the composition is already one of the main sources of the uncertainty in the spectrally-averaged opacity. By varying both composition and overall opacity, we are therefore more likely to over- than to underestimate the axion flux uncertainty. However, this only holds true for smooth parts of the spectrum, where it is reasonable to expect that the uncertainty of the monochromatic opacity contribution from one element is not much larger than the uncertainty of the Rosseland mean. The story is very different at and around the peaks from bb~transitions in the atomic shell. At these energies, the different opacity codes produce noticeably diverging results. This issue was already discussed in the context of the solar abundance problem~\cite{Mendoza:2018iyx} but it also has consequences for solar axion searches: for example, even if the peaks can be experimentally resolved, the uncertainty in the opacity calculation causes problems for abundance measurements of solar metals~\cite{1908.10878}. Since this is an unresolved issue, we can still only estimate the uncertainty in the solar axion flux around the peaks by comparing different opacity codes. Note, however, that the influence of this uncertainty on the \emph{integrated} axion flux is small since the power in all of the peaks is negligible -- the only possible exception being the iron peak at $\sim \SI{6.5}{\keV}$, which can constitute up to 1\%~of the total flux. Hence, this locally large uncertainty in the spectrum is irrelevant for many solar axion searches with limited spectral resolution.

\subsection{Results}
\begin{figure}
	\centering
    \includegraphics[width=6in]{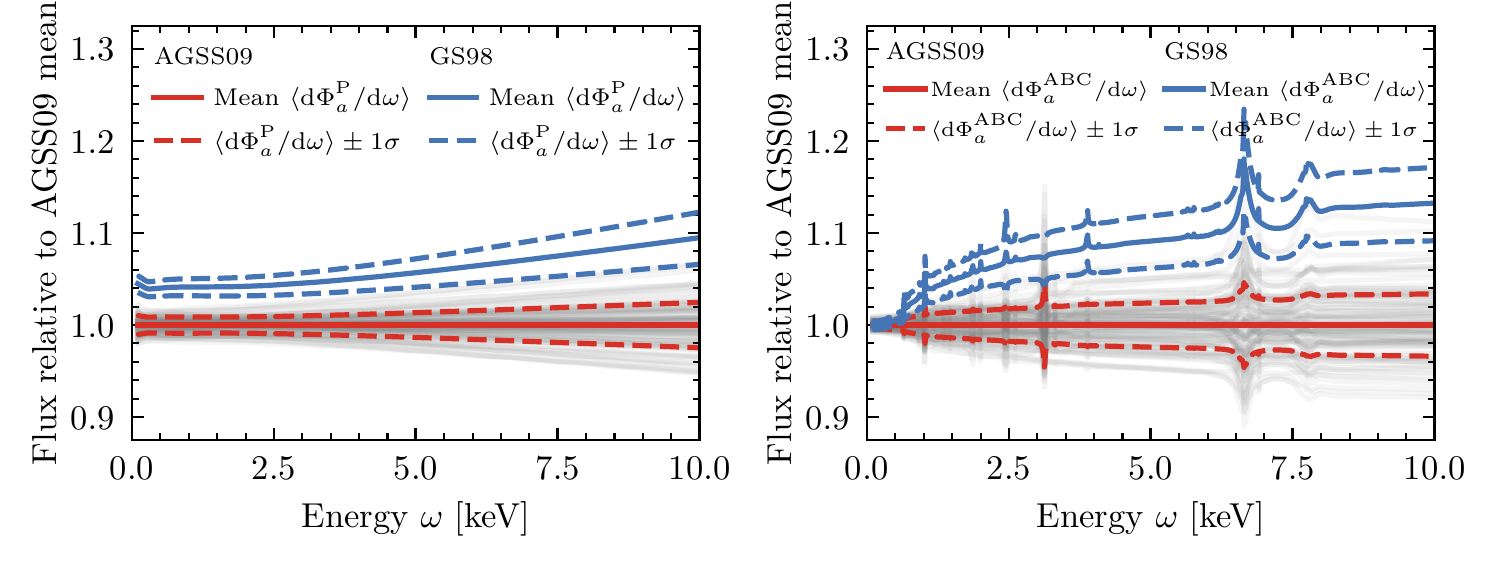}
	\caption{Solar axion spectra obtained from Monte Carlo simulations, relative to the mean value of the AGSS09 model. We show the mean values~(solid line) and $\pm 1\sigma$ bands~(dotted lines) for axion-photon~(left panel) and axion-electron interactions~(right panel), and AGSS09~(blue) and GS98~(red) solar models. The transparent grey lines are 100~randomly chosen Monte Carlo samples of the AGSS09 model. \label{fig:10k_spectra}}
\end{figure}
Following the procedure outlined before, we calculate the axion fluxes for all MC realisations. The outcome for both AGSS09 and GS98 models are shown in figure~\ref{fig:10k_spectra}. Echoing the trend for the systematic effects, the relative statistical fluctuations are also larger for higher energies. However, the overall level of the uncertainties is noticeably smaller, usually at the few-percent level. Owing to the greater complexity and higher number of processes involved, the typical uncertainties for the \ABC~spectrum are generally larger than for the Primakoff flux.

\begin{figure}
	\centering
	\includegraphics[width=6in]{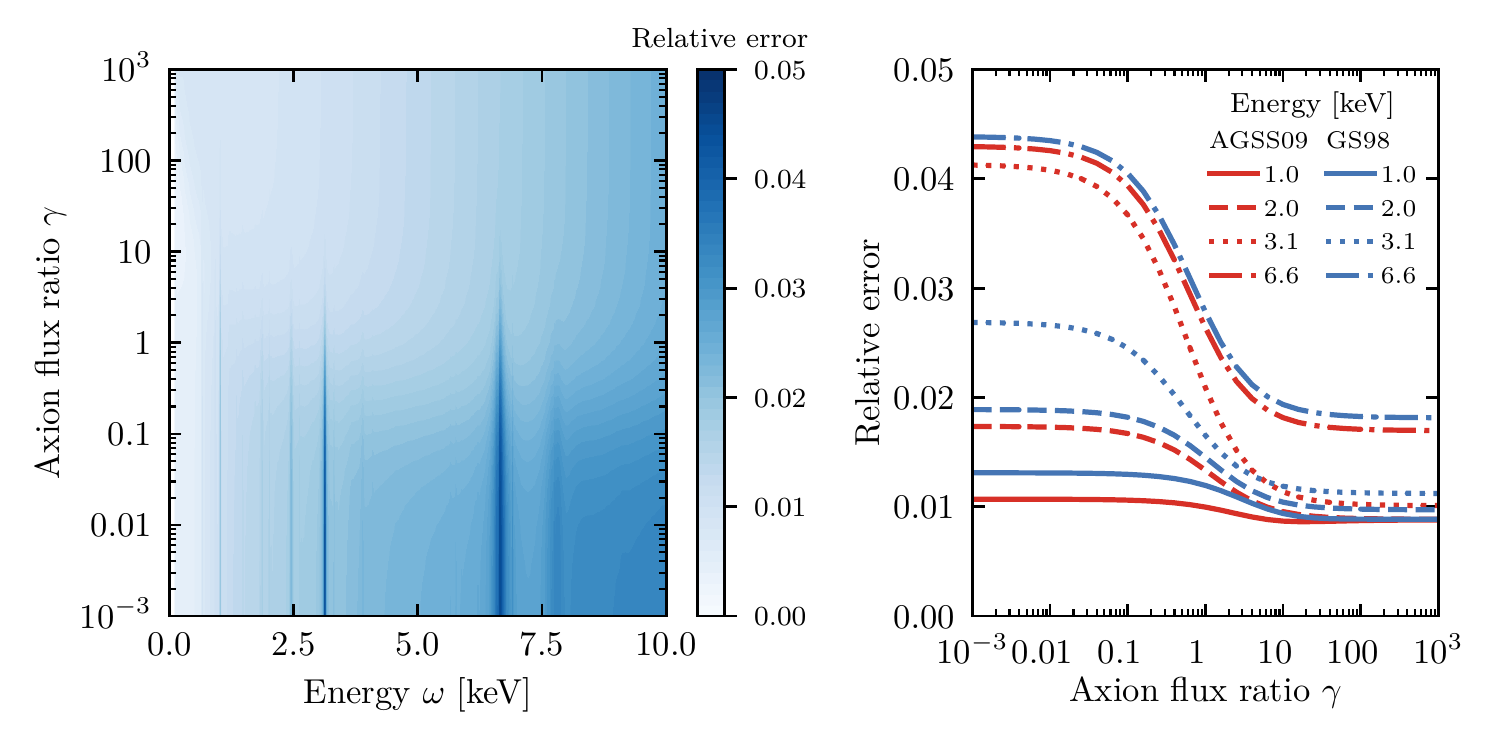}
	\caption{Relative \emph{statistical} errors of the solar axion flux as a function of energy~$\omega$ and ratio of Primakoff to \ABC flux~$\gamma$. We show the corresponding density plot for AGSS09 solar models in the left panel. The right panel compares AGSS09~(red lines) and GS98~(blue lines) solar models. Note that the systematic uncertainties around the peaks can be notably larger than a few percent~(cf.\ \reffig{fig:opacity_codes_deviations}).\label{fig:10k_spectra_stat_errors}}
\end{figure}
To visualise the difference between the different interactions, let us define the quantity~$\gamma$, which represents the ratio of Primakoff and \ABC fluxes as a proxy for relative strength and importance of axion interactions: 
\begin{equation}
    \gamma \equiv \frac{\Phiag}{\Phiae} = \num{1.32} \, \left(\frac{\gagg}{\SI{e-10}{\GeV^{-1}}}\right)^2 \left(\frac{\num{e-12}}{\gaee}\right)^2 \, ,
\end{equation}
where the numerical prefactor is obtained from integrating the rates over the whole Sun and the resulting spectrum between~\SI{0.1}{\keV} and~\SI{10}{\keV} for the AGSS09 model and OP~opacities. If $\gamma \ll 1$~($\gamma \gg 1$), then axion-electron~(axion-photon) interactions are the dominant contribution to the solar axion flux.

It is useful to look at the results with this definition in mind, which is what we do in figure~\ref{fig:10k_spectra_stat_errors}. In the left panel, we show the relative $1\sigma$~uncertainty of the \emph{total} solar axion flux from AGSS09 models. We confirm that typical uncertainties are of the order of a few percent, with the largest uncertainties of about~5\% occurring around the peaks in the \ABC~spectra, where the systematic uncertainties are larger anyway. Away from the peaks, the differences between AGSS09 (low-$Z$) and GS98 (high-$Z$) models are smaller. From the right panel in figure~\ref{fig:10k_spectra_stat_errors} we can also clearly see that the uncertainties approach asymptotic values as $\gamma \ll 1$ or $\gamma \gg 1$, indicating that one type of axion interactions dominates and confirming that our definition of~$\gamma$ to quantify this is indeed sensible.

\subsection{Additional sources of uncertainty and limitations of our approach}\label{sec:limitations}
By using solar model samples from MC simulations whilst also simulating the monochromatic opacity uncertainties, we can address the statistical sources of uncertainty. In addition, figures~\ref{fig:solar_models_overview} to~\ref{fig:opacity_codes_deviations} illustrate the systematic differences between solar models and opacity codes. However, there are further systematic uncertainties and corrections. These originate from the approximations employed in sections~\ref{sec:primakoff} and~\ref{sec:axionelectron_interactions}. All possible corrections that we consider are listed in table~\ref{tab:corrections}, for comparison together with our other findings.

\paragraph{Primakoff flux.} The Primakoff flux may receive corrections from \updated{Compton scattering, bremsstrahlung or} atomic interactions that emit a photon and an axion instead of only a photon. The rate of such processes is highly suppressed, and we can approximate it with an expression analogous to eq.~\eqref{eq:axion_photon_ratio},
\begin{equation}
      \frac{\Gamma^{i}_{\gamma + a}}{\Gamma^{i}_\gamma } \sim \gagg^2 \, \omega^2 \, , \label{eq:primakoff_correction}
\end{equation}
\updated{where $i = \text{bb},\text{fb},\text{ff},\text{B},\text{C}$.} Since eq.~\eqref{eq:primakoff_correction} has the same $\omega$~dependence as the \ABC flux, we can immediately see that this correction will be largest for lower axion energies. At \SI{1}{\keV}, we expect an effect of a few percent, while the effect on the overall flux becomes negligible at a relative size of less than~0.2\%.

When the ions involved in Primakoff interactions are not fully ionised, the Primakoff process can be inelastic in the sense that energy may be transferred by exciting or de-exciting an electron in the shell.\footnote{We thank Georg Raffelt for pointing out this contribution.\label{footthanks}}  This effect is, however, negligibly small. To see this, we follow ref.~\cite{Abe:2021ocf}, which investigated inelastic scattering in the context of the inverse Primakoff effect. While the elastic Primakoff cross section is proportional to the charge of the ion or electron, ref.~\cite{Abe:2021ocf} showed that the inelastic Primakoff effect is proportional to the incoherent scattering function $S(q)$, whose maximal value is given by the number of bound electrons $z-Q_z$~\cite{doi:10.1063/1.555523}. The inelastic Primakoff cross section in the plasma is hence suppressed by
\begin{equation}
    \frac{ \sum_z \, n_z \, (z-Q_z)^2}{n_e +\bar{n}} \lsim 0.1\% \, ,
\end{equation}
where $z$ is the atomic number. One may wonder whether the energy transfer in inelastic scattering can lead to a resonant enhancement that potentially compensates for this suppression. However, we have checked that no such resonance occurs when taking into account the plasma frequency as an effective photon mass.\footnote{Here, we point out again, that a more systematic treatment of screening effects would be useful.}

Furthermore, we do not include the electro-Primakoff process~\cite{Raffelt:1985nk}\footref{footthanks}
i.e.\ an electron scattering off an ion or electron, with an axion emitted from the virtual photon propagator. This is because we can use the total energy emission of this process, as calculated in ref.~\cite{Raffelt:1985nk}, to estimate the resulting average correction of the Primakoff flux. In line with their conclusion, we find it to be completely negligible at the order of \num{4e-5}. To provide an upper bound on the correction at specific energies, we allow for a relative scaling between Primakoff and electro-Primakoff of~$\omega^2$. As a result, the correction between \SIrange[range-phrase=\text{~and~},range-units=single]{1}{10}{\keV} is at most~0.4\%.

The Primakoff flux may also receive corrections from higher-order QED~diagrams, which feature an additional factor of~$\alphaEM$. Also recall that we assume the axion mass to be negligible since solar axion searches are typically not sensitive to axions heavier than \SI{1}{\eV}. Corrections associated with a non-vanishing axion mass are expected to be smaller than $\ma/\omega \lsim 0.1\%$.

\updated{At low energies near the plasma frequency, the dispersion relation in eq.~\eqref{eq:dispersion}, and hence the Primkoff flux, may receive further corrections because we worked in the limit of non-relativistic and non-degenerate electrons. However, evaluating the full expression as given in ref.~\cite{Raffelt:1996wa} only changes the prefactor of $\ompl^2$ by 0.3\%. This results in a tiny change of the low-energy cutoff but does not have any significant effect on the flux in our energy range of interest between 1 and \SI{10}{\keV} (see relative uncertainties given in table~\ref{tab:corrections}).}

\updated{Another potential source of uncertainty is the correct screening prescription. As Raffelt pointed out in ref.~\cite{Raffelt:1985nk}, the time it takes for one plasmon to cross the scattering potential is much smaller than the typical time it takes for an electron to cover the same distance. This is the justification for deriving the effective Primakoff form factor in the static limit, which amounts to first squaring the matrix elements and then averaging over different charge distributions. The resulting form factor is included in the Primakoff rates in eqs.~\eqref{eq:PrimaRate} and~\eqref{eq:PrimaRate_full}. To quantify the uncertainty associated with this static limit, we generalised Raffelt's calculation to charges moving at constant velocities in appendix~\ref{appendix:formfactor}. We find the relative corrections to be of $\mathcal{O}(\num{e-4})$, and thus negligible.}

Finally, note that we do not include the contribution from the conversion of plasmons as described in refs.~\cite{2005.00078,2006.10415,2010.06601} because they depend on the solar magnetic field, which is not tabulated in our solar models. As discussed before, the flux component from longitudinal plasmons is only present at energies $\lsim\SI{0.2}{\keV}$ and can therefore be separated from the Primakoff flux with spectral information. That is why we do not regard it as a source of uncertainty in this work. The flux from non-resonant conversions of transverse plasmons is not equally well separated from the Primakoff spectrum. However, it has a similar dependence on the large scale magnetic field, which means that its relevance can be inferred from the measurement of the resonant longitudinal plasmon flux. Hence, it is also not treated as a correction to the Primakoff flux.

\paragraph{\ABC flux.}$\!\!\!\!\!$\footnote{We are very grateful to Javier Redondo for allowing us to study his more detailed notes and calculations for ref.~\cite{1310.0823}.}\quad Some of the Primakoff flux corrections also apply to the \ABC flux, but there are a few additional ones. For instance, three simplifications were used in the derivation of eq.~\eqref{eq:axion_photon_ratio}: a non-relativistic expansion of the interaction Hamiltonian for the electron, a leading-order multipole expansion of the transition amplitudes, and a separation of the wave functions in spin and spatial parts~\cite{1310.0823}.

The non-relativistic expansion of the matrix elements entering the derivation of eqs.~\eqref{eq:ffRate} and \eqref{eq:eeRate} discards terms that are suppressed by the electron velocity. To get an upper bound on the resulting error, we do a full relativistic calculation  of the bremsstrahlung rates~(as described in refs.~\cite{Raffelt:1996wa,Carenza:2021osu}) in the core of the Sun, where relativistic effects are expected to be most significant. For the spectral average, we arrive at an uncertainty of about~3\%. This can be rescaled to obtain an estimate for the high-energy tail, where we expect relativistic corrections to be less than~6\%.

The multipole expansion of the transition amplitudes for both axion and photon production is outlined in ref.~\cite{1310.0823}. Each higher order comes with an additional product of the photon/axion momentum $k$ and the electron's position operator. We can estimate the size of these contributions to the transition amplitudes by a factor of $\omega/(m_e Z \alphaEM)$, for which we used $k\sim \omega$ and approximated the distance by the Bohr radius of a hydrogen-like atom. Since this only applies to fb and bb~transitions, and since higher-order multipoles do not play a role in ff~interactions, the size of this correction relative to the overall flux is limited to few percent for most of the energy range.

A final simplification, entering eq.~\eqref{eq:axion_photon_ratio}, is the separability of spin and spatial wave functions. This effectively neglects spin-orbit interactions in bb and fb~transitions, and has two effects. To obtain eq.~\eqref{eq:axion_photon_ratio}, an average over the spin states is used. However, spin-orbit couplings split the energy levels according to their total angular momentum. The individual energy levels from the fine structure may contribute to the axion emission with a rate different from the average. Equation~\eqref{eq:axion_photon_ratio} is therefore only valid if the fine structure is not resolved. This effectively limits the energy resolution for which one can employ this approximation.
For the prominent \ce{Fe}~peak, this corresponds to an energy resolution of $\Delta E\sim \SI{25}{\eV}$~(to estimate this we consider the spread between different fine structure states of transitions relevant to the axion emission using the NIST database~\cite{NIST_ASD}). Second, the matrix element explicitly depends on the energy of the transition and will receive relative corrections of the order $\Delta E/\omega$.

As already mentioned in section~\ref{sec:axionelectron_interactions}, the \ABC flux is affected by the degeneracy of electrons: first through the dependence of eqs.~\eqref{eq:ffRate} and~\eqref{eq:eeRate} on the screening scale and, second, due to Pauli blocking. While degeneracy effects are included in our calculations, we estimate that Pauli blocking of the same bremsstrahlung processes can reduce the \ABC flux by at most~7\%. To arrive at this upper bound, we evaluate the full ion-bremsstrahlung phase space integrals with and without Pauli blocking around the solar core. The resulting suppression factor can be used for all radii and squared due to the two electrons in the final state of electron-electron bremsstrahlung. Since degeneracy effects are expected to be strongest at the core, our estimate should be conservative. In principle, it would be possible to replace eqs.~\eqref{eq:ffRate} and~\eqref{eq:eeRate} by full phase space integrals, including Pauli blocking. However, since the \ABC flux is affected by several other large uncertainties, our focus on KSVZ~axion models, and due to the computational cost of such high-dimensional integrals, we do not perform the full phase-space integrals in general. Note, however, that the full calculation would be required for a high-precision measurement of the \ABC~flux.

As also discussed in section~\ref{sec:axionelectron_interactions}, different screening prescriptions lead to additional uncertainties of the order of at least a few percent.

Furthermore, the bremsstrahlung rates in eqs.~\eqref{eq:ffRate} and~\eqref{eq:eeRate} are computed in the Born approximation, which assumes plain wave solutions for the incident electrons. However, long-range attractive forces can significantly enhance the scattering amplitude -- an effect that is often referred to as Sommerfeld enhancement. For a Coulomb potential, the size of this effect can be well approximated by including the Elwert factor~\cite{Elwert1939,Vitagliano:2017odj} in the phase space integral, as was done for neutrino emission in ref.~\cite{Vitagliano:2017odj}.
We find that including the Elwert factor enlarges the results of eqs.~\eqref{eq:ffRate} and~\eqref{eq:eeRate} by approximately~20\%. Note that this estimate should be understood as an upper bound only: as mentioned above, long-range Coulomb interactions are screened in the solar plasma, thus reducing the Sommerfeld enhancement. We use the estimate provided in ref.~\cite{Vitagliano:2017odj} and conclude that expressions~\eqref{eq:ffRate} and~\eqref{eq:eeRate} should underestimate the true value by less than \updated{approximately} 10\%.

As Redondo pointed out~\cite{1310.0823}, the OP opacities employ Coulomb wave functions in their calculations. This means that the effect of enhanced wave functions towards the centre of the scattering potential is overestimated since charge screening is not taken into account for the incident waves. The OP's approximation should, however, become more precise with larger charges of the target nuclei because Coulomb wave functions and plain waves only differ at radii given by the size of electronic orbitals around the nucleus. This is yet another reason to use the analytical results in eqs.~\eqref{eq:ffRate} and~\eqref{eq:eeRate} for the scattering of light nuclei or two electrons, and to apply eq.~\eqref{eq:totalrate} with the OP data only for heavier elements.

In conclusion, our calculation in section~\ref{sec:axionelectron_interactions} underestimates the bremsstrahlung rates for light nuclei or two electrons. This is due to using the Born approximation with incident plain waves. On the other hand, the Coulomb wave functions employed by the OP overestimate the rates of electrons scattering off heavier nuclei -- even though to a smaller degree. We conservatively estimate that the systematic uncertainty of the total \ABC rate resulting from both effects is smaller than~$\sim$10\%.

\updated{\paragraph{Interference effects.}
Interference between two axion production processes may only occur if both have identical initial and final states.
It may also occur for processes that are mediated by different couplings such that interference terms can in principle prevent us from strictly separating Primakoff and \ABC fluxes.
In particular, as there can be a relative sign between photon and electron couplings, the interference between $\gagg$ and $\gaee$ interactions can be either negative or positive.}

\updated{Out of all the diagrams in figure~\ref{fig:feynman_diagrams}, only the Primakoff and Compton processes can interfere. As pointed out in ref.~\cite{Raffelt:1985nk}, this interference is suppressed by $\omega/m_e$ because in the non-relativistic limit the two final state electrons have opposing spin. Therefore, the interference can only give a significant contribution to the Primakoff flux when $\gaee$ is very large, compensating for the $\omega/m_e$ suppression and the fact that the Compton process only contributes a small fraction to the \ABC flux. Such a large electron coupling would necessarily result in a dominant \ABC flux i.e.\ $\gamma \ll 1$. For our further investigations in section~\ref{sec:solar-vs-axion}, we require that Primakoff interactions dominate, $\gamma \gg 1$, such that interference effects will not have a significant effect on the Primakoff flux.

There is further a possible interference between electro-Primakoff, whose rate we found to be negligible in agreement with~\cite{Raffelt:1985nk}, and bremsstrahlung processes. Here, we also find a $\omega/m_e$ suppression because the spin of final state electrons in the non-relativistic limit differs between the two processes.}

\paragraph{Sun-Earth distance \updated{and positioning}.} To conclude this section, note the dependence of the solar axion flux in eq.~\eqref{eq:integration_master} on the distance between Sun and Earth~$\dE$. Over the course of a year, $\dE$~oscillates around its average by about 1.7\%~\cite{NASA}. Consequently, this effect has to be taken into account when a sizable axion signal is detected. Since~$\dE$ can, in principle, be measured with high precision, we do not consider this as a source of uncertainty. Besides, our results for the reference value of $\dE = \SI{1}{\astrounit}$ can simply be rescaled to the measured value.

\updated{A similar, biannual effect applies to the relative positioning of Earth w.r.t.\ the Sun's equator, which affects the plasmon flux since the large-scale solar magnetic fields are not spherically symmetric~(see e.g.\ Fig.~4 of ref.~\cite{2006.10415}). The resulting geometrical factor may be determined for the data taking times of an experiment, and then be used to rescale the equations presented in this work.}

\section{Consequences for axion detection}\label{sec:axionmodels}
While more and more axion searches are setting out to explore large parts of the presently untested parameter space, it is often unclear how much information could be gained from a possible detection. Determining as many properties of the detected particle as possible is crucial for learning about the underlying UV~model. The ultimate proof for the existence of QCD~axions would be a measurement of the axion-gluon coupling, which would confirm the PQ~solution of the Strong~CP problem. Since helioscopes rely on the axion-photon coupling for detection, they cannot achieve this. However, helioscopes are in principle able to find or constrain three axion parameters -- $\ma$, $\gagg$, and $\gaee$ -- from the strength and shape of the signal~\cite{1811.09290,1811.09278}. This means that they can locate the axion within the so-called QCD~axion band, which is defined by a range of proportionality constants between the mass $\ma$ and the photon interaction $\gagg$ of an axion. In addition, helioscopes can also be seen as a new way of observing the Sun. It has already been suggested that they could probe individual heavy element abundances~\cite{1908.10878} and the solar magnetosphere~\cite{2006.10415}. Having discussed the systematic and statistical solar axion flux uncertainties in the previous sections, we can now investigate how they affect these kinds of axion parameter measurements.

\subsection{Uncertainties on the axion signal}
\begin{table}
	\caption{Statistical uncertainties~(``Error'') of the integrated Primakoff and \ABC fluxes, respectively. We normalise the mean values~(``Mean'') to the corresponding mean values of the AGSS09 model.\label{tab:uncertainties_overview}}
	\centering
	\begin{tabular}{lcccc}
		\toprule
		\textbf{Model} & \multicolumn{2}{c}{$\boldsymbol{N_a^{\scriptstyle\textbf{P}}}$} & \multicolumn{2}{c}{$\boldsymbol{N_a^{\scriptscriptstyle \textbf{ABC}}}$}\\
		\cmidrule(rl){2-3}\cmidrule(rl){4-5}
		& Mean & Error & Mean & Error \\
		\midrule
		AGSS09 & 1 & 1.2\% & 1 & 1.4\%\\
		GS98 & 1.052 & 1.3\% & 1.051 & 1.6\%\\
		\bottomrule
	\end{tabular}
\end{table}
To gain an intuition for the overall statistical errors, one can e.g.\ consider the uncertainties on the integrated solar axion fluxes in the relevant energy range of $\omega \in [\SI{0.1}{\keV},\SI{10}{\keV}]$. We quote these uncertainties in table~\ref{tab:uncertainties_overview}. As noticed before, the overall statistical uncertainty of the axion flux for each solar model is only at the percent~level. However, for both the axion-photon and \ABC~flux, the resulting counts are systematically higher by about 5\% for photospheric, GS98-type models.

\begin{figure}
	\centering
	\includegraphics[width=6in]{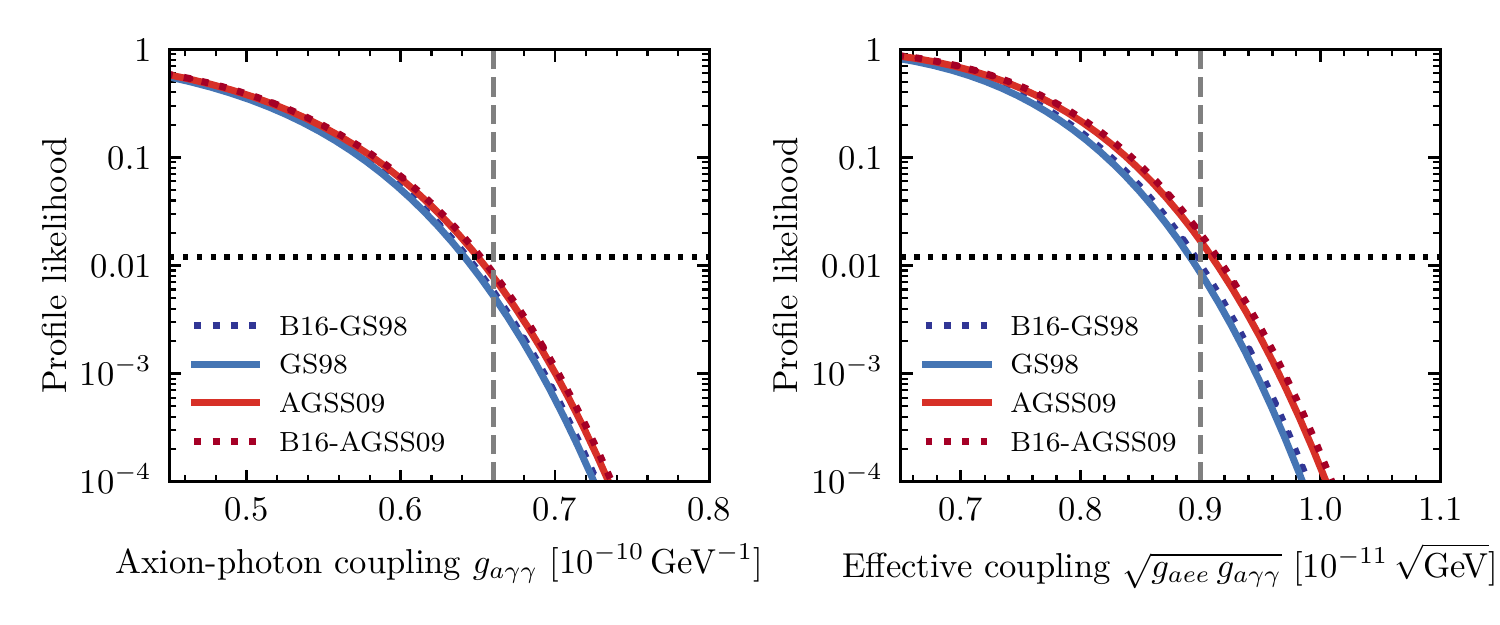}
	\caption{Profile likelihoods of the axion-photon coupling~(\textit{left}; assuming $\gamma \gg 1$) and the product $\sqrt{\gaee\,\gagg}$~(\textit{right}; assuming $\gamma \ll 1$) for $m_a = 0$ and using CAST 2017 data~\cite{1705.02290} (based on the implementation in ref.~\cite{1810.07192}). We show these for different solar models~(different coloured lines). For reference, we also indicate the 95\% CLs~(dashed red lines) and the corresponding nominal likelihood thresholds~(dotted black lines). Note that the CAST 2017 limit on $\gagg$~\cite{1705.02290} uses the BP04 model, and that the reference limit for $\sqrt{\gaee\,\gagg}$ was only derived for older data by the CAST~collaboration~\cite{1302.6283}. \label{fig:exclusion_curves}}
\end{figure}
To consider a more concrete example, we take the latest data and limits from the CAST experiment~\cite{1705.02290}. In figure~\ref{fig:exclusion_curves}, we show the profile likelihood ratios resulting from this measurement using different solar models (the axions are treated as effectively massless). Clearly, the overall effect of the systematic differences on the limits from the different models is rather small, and the statistical differences~(not shown) are smaller still.
We conclude that the impact of the uncertainty on the exclusion limits is negligible. This is not unexpected, as the uncertainties only enters as the fourth root in the effective coupling parameter.

\subsection{Disentangling axion and solar models}
\label{sec:solar-vs-axion}
We saw in table~\ref{tab:uncertainties_overview} that the statistical uncertainty of each model is typically much smaller than the difference between high-$Z$ and low-$Z$ solar models.
This means that it is in principle possible to distinguish the two cases. The problem with this {na\"ive} observation is a degeneracy between the solar metallicity and the axion coupling constants. Smaller $Z$~values result in a smaller axion flux, which may always be compensated by a larger value of the unknown coupling constant.

To see how this degeneracy can be circumvented, we note a crucial difference between the two axion coupling constants $\gaee$ and $\gagg$. In DFSZ models, for example, $\gaee$~depends on the mixing angle between two Higgs fields. As a result, $\gaee$~can take any value within a range only constrained by perturbative unitarity~\cite{1708.02111}; the latter implying that the ratio of the Higgs vevs, usually called $\tan(\beta)$, is constrained by $0.28 < \tan(\beta) < 140$~\cite{1301.0309}. Measuring~$m_a$ relatively precisely, and combining with the restriction on $\tan(\beta)$, may reduce this degeneracy.

In contrast, $\gagg$~is generated from an anomaly, and it takes on discrete values for different viable axion models. A~comprehensive list of values of $\gagg$ in hadronic~(also known as KSVZ~\cite{1979_kim_ksvz,1980_shifman_ksvz}) axion models with a single additional coloured fermion~(``heavy quark'') is given in refs.~\cite{1610.07593,1705.05370}. The authors consider KSVZ~models to be \textit{preferred} if the additional heavy quark is sufficiently unstable and if the model has no Landau pole below the Planck scale. Considering hadronic axions with the above properties, which we call the \textit{preferred axion window}, we can break the degeneracy between the solar metallicity and the coupling constant. This is because $\gagg$ cannot be continuously adjusted to compensate for the uncertainty of the metallicity.

Before continuing, let us make three comments regarding the \textit{preferred models} of refs.~\cite{1610.07593,1705.05370}. First, a specific $E/N$~value does in general not correspond to a unique UV model. For instance, among \updated{the models that we have considered}, there are multiple ones with $E/N = 2/3$ and $E/N = 8/3$.
Second, note that the Landau pole criterion imposed in ref.~\cite{1705.05370} eliminates five hadronic models, for which we calculate $E/N$~values of~$5/12$, $23/3$, $35/3$, $47/3$, and $62/3$.\footnote{We thank Vaisakh Plakkot for confirming these values with an independent calculation.} Lifting the Landau pole criterion therefore allows for additional values of~$E/N$, potentially making the model distinction more difficult but also potentially allowing us to explain axion detection corresponding to higher $E/N$~values.
Finally, one could extend the models under consideration to KSVZ axion models with more than one new heavy fermion, DFSZ models, models with non-universal couplings, or even ALPs. This would drastically increase the possible $E/N$ ratios and, hence, the model density in the axion band~(see e.g.\ ref.~\cite{2003.01100} for a recent review on the landscape of axion models). This implies the need for complementary information, e.g.\ on $\gaee$, in order to distinguish these. Nevertheless, a measurement of $E/N$ would be a severe restriction on possible UV models and going beyond typical values for $E/N$, which we are considering, one often runs into cosmological problems or requires an extended UV sector.

With all these considerations in mind, we want to answer two questions for the specific case of solar axion detection with IAXO.
First, can IAXO distinguish the different types of hadronic models within the axion window?
And, second, could IAXO find a preferred combination of a solar and an axion model that is compatible with the observed signal?
To answer both these questions, we must carefully take into account all uncertainties which are relevant for the detection of a Primakoff flux in helioscopes.

Since hadronic axion models have a negligible axion-electron coupling, $\gamma \gsim 10^7$, we can assume that the resulting solar axion spectrum will be Primakoff-dominated.\footnote{We remind the reader of our assumption that the flux generated by plasmon conversions in large scale magnetic fields is small or separately detectable.}
Because the relative systematic~(left panel of figure~\ref{fig:solar_models_deviations}) and statistical~(left panel of figure~\ref{fig:10k_spectra}) uncertainties of the Primakoff flux are approximately constant for all relevant energies,\footnote{While not exactly true, this is an acceptable approximation for our purposes. In case of an experimental detection, the data analysis~(in the $m_a$-$\gagg$ plane) needs to be repeated with various solar models (not by a simple rescaling) in order to make statements about the nature of the signal. Analyses in the $m_a$-$\gagg$ plane lead to multiple confidence regions from the different solar models, which can be compared to a single band for each KSVZ~model; this in contrast to the $m_a$-$C\gagg$ plane~(single region, multiple bands).} we can model the solar uncertainty by a simple re-scaling of the axion spectrum. To this end, we introduce a scaling constant~$C$ such that the overall signal or, equivalently, the expected number of detected photons~$N_\gamma$ is proportional to $(C\gagg)^4$, i.e.\ $N_\gamma \propto (C\gagg)^4$. $C$ is normalised to unity for the B16-AGSS09 model, which is our preferred low-$Z$ solar model. With these definitions, IAXO does not directly constrain~$\gagg$ but $C\gagg$ instead.

\begin{figure}
    \centering
    \includegraphics[width=6in]{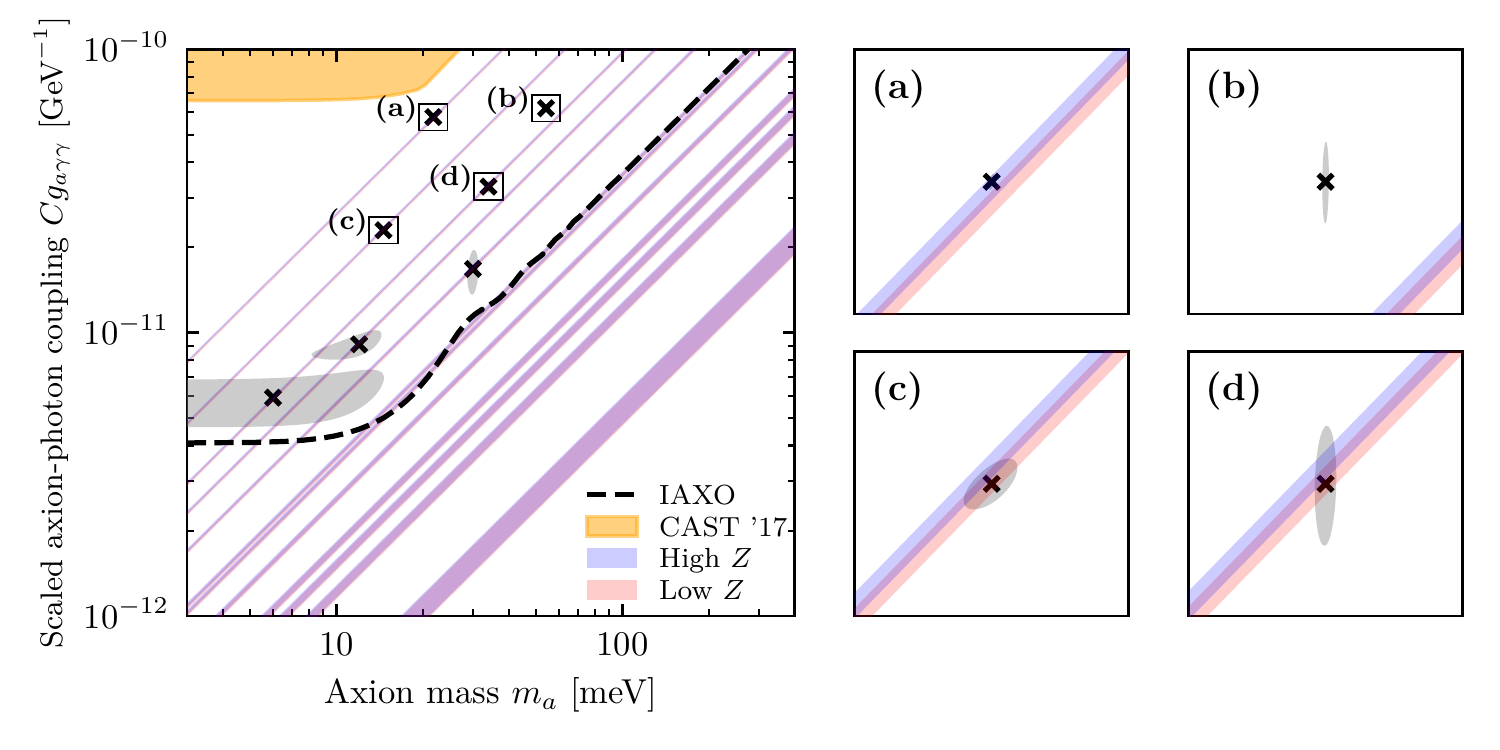}
    \caption{IAXO sensitivity and benchmark cases in the $\ma$-$C\gagg$ parameter plane. We show hadronic axion models in the \textit{preferred axion window} for high- and low-$Z$ solar models~(blue and red shading, respectively; their overlaps appear in a darker, purple shading). \textit{Left:} We select seven benchmark points (black crosses) and determine the expected 95\% CLs~(grey-shaded areas) of a detection with the baseline IAXO setup~\cite{1904.09155}. We show the corresponding IAXO sensitivity~(dashed black line) and include the region already excluded by CAST~\cite{1705.02290}~(yellow region). \textit{Right:} We provide zoom-ins for the most precise measurements, labeled~(a) to~(d).\label{fig:solar-vs-axion}}
\end{figure}

This definition allows us to make a plot of IAXO's predicted sensitivity in the plane spanned by $\ma$ and $C\gagg$. Since $C$ is different for high-$Z$ and low-$Z$ solar models, each combination of a solar model with a hadronic axion model can be plotted separately. This results in low-$Z$~(red) and high-$Z$~(blue) bands in figure~\ref{fig:solar-vs-axion}, which are only clearly visible as individual bands in the zoom-ins~(right panel). Overlapping bands appear in a darker, purple shading.

The underlying uncertainties in the axion model parameters are the reason why each combination of a solar and an axion model defines a narrow band rather than a line. The width of each band includes the statistical uncertainty of the solar model in addition to the QCD~uncertainties of $\ma$ and $\gagg$:
\begin{align}
    \ma &= \frac{\Lambda_\chi^2}{f_a} = \SI{5.69(5)}{\micro\eV} \left(\frac{\SI{e12}{\GeV}}{f_a}\right) \, , \label{eq:axion_mass}\\
    \gagg &= \frac{\alphaEM}{2\pi f_a}\left(\frac{E}{N} - \widetilde{C}_{a\gamma\gamma} \right) = \frac{\alphaEM \, m_a}{2\pi \, \Lambda_\chi^2}\left(\frac{E}{N} - \widetilde{C}_{a\gamma\gamma} \right) \nonumber \\  & = \SI{2.04(2)e-10}{\GeV^{-1}} \, \left(\frac{m_a}{\si{\eV}}\right) \left(\frac{E}{N} - 1.93(3) \right) \, . \label{eq:def_gagamma}
\end{align}
To derive the errors on $\Lambda_\chi = \SI{75.4 \pm 0.3}{\MeV}$ and $\widetilde{C}_{a\gamma\gamma} = \num{1.93 \pm 0.03}$, we follow refs.~\cite{1511.02867,1812.01008} and use updated numbers and additional corrections from ref.~\cite{1812.01008} where available. Note that, mostly through their dependence on the up-\ and down-quark mass ratio, the errors in eq.~\eqref{eq:def_gagamma} are correlated. By propagating the errors with MC simulations, we estimate the correlation coefficient between $\Lambda_\chi$ and $\widetilde{C}_{a\gamma\gamma}$ to be about~$-0.098$. In what follows, we take this correlation into account. The uncertainty from $\widetilde{C}_{a\gamma\gamma}$ dominates the total theoretical uncertainty of axion models when $E/N$ is close to $\widetilde{C}_{a\gamma\gamma}$ i.e.\ for models with small couplings to photons. This effect is clearly visible in figure~\ref{fig:solar-vs-axion}, where the bands with smallest couplings are also the widest.

It is now interesting to look at how the width and the distance between the model bands compare to the discovery potential of IAXO. To estimate the latter, we perform a likelihood ratio analysis of the Asimov data set to find the expected exclusion contours at the 95\% confidence limit~(CL) for each point in the ($\ma$,\,$C\gagg$)~parameter space. Similar studies have been performed in refs.~\cite{1811.09290,1811.09278}. We apply the same method for benchmark values of $m_a$ and $\gagg$, using the baseline IAXO parameters as given in ref.~\cite{1904.09155}. We choose seven benchmark points to illustrate qualitatively-different potential experimental outcomes. They are marked with black crosses in figure~\ref{fig:solar-vs-axion} and are surrounded by their respective expected 95\% CL contour, shaded in grey.\footnote{For simplicity, we only simulate an evacuated IAXO setup. If, instead, the experimental setup is filled with a buffer gas, the experimental sensitivity would improve for higher axion masses~\cite{1904.09155,2010.12076}.}

Looking at the two points at the bottom with the smallest coupling constants, it becomes clear that in order to locate the true axion values within a specific model band, we necessarily require a mass measurement. For IAXO this is only possible if the axions mass is at least a few meV and the coupling is sufficiently large~\cite{1811.09290,1811.09278}. A~detailed discussion, and the full region of parameter space where a mass measurement is possible, is given in ref.~\cite{1811.09290}.

In all cases with $\gagg \gsim \SI{e-11}{\GeV^{-1}} $ and $\ma \gsim \SI{10}{\meV}$, the axion parameters can be located on a specific axion band, which means that there is only one possible value of $E/N$ within the \textit{preferred axion window} that would explain the observation. This is of course only true if the observed particle actually falls on such a band.

At large coupling values, the experimental precision becomes so good that the expected 95\% CL contour is only visible when we zoom in on the corresponding region~(right part of figure~\ref{fig:solar-vs-axion}). In these zoom-ins, we can also clearly make out the high-~(blue) and low-metallicity~(red) solar models. The two bands overlap due to the combination of statistical solar model and QCD~uncertainties, recalling that the latter dominate the overall uncertainty. In fact, had we excluded the QCD~uncertainties, the two bands would not overlap.

As we can see from the grey region in figure~\ref{fig:solar-vs-axion}, it is possible to obtain a hint from IAXO towards higher or lower metallicity in some serendipitous cases. For instance, a measurement as indicated in subpanel~(a) of figure~\ref{fig:solar-vs-axion} could only be explained by the combination of a specific value of $E/N$ and a high-$Z$ solar model. In contrast, example~(b) shows the case where IAXO makes a detection that cannot be explained by any of the \textit{preferred hadronic axion models}. Example~(c) could also be interpreted as a hint towards lower $Z$ even though the picture is less clear in this case. In example~(d), the experimental uncertainty smears out any potential information on the solar metallicity.

With these observations, we can answer the two questions stated above: IAXO can indeed differentiate between different hadronic axion models in the \textit{preferred axion window}. This is approximately possible for all parameters in the mass-coupling plane, for which IAXO can determine the (non-zero) axion mass, as described in refs.~\cite{1811.09278,1811.09290}. In the case of a strong signal in IAXO, it may also be possible to infer a combination of a solar model and axion model at the same time, cf.\ example~(a) in figure~\ref{fig:solar-vs-axion}.
Indeed for such a fortuitous case, and making the assumption of a reasonably simple KSVZ model, it is possible to distinguish low-$Z$ and high-$Z$ models without making use of the $g_{aee}$~coupling (which was used in ref.~\cite{1908.10878}).

It is, however, important to note that a detected particle could also not be in the \textit{preferred axion window}, example~(b), and hence may not fall on one of the bands in figure~\ref{fig:solar-vs-axion}. This could for instance be an axion model which includes multiple heavy coloured fermions in various representations of $\text{SU}(3)$~\cite{1610.07593}. In such a case, it would still be interesting to find out that a simple realisation of an hadronic axion is not able to explain the signal. The degeneracy between metallicity and the axion coupling would still persist.

\section{Conclusions}\label{sec:conclusions}
In light of the recently growing interest in solar axions, we have revisited the processes generating the solar axion flux to assess their uncertainties. This is motivated by the crucial influence that they can have when measuring QCD~axion or axion-like particle properties, such as their mass or their couplings to electrons and photons. 

The solar axion flux depends on solar properties and therefore on solar models. We find that the uncertainties of the axion flux resulting from the statistical variation within solar models of a given type are relatively modest in size. 
This shifts the focus to a better determination of systematic errors. One example is the metallicity problem, which yields flux uncertainties at the level of 10\% or more across a large energy range. 

To determine properties of underlying axion models it would be desirable to achieve percent-level accuracy.
Reaching this target will not only depend on better astrophysical modelling but also requires improving the theoretical inputs.
Calculations of the solar axion flux use several approximations, some of which may need to be improved in order to reduce the systematic theory errors.
An important example is the partial degeneracy of electrons in the solar core (cf., e.g., ref.~\cite{Raffelt:1996wa}), which we included in section~\ref{sec:primakoff_deg} for the Primakoff emission.
Recently, this effect has also been investigated for the case of electron-ion bremsstrahlung~\cite{Carenza:2021osu} while work on other contributions to the \ABC flux is still underway.

The \ABC flux is also affected by uncertainties originating from variations between different opacity codes.
Achieving percent-level accuracy here will likely require going beyond the approximation where the axion emission is directly proportional to the photon opacity and accounting for the differences between the emission of axions and photons from each atomic transition. Moreover, a better treatment of screening effects as well as non-trivial electron wave-functions (Sommerfeld enhancement) using a systematic, first-principles calculation would be very useful.

Beyond uncertainties in the standard Primakoff and \ABC fluxes, there are additional production mechanisms such as the plasmon conversion in the Sun's magnetic field. The latter could add a noticeable flux contribution in our energy range of interest \emph{if} the magnetic fields are at the higher end of the estimated ranges. To our knowledge, the uncertainty of the magnetic field has not yet been statistically quantified, which would be a logical next step for further investigating the plasmon conversion.

The inference of underlying axion model parameters is also limited by the current knowledge of low-energy axion parameters, such as the mass and the model-independent contribution to the axion-photon coupling (see e.g.\ refs.~\cite{1511.02867,1812.01008}).
Addressing these effects will be important in order to accurately determine the properties of axion models with helioscopes.

Having said all this, the Primakoff flux calculation is less affected by uncertainties, which makes it suitable for a case study on the determination of axion model parameters.
As a concrete example for such a study, we considered benchmark cases of a potential future axion detection in section~\ref{sec:solar-vs-axion}.
We find that the known uncertainties on the solar Primakoff flux are sufficiently small that (an upgraded version of) IAXO may indeed be able to identify and discriminate between different KSVZ axion models.
Moreover, it may even be possible to address solar physics questions by distinguishing between low-$Z$ and high-$Z$ solar models.

In summary, the present work highlights the potential of helioscopes as high-precision axion probes and -- in case of a discovery -- also solar probes.
Fully realising this potential will require additional theoretical work on the calculation of the axion flux as well as the low-energy axion parameters.

\acknowledgments
{\small We thank Valerie Domcke, Alexander J.\ Millar, Georg Raffelt, Javier Redondo, Kai Schmitz, Aldo Serenelli, and the anonymous referee for helpful discussions and/or suggestions.
In particular, we are very grateful to Georg Raffelt for valuable comments that prompted us to take a closer look at several further relevant contributions and uncertainties affecting the flux.
We are also indebted to Aldo Serenelli for making the Monte~Carlo samples of solar models available to us, which allowed us to compute the statistical uncertainties in solar models in a rigorous and straightforward fashion, and to Javier Redondo for sharing his notes which greatly helped in obtaining estimates for the corrections to the \ABC flux.
We thank Christophe Blancard and the OPAS team for providing us with more detailed data and Vaisakh Plakkot for independently checking the list of relevant $E/N$~values.
S.H.\ is funded by the Alexander von Humboldt Foundation and the German Federal Ministry of Education and Research.
L.T.\ is funded by the \textit{Graduiertenkolleg} ``Particle physics beyond the Standard Model''~(GRK 1940).
We used the Scientific Computing Cluster at GWDG, the joint data centre of Max Planck Society for the Advancement of Science~(MPG) and the University of G\"ottingen.}

\appendix

\section{\updated{Corrections to the Primakoff form factor}}
\label{appendix:formfactor}
The derivations of the Primakoff rates $\GPrim$ in eqs.~\eqref{eq:PrimaRate} and \eqref{eq:PrimaRate_full} assume a form factor for the scattering potential of the form~\cite{Raffelt:1985nk}
\begin{equation}
    |F_\text{eff}|^2=Z^2\frac{q^2}{q^2+\ks^2}\, , \label{eq:eff_formfactor_static}
\end{equation}
where $Z$ is the charge in units of the elementary charge, $q = |\vc{q}|$ is the magnitude of the transferred momentum~$\vc{q}$, and $\ks$ is the usual screening scale. 

This form factor was derived in ref.~\cite{Raffelt:1985nk} by applying the strict static limit, hence assuming that the scattering potential does not change during the time it takes for one plasmon to cross the potential. Since charges are screened at typical distances of $1/\ks$, the corresponding time scale is given by $t \sim 1/\ks$.

The opposing limit would be to assume a Yukawa potential for every scattering plasmon. Since the results of these two limiting cases differ by a large factor $\sim 100$, it is necessary to investigate the size of the expected corrections once we drop the assumption of static point charges. To do this, we briefly recap the calculation in the static limit as presented in ref.~\cite{Raffelt:1985nk}, and then modify it to include point charges moving at constant velocity. 

\subsection{\updated{Static limit}}
As a starting point, let us recall the standard result for a situation where the plasmon encounters a set of non-moving charges $Z_{i}$. This was first discussed in~\cite{Raffelt:1985nk} whom we follow closely.

The form factor of $N$ particles with charges $Z_i$ at positions $\vc{r}_i$ is given by
\begin{equation}
    F_N(\vc{q}) = \sum_{i=1}^N Z_i \,  \ee^{\ii \, \vc{q}\cdot \vc{r}_i}\,,
\end{equation}
and hence
\begin{equation}
    |F_N(\vc{q})|^2 = \sum_i Z_i^2 + \sum_{\substack{i,j \\i\neq j}} Z_i Z_j\cos(\vc{q}\cdot \vc{r}_{ij})\, ,
\end{equation}
where $\vc{r}_{ij}$ are the interparticle distances. 
To obtain the result in the plasma we now have to average this over the locations of the charges. Using the probability density~\cite{Landau:1968,Raffelt:1985nk}
\begin{equation}
    p_{ij} = \frac{1}{V} \left( \ee^{-\Lambda r}- \frac{Z_i Z_j \alphaEM}{T} \frac{\ee^{-\ks r}}{r}\right) \, ,
\end{equation}
where $r$ is the distance between particles $i$ and $j$, we can take into account that the locations of particles $i$ and $j$ are not entirely independent. To simplify the calculation of volume integrals we have explicitly included
a regulator $\Lambda$ which will be taken to 0 at the end. $V$ is a volume normalisation factor that converges to the full volume in the limit of $\Lambda \rightarrow 0$. The screening scale as in eq.~\eqref{eq:screening_nondeg} can then be written as 
\begin{equation}
    \ks^2 = \frac{4 \pi \alphaEM}{T} \sum_i \frac{Z_i^2}{V} \, .
\end{equation}
Averaging $|F_N|^2$ over the interparticle distances yields
\begin{align}
    \langle |F_N(\vc{q})|^2\rangle_r =& \sum_i Z_i^2 + \sum_{\substack{i,j \\i\neq j}} Z_i Z_j \int \! \diff r^3 \, p_{ij}(r) \cos(\vc{q}\cdot \vc{r}) \nonumber \\ 
    =& \sum_i Z_i^2 +4\pi\sum_{\substack{i,j \\i\neq j}}Z_i Z_j \int_0^\infty \! \diff r \, r^2 \frac{1}{V}\left( \ee^{-\Lambda r} - \frac{Z_i Z_j \alphaEM}{T} \frac{\ee^{-\ks r}}{r}\right) \frac{\sin(q\,r)}{q\,r} \, . \label{eq:uptohere}
\end{align}
The first term in parantheses vanishes in the limit $\Lambda \rightarrow 0$, while the second one gives a finite contribution. For the second term, we have to evaluate the sum
\begin{equation}
    \sum_{\substack{i,j \\i\neq j}}Z_i^2 Z_j^2 = \sum_i Z_i^2 \, \sum_{j\neq i} Z_j^2 \simeq \sum_i Z_i^2 \frac{\kappa^2 TV}{4\pi\alphaEM} \, ,
\end{equation}
where the last equality is exact in the large volume limit. Putting everything together, we recover the result of ref.~\cite{Raffelt:1985nk}, viz.\ eq.~\eqref{eq:eff_formfactor_static}:
\begin{equation}
    \langle |F_N(\vc{q})|^2\rangle_r = \sum_i Z_i^2\left(1- \frac{\ks^2}{\ks^2 + q^2}\right) =\sum_i Z_i^2\frac{q^2}{\ks^2+q^2} \, .
\end{equation}

\subsection{\updated{Beyond the static limit}}
To see how decoherence can affect the expected form factor, we now drop the strict static limit and assume that each particle~$i$ moves with a constant velocity~$\vc{v}_i$. As a result, the form factor is now time-dependent,
\begin{equation}
    F_N(\vc{q}) = \sum_i Z_i \, \ee^{\ii \, \vc{q}\cdot (\vc{r}+\vc{v}_i t)} \, ,
\end{equation}
and we can average this form factor over the time $t \sim 1/\ks$ of one scattering event,
\begin{equation}
    \langle F_N(\vc{q})\rangle_t = \frac{1}{t}\int_{-\frac{t}{2}}^{\frac{t}{2}} \dd t' \, \sum_i Z_i \, \ee^{\ii \, \vc{q}\cdot (\vc{r}+\vc{v}_i t')} = \sum_i Z_i\  \mathrm{sinc}(\Delta\phi_i) \, \ee^{\ii\,\vc{q}\cdot \vc{r}} \, ,
 \end{equation}
 where we have defined the phase shift $\Delta\phi_i \equiv \frac{1}{2} \vc{q}\cdot\vc{v}_it$. 
 
 The average over the interparticle distance can now be evaluated exactly as before, up to the point of eq.~\eqref{eq:uptohere}, by just carrying along the factors of $\mathrm{sinc}(\Delta\phi_i)$:
 \begin{align}
      \langle|\langle F_N(\vc{q})\rangle_t|^2\rangle_r
      &= \sum_i Z_i^2 \, \mathrm{sinc}^2(\Delta\phi_i)  \nonumber\\
      &+ \frac{4\pi}{V}\sum_{\substack{i,j \\i\neq j}}Z_i Z_j\mathrm{sinc}(\Delta\phi_i)\mathrm{sinc}(\Delta\phi_j) \int_0^\infty \!\! \dd r \, r^2 \left(\ee^{-\Lambda r}- \frac{Z_i Z_j \alphaEM}{T} \frac{\ee^{-\ks r}}{r}\right) \frac{\sin(q\,r)}{q\,r} \, .
 \end{align}
As before, the first term in parentheses vanishes in limit $\Lambda\rightarrow 0$ while the second term is finite. The sum over pairs of charges is modified by the suppression factors.
\begin{align}    
    \sum_{\substack{i,j \\i\neq j}}\mathrm{sinc}(\Delta\phi_i)Z_i^2 \mathrm{sinc}(\Delta\phi_j)Z_j^2
    &= \sum_i\mathrm{sinc}(\Delta\phi_i) Z_i^2 \sum_{\substack{j \\j\neq i}}\mathrm{sinc}(\Delta\phi_j) Z_j^2 \\
    &\simeq \sum_i\mathrm{sinc}(\Delta\phi_i) Z_i^2 \frac{\kappa_{\text{eff}}^2 TV}{4\pi \alphaEM}\, ,
\end{align}
where in the last line we have defined $\kappa^2_\mathrm{eff}$ as
\begin{equation}
\label{eq:effective_screening}
     \kappa^2_\text{eff} \equiv \frac{4 \pi \alphaEM}{T}\sum_i \frac{\mathrm{sinc}(\Delta\phi_i) Z_i^2}{V}.
\end{equation}
Inserting these results into the expression for $\langle|\langle F_N(\vc{q})\rangle_t|^2\rangle_{r}$, we get
\begin{align}
  \langle|\langle F_N(\vc{q})\rangle_t|^2\rangle_r &= \sum_i \mathrm{sinc}(\Delta\phi_i)Z_i^2\left( \mathrm{sinc}(\Delta\phi_i) - \kappa^2_\mathrm{eff} \int_0^\infty \mathrm{d} r \, \ee^{-\ks r} \frac{\sin(q\,r)}{q}\right)\\
  &=  \sum_i Z_i^2 \,  \mathrm{sinc}(\Delta\phi_i)\left( \mathrm{sinc}(\Delta\phi_i) - \frac{\kappa^2_\mathrm{eff}}{\ks^2 +q^2}\right) \\
  &= \sum_i Z_i^2 \, \mathrm{sinc}(\Delta\phi_i)\left( \frac{\mathrm{sinc}(\Delta\phi_i)\ks^2-\kappa^2_\mathrm{eff}+\mathrm{sinc}(\Delta\phi_i)q^2}{\ks^2 + q^2}\right) \, . \label{eq:formfactor_nonstatic}
\end{align}
As one would expect, this result converges to our previous one in the limit $\Delta\phi_i\rightarrow 0$ where $\kappa_\mathrm{eff}\rightarrow\ks$.
 
To evaluate the relative correction of the new result compared to the static limit, we need to evaluate the thermal averages of the suppression factors $\mathrm{sinc}(\Delta\phi_i)$ and $\mathrm{sinc}^2(\Delta\phi_i)$. For simplicity, we assume a Maxwell-Boltzmann distribution for electrons and use typical values for the solar core temperature. By expanding the $\mathrm{sinc}(\cdot)$ function, we arrive at the estimates
\begin{align}
  \label{eq:sinc_suppression}
     \left. \langle \mathrm{sinc}(\Delta\phi_i) \rangle_{v_i}\right|_e 
     &= \int \!\dd^3v \, \mathrm{sinc}(\Delta\phi_i) \left( \frac{m_e}{2 \pi T}\right)^\frac{3}{2} \ee^{-\frac{m_ev^2}{2T}}
     \sim 1- \left(\num{e-4}\, \frac{q^2}{\ks^2}\right) \\
     \label{eq:sinc2_suppression}
     \left.\langle \mathrm{sinc}^2(\Delta\phi_i) \rangle_{v_i}\right|_e
     &= \int \! \dd^3v \, \mathrm{sinc}^2(\Delta\phi_i) \left( \frac{m_e}{2 \pi T}\right)^\frac{3}{2} \ee^{-\frac{m_ev^2}{2T}}
     \sim 1- \left( \num{2e-4}\,\frac{q^2}{\ks^2}\right)
\end{align}
These factors for electrons are already very close to unity and it is safe to neglect the equivalent suppression factors for the significantly heavier ions.
 
The effective screening scale as defined in eq.~\eqref{eq:effective_screening} still depends on the velocities of particles. But using eq.~\eqref{eq:sinc_suppression}, and the fact that electrons contribute at most 45\% to the square of the screening scale~(as shown in fig.~\ref{fig:el_contrib_to_ks}), we find that the thermal average of the difference between $\ks^2$ and $\kappa^2_\mathrm{eff}$ is given by
\begin{equation}
    \ks^2 - \langle \kappa^2_\mathrm{eff} \rangle_{v_i} \lsim \num{0.45e-4}\; q^2\, .
\end{equation}
This means that the resulting relative correction to the square of the form factor in eq.~\eqref{eq:formfactor_nonstatic} does not exceed~\num{5e-5}.
Furthermore, eqs.~\eqref{eq:sinc_suppression} and~\eqref{eq:sinc2_suppression} indicate that the $\mathrm{sinc}(\cdot)$ factors in eq.~\eqref{eq:formfactor_nonstatic} result in corrections by at most \num{e-4} when we conservatively assume that $q \lesssim \ks$.
 
In conclusion, we expect the square of the form factor, and thereby the total Primakoff rate, to only change by $\lsim 0.02\%$ when we drop the assumption of a static distribution of point-like charges.

\section{Description of the SolarAxionFlux library Python wrapper}\label{appendix:solaxlib}
Here we describe the Python frontend for our \texttt{SolarAxionFlux} \cpp library, which we developed based on existing code from our previous work~\cite{1810.07192,1908.10878}. The library is available at \url{https://github.com/sebhoof/SolarAxionFlux} under the BSD 3-clause license. The most up-to-date information, including e.g.\ a description on how to install the library, can be found following the Github link. The version used for this paper has been tagged as \ver.

To facilitate the reproducibility of this paper, the Python frontend wraps the underlying \cpp functions relevant for this paper. However, note that the underlying \cpp code and development branches contain more functions and objects that are not explicitly documented here.

\begin{table}
	\caption{Functions of the Python wrapper for \texttt{SolarAxionLib}. Arguments highlighted \texttt{\textbf{in bold}} must be lists or arrays; the default values of optional arguments are highlighted \texttt{\color{gray}in grey}. See the text for further explanations.\label{tab:library_capabilities}}
	\centering
    \renewcommand\cellalign{lc}
    \renewcommand\theadalign{lc}
    \setcellgapes{2.5pt}
    \makegapedcells
    {\small
	\begin{tabular}{ll}
		\toprule
		\makecell[l]{\textbf{Python function}} & \textbf{Arguments} \\
		\midrule
		\texttt{module\_info} & none \\
		\texttt{test\_module} & none \\
		\texttt{save\_solar\_model} & \texttt{\textbf{ergs}}, \texttt{solar\_model\_file}, \texttt{output\_file\_root}, \texttt{n\_radii\color{gray}=1000} \\
		\texttt{calculate\_spectra} & \makecell{\texttt{\textbf{ergs}},  \texttt{\textbf{radii}}, \texttt{solar\_model\_file}, \texttt{output\_file\_root}, \\ \texttt{process\color{gray}="Primakoff"}, \texttt{op\_code\color{gray}="OP"}} \\
		\texttt{calculate\_varied\_spectra} & \makecell{\texttt{\textbf{ergs}}, \texttt{solar\_model\_file}, \texttt{output\_file\_root}, \texttt{a{\color{gray}=0}}, \texttt{b{\color{gray}=0}},\\ \texttt{\textbf{c}{\color{gray}=[3e3,40,3]}}} \\
		\texttt{calculate\_reference\_counts} & \makecell{\texttt{\textbf{masses}}, \texttt{dataset}, \texttt{spectrum\_file\_P}, \texttt{spectrum\_file\_ABC{\color{gray}=""}},\\ \texttt{output\_file\_name{\color{gray}=""}}} \\
		\bottomrule
	\end{tabular}
	}
\end{table}
Table~\ref{tab:library_capabilities} provides an overview of the functions and their function arguments described in what follows. Note that \texttt{SolarAxionLib} ships with a number of standard solar models, which can be found in the \texttt{data/solar\_models/} folder. To use the Python frontend, the library has to be installed with Python support (turned on by default), which requires Python~3 and \href{https://pybind11.readthedocs.io}{the \texttt{pybind11} library} and \href{https://cython.org/}{the \texttt{cython} extension/package}.\footnote{The \texttt{SolarAxionFlux} library itself depends on the popular \href{https://www.gnu.org/software/gsl/}{\texttt{GSL} library}. More information on how to obtain all dependencies and how to install \texttt{SolarAxionLib} can be found at \url{https://github.com/sebhoof/SolarAxionFlux}.}

\subsection*{\texttt{module\_info()}}
Prints a string that contains the library version and the directory where it was installed. This function has no arguments.

\subsection*{\texttt{test\_module()}}
Runs a series of test routines for the library, which can take up to 10--15 minutes. The source code of this routine can be found in \texttt{include/solaxlib/tests.hpp} and should provide a good starting point for understanding the \cpp code of the library. This function has no arguments.

\subsection*{\texttt{save\_solar\_model(\texttt{\textbf{ergs}, solar\_model\_file, output\_file\_root, n\_radii\color{gray}=1000})}}
Saves the solar model from file \texttt{solar\_model\_file} as a table (in the file \texttt{output\_file\_root} plus the suffix \texttt{"\_model.dat"}) for \texttt{n\_radii} equally-spaced values of the allowed radius range. The file only contains all relevant quantities for axion physics. A separate table (suffix \texttt{"\_opacities.dat"}) stores the total opacity values for all combinations of the \texttt{n\_radii} radius values and the energies \texttt{\textbf{ergs}}. The headers of the output files contain information regarding the format.

\subsection*{\texttt{calculate\_spectra(ergs, radii, solar\_model\_file, output\_file\_root,\\ \phantom{\_\_}process{\color{gray}="Primakoff"}, op\_code{\color{gray}="OP"})}}
Saves files that contain the radius values on the solar disc (if \texttt{\textbf{radii}} contains more than one value), energy values (specified in \texttt{\textbf{ergs}}), and the corresponding differential axion fluxes as columns. The strings \texttt{solar\_model\_file} and \texttt{output\_file\_root} are paths to the solar model and output files (saved as \texttt{output\_file\_root} plus suffix \texttt{"\_plasmon.dat"}, \texttt{"\_P.dat"}, or \texttt{"\_ABC.dat"} for the plasmon, Primakoff, or \ABC flux, respectively). The string \texttt{process} can be \texttt{"plasmon"}, \texttt{"Primakoff"}, \texttt{"ABC"}, or \texttt{"all"}. The string \texttt{op\_code} specifies the opacity code to be used (can be \texttt{"OP"}, \texttt{"OPAS"}, \texttt{"ATOMIC"}, or \texttt{"LEDCOP"}). The headers of the output files contain information regarding the format.

\subsection*{\texttt{calculate\_varied\_spectra(ergs, solar\_model\_file, output\_file\_root, a{\color{gray}=0}, b{\color{gray}=0},\\ \phantom{\_\_}c{\color{gray}=[3e3,40,3]})}} 
Saves files, which contain the energy values (specified in \texttt{\textbf{ergs}}) and the corresponding differential plasmon, Primakoff and ABC fluxes as columns (integrated over the entire Sun). The float variables \texttt{a} and \texttt{b} refer to the opacity variation parameters in eq.~\eqref{eq:opacity_variation}, the array \texttt{c} contains the normalisations $B_\text{rad}$, $B_\text{tach}$, and $B_\text{outer}$ in eq.~\eqref{eq:solar_b_fields}. The remaining function arguments are identical to those in \texttt{calculate\_spectra}.

\subsection*{\texttt{calculate\_reference\_counts(masses, dataset, spectrum\_file\_P,\\
\phantom{\_\_}spectrum\_file\_ABC{\color{gray}=""}, output\_file\_name{\color{gray}=""})}}
Returns an array whose columns correspond to the axion mass values (in \si{\eV}) contained in \texttt{\textbf{masses}}, the central energy of each bin for the experimental setup \texttt{dataset}, the reference photon counts obtained by integrating a tabulated Primakoff and/or plasmon spectrum (from file \texttt{spectrum\_file\_P}) and, optionally, a tabulated \ABC spectrum (from file \texttt{spectrum\_file\_ABC}). Note that the distinction is important because the code will use a (slower but more accurate) integrator for the ABC spectrum. The variable \texttt{output\_file\_name}, optionally, specifies the full name of the output file, which has the same format as the returned array. Currently allowed options for \texttt{dataset} are \texttt{"CAST2007"} and \texttt{"CAST2017\_A"}, \texttt{"CAST20
2
17\_B"}, \ldots, \texttt{"CAST2017\_L"}, corresponding to data from refs.~\cite{hep-ex/0702006} and~\cite{1705.02290}~(see ref.~\cite{1810.07192} for details on the implementation).

\renewcommand{\baselinestretch}{1}\normalfont
\setlength{\bibsep}{0.5em plus 0.3ex}
\bibliographystyle{apsrev_mod}
\bibliography{bibliography}

\begin{thebibliography}{100}%
\makeatletter
\providecommand \@ifxundefined [1]{%
 \@ifx{#1\undefined}
}%
\providecommand \@ifnum [1]{%
 \ifnum #1\expandafter \@firstoftwo
 \else \expandafter \@secondoftwo
 \fi
}%
\providecommand \@ifx [1]{%
 \ifx #1\expandafter \@firstoftwo
 \else \expandafter \@secondoftwo
 \fi
}%
\providecommand \natexlab [1]{#1}%
\providecommand \enquote  [1]{``#1''}%
\providecommand \bibnamefont  [1]{#1}%
\providecommand \bibfnamefont [1]{#1}%
\providecommand \citenamefont [1]{#1}%
\providecommand \href@noop [0]{\@secondoftwo}%
\providecommand \href [0]{\begingroup \@sanitize@url \@href}%
\providecommand \@href[1]{\@@startlink{#1}\@@href}%
\providecommand \@@href[1]{\endgroup#1\@@endlink}%
\providecommand \@sanitize@url [0]{\catcode `\\12\catcode `\$12\catcode
  `\&12\catcode `\#12\catcode `\^12\catcode `\_12\catcode `\%12\relax}%
\providecommand \@@startlink[1]{}%
\providecommand \@@endlink[0]{}%
\providecommand \url  [0]{\begingroup\@sanitize@url \@url }%
\providecommand \@url [1]{\endgroup\@href {#1}{\urlprefix }}%
\providecommand \urlprefix  [0]{URL }%
\providecommand \Eprint[0]{\href }%
\providecommand \doibase [0]{http://dx.doi.org/}%
\providecommand \selectlanguage [0]{\@gobble}%
\providecommand \bibinfo  [0]{\@secondoftwo}%
\providecommand \bibfield  [0]{\@secondoftwo}%
\providecommand \translation [1]{[#1]}%
\providecommand \BibitemOpen [0]{}%
\providecommand \bibitemStop [0]{}%
\providecommand \bibitemNoStop [0]{.\EOS\space}%
\providecommand \EOS [0]{\spacefactor3000\relax}%
\providecommand \BibitemShut  [1]{\csname bibitem#1\endcsname}%
\let\auto@bib@innerbib\@empty
\bibitem [{\citenamefont {{Peccei}}\ and\ \citenamefont
  {{Quinn}}(1977{\natexlab{a}})}]{1977_pq_axion1}%
  \BibitemOpen
  \bibfield  {author} {\bibinfo {author} {\bibfnamefont {R.~D.}\ \bibnamefont
  {{Peccei}}}\ \bibnamefont {and}\ \bibinfo {author} {\bibfnamefont {H.~R.}\
  \bibnamefont {{Quinn}}},\ }\bibfield  {title} {\enquote {\bibinfo {title}
  {{CP conservation in the presence of pseudoparticles}},}\ }\href {\doibase
  10.1103/PhysRevLett.38.1440} {\bibfield  {journal} {\bibinfo  {journal}
  {Physical Review Letters}\ }\textbf {\bibinfo {volume} {38}},\ \bibinfo
  {pages} {1440} (\bibinfo {year} {1977}{\natexlab{a}})}\BibitemShut {NoStop}%
\bibitem [{\citenamefont {{Peccei}}\ and\ \citenamefont
  {{Quinn}}(1977{\natexlab{b}})}]{1977_pq_axion2}%
  \BibitemOpen
  \bibfield  {author} {\bibinfo {author} {\bibfnamefont {R.~D.}\ \bibnamefont
  {{Peccei}}}\ \bibnamefont {and}\ \bibinfo {author} {\bibfnamefont {H.~R.}\
  \bibnamefont {{Quinn}}},\ }\bibfield  {title} {\enquote {\bibinfo {title}
  {{Constraints imposed by CP conservation in the presence of
  pseudoparticles}},}\ }\href {\doibase 10.1103/PhysRevD.16.1791} {\bibfield
  {journal} {\bibinfo  {journal} {\prd}\ }\textbf {\bibinfo {volume} {16}},\
  \bibinfo {pages} {1791} (\bibinfo {year} {1977}{\natexlab{b}})}\BibitemShut
  {NoStop}%
\bibitem [{\citenamefont {{Weinberg}}(1978)}]{1978_weinberg_axion}%
  \BibitemOpen
  \bibfield  {author} {\bibinfo {author} {\bibfnamefont {S.}~\bibnamefont
  {{Weinberg}}},\ }\bibfield  {title} {\enquote {\bibinfo {title} {{A new light
  boson?}}}\ }\href {\doibase 10.1103/PhysRevLett.40.223} {\bibfield  {journal}
  {\bibinfo  {journal} {Physical Review Letters}\ }\textbf {\bibinfo {volume}
  {40}},\ \bibinfo {pages} {223} (\bibinfo {year} {1978})}\BibitemShut
  {NoStop}%
\bibitem [{\citenamefont {{Wilczek}}(1978)}]{1978_wilczek_axion}%
  \BibitemOpen
  \bibfield  {author} {\bibinfo {author} {\bibfnamefont {F.}~\bibnamefont
  {{Wilczek}}},\ }\bibfield  {title} {\enquote {\bibinfo {title} {{Problem of
  strong P and T invariance in the presence of instantons}},}\ }\href {\doibase
  10.1103/PhysRevLett.40.279} {\bibfield  {journal} {\bibinfo  {journal}
  {Physical Review Letters}\ }\textbf {\bibinfo {volume} {40}},\ \bibinfo
  {pages} {279} (\bibinfo {year} {1978})}\BibitemShut {NoStop}%
\bibitem [{\citenamefont {{Kim}}(1987)}]{Kim:1986ax}%
  \BibitemOpen
  \bibfield  {author} {\bibinfo {author} {\bibfnamefont {J.~E.}\ \bibnamefont
  {{Kim}}},\ }\bibfield  {title} {\enquote {\bibinfo {title} {{Light
  pseudoscalars, particle physics and cosmology.}}}\ }\href {\doibase
  10.1016/0370-1573(87)90017-2} {\bibfield  {journal} {\bibinfo  {journal}
  {\physrep}\ }\textbf {\bibinfo {volume} {150}},\ \bibinfo {pages} {1}
  (\bibinfo {year} {1987})}\BibitemShut {NoStop}%
\bibitem [{\citenamefont {{Jaeckel}}\ and\ \citenamefont
  {{Ringwald}}(2010)}]{1002.0329}%
  \BibitemOpen
  \bibfield  {author} {\bibinfo {author} {\bibfnamefont {J.}~\bibnamefont
  {{Jaeckel}}}\ \bibnamefont {and}\ \bibinfo {author} {\bibfnamefont
  {A.}~\bibnamefont {{Ringwald}}},\ }\bibfield  {title} {\enquote {\bibinfo
  {title} {{The Low-Energy Frontier of Particle Physics}},}\ }\href {\doibase
  10.1146/annurev.nucl.012809.104433} {\bibfield  {journal} {\bibinfo
  {journal} {Annual Review of Nuclear and Particle Science}\ }\textbf {\bibinfo
  {volume} {60}},\ \bibinfo {pages} {405} (\bibinfo {year} {2010})},\
  \Eprint{http://arxiv.org/abs/1002.0329}{arXiv:1002.0329 [hep-ph]}\BibitemShut
  {NoStop}%
\bibitem [{\citenamefont {{Preskill}}\ \emph {et~al.}(1983)\citenamefont
  {{Preskill}}, \citenamefont {{Wise}},\ and\ \citenamefont
  {{Wilczek}}}]{Preskill:1982cy}%
  \BibitemOpen
  \bibfield  {author} {\bibinfo {author} {\bibfnamefont {J.}~\bibnamefont
  {{Preskill}}}, \bibinfo {author} {\bibfnamefont {M.~B.}\ \bibnamefont
  {{Wise}}},\ \bibnamefont {and}\ \bibinfo {author} {\bibfnamefont
  {F.}~\bibnamefont {{Wilczek}}},\ }\bibfield  {title} {\enquote {\bibinfo
  {title} {{Cosmology of the invisible axion}},}\ }\href {\doibase
  10.1016/0370-2693(83)90637-8} {\bibfield  {journal} {\bibinfo  {journal}
  {Physics Letters B}\ }\textbf {\bibinfo {volume} {120}},\ \bibinfo {pages}
  {127} (\bibinfo {year} {1983})}\BibitemShut {NoStop}%
\bibitem [{\citenamefont {{Abbott}}\ and\ \citenamefont
  {{Sikivie}}(1983)}]{Abbott:1982af}%
  \BibitemOpen
  \bibfield  {author} {\bibinfo {author} {\bibfnamefont {L.~F.}\ \bibnamefont
  {{Abbott}}}\ \bibnamefont {and}\ \bibinfo {author} {\bibfnamefont
  {P.}~\bibnamefont {{Sikivie}}},\ }\bibfield  {title} {\enquote {\bibinfo
  {title} {{A cosmological bound on the invisible axion}},}\ }\href {\doibase
  10.1016/0370-2693(83)90638-X} {\bibfield  {journal} {\bibinfo  {journal}
  {Physics Letters B}\ }\textbf {\bibinfo {volume} {120}},\ \bibinfo {pages}
  {133} (\bibinfo {year} {1983})}\BibitemShut {NoStop}%
\bibitem [{\citenamefont {{Dine}}\ and\ \citenamefont
  {{Fischler}}(1983)}]{Dine:1982ah}%
  \BibitemOpen
  \bibfield  {author} {\bibinfo {author} {\bibfnamefont {M.}~\bibnamefont
  {{Dine}}}\ \bibnamefont {and}\ \bibinfo {author} {\bibfnamefont
  {W.}~\bibnamefont {{Fischler}}},\ }\bibfield  {title} {\enquote {\bibinfo
  {title} {{The not-so-harmless axion}},}\ }\href {\doibase
  10.1016/0370-2693(83)90639-1} {\bibfield  {journal} {\bibinfo  {journal}
  {Physics Letters B}\ }\textbf {\bibinfo {volume} {120}},\ \bibinfo {pages}
  {137} (\bibinfo {year} {1983})}\BibitemShut {NoStop}%
\bibitem [{\citenamefont {{Turner}}(1983)}]{Turner:1983he}%
  \BibitemOpen
  \bibfield  {author} {\bibinfo {author} {\bibfnamefont {M.~S.}\ \bibnamefont
  {{Turner}}},\ }\bibfield  {title} {\enquote {\bibinfo {title} {{Coherent
  scalar-field oscillations in an expanding universe}},}\ }\href {\doibase
  10.1103/PhysRevD.28.1243} {\bibfield  {journal} {\bibinfo  {journal} {\prd}\
  }\textbf {\bibinfo {volume} {28}},\ \bibinfo {pages} {1243} (\bibinfo {year}
  {1983})}\BibitemShut {NoStop}%
\bibitem [{\citenamefont {Turner}(1986)}]{Turner:1985si}%
  \BibitemOpen
  \bibfield  {author} {\bibinfo {author} {\bibfnamefont {M.~S.}\ \bibnamefont
  {Turner}},\ }\bibfield  {title} {\enquote {\bibinfo {title} {Cosmic and local
  mass density of ``invisible'' axions},}\ }\href {\doibase
  10.1103/physrevd.33.889} {\bibfield  {journal} {\bibinfo  {journal} {Physical
  Review D}\ }\textbf {\bibinfo {volume} {33}},\ \bibinfo {pages} {889}
  (\bibinfo {year} {1986})}\BibitemShut {NoStop}%
\bibitem [{\citenamefont {{Arias}}\ \emph {et~al.}(2012)\citenamefont
  {{Arias}}, \citenamefont {{Cadamuro}}, \citenamefont {{Goodsell}},
  \citenamefont {{Jaeckel}}, \citenamefont {{Redondo}},\ and\ \citenamefont
  {{Ringwald}}}]{1201.5902}%
  \BibitemOpen
  \bibfield  {author} {\bibinfo {author} {\bibfnamefont {P.}~\bibnamefont
  {{Arias}}}, \bibinfo {author} {\bibfnamefont {D.}~\bibnamefont {{Cadamuro}}},
  \bibinfo {author} {\bibfnamefont {M.}~\bibnamefont {{Goodsell}}},
  \bibnamefont {et~al.},\ }\bibfield  {title} {\enquote {\bibinfo {title}
  {{WISPy cold dark matter}},}\ }\href {\doibase 10.1088/1475-7516/2012/06/013}
  {\bibfield  {journal} {\bibinfo  {journal} {\jcap}\ }\textbf {\bibinfo
  {volume} {6}},\ \bibinfo {eid} {013} (\bibinfo {year} {2012})},\
  \Eprint{http://arxiv.org/abs/1201.5902}{arXiv:1201.5902 [hep-ph]}\BibitemShut
  {NoStop}%
\bibitem [{\citenamefont {{Giannotti}}\ \emph {et~al.}(2016)\citenamefont
  {{Giannotti}}, \citenamefont {{Irastorza}}, \citenamefont {{Redondo}},\ and\
  \citenamefont {{Ringwald}}}]{1512.08108}%
  \BibitemOpen
  \bibfield  {author} {\bibinfo {author} {\bibfnamefont {M.}~\bibnamefont
  {{Giannotti}}}, \bibinfo {author} {\bibfnamefont {I.}~\bibnamefont
  {{Irastorza}}}, \bibinfo {author} {\bibfnamefont {J.}~\bibnamefont
  {{Redondo}}},\ \bibnamefont {and}\ \bibinfo {author} {\bibfnamefont
  {A.}~\bibnamefont {{Ringwald}}},\ }\bibfield  {title} {\enquote {\bibinfo
  {title} {{Cool WISPs for stellar cooling excesses}},}\ }\href {\doibase
  10.1088/1475-7516/2016/05/057} {\bibfield  {journal} {\bibinfo  {journal}
  {\jcap}\ }\textbf {\bibinfo {volume} {5}},\ \bibinfo {eid} {057} (\bibinfo
  {year} {2016})},\ \Eprint{http://arxiv.org/abs/1512.08108}{arXiv:1512.08108
  [astro-ph.HE]}\BibitemShut {NoStop}%
\bibitem [{\citenamefont {{Giannotti}}\ \emph {et~al.}(2017)\citenamefont
  {{Giannotti}}, \citenamefont {{Irastorza}}, \citenamefont {{Redondo}},
  \citenamefont {{Ringwald}},\ and\ \citenamefont {{Saikawa}}}]{1708.02111}%
  \BibitemOpen
  \bibfield  {author} {\bibinfo {author} {\bibfnamefont {M.}~\bibnamefont
  {{Giannotti}}}, \bibinfo {author} {\bibfnamefont {I.~G.}\ \bibnamefont
  {{Irastorza}}}, \bibinfo {author} {\bibfnamefont {J.}~\bibnamefont
  {{Redondo}}}, \bibnamefont {et~al.},\ }\bibfield  {title} {\enquote {\bibinfo
  {title} {{Stellar recipes for axion hunters}},}\ }\href {\doibase
  10.1088/1475-7516/2017/10/010} {\bibfield  {journal} {\bibinfo  {journal}
  {\jcap}\ }\textbf {\bibinfo {volume} {10}},\ \bibinfo {eid} {010} (\bibinfo
  {year} {2017})},\ \Eprint{http://arxiv.org/abs/1708.02111}{arXiv:1708.02111
  [hep-ph]}\BibitemShut {NoStop}%
\bibitem [{\citenamefont {{Vysotsskii}}\ \emph {et~al.}(1978)\citenamefont
  {{Vysotsskii}}, \citenamefont {{Zel'dovich}}, \citenamefont {{Khlopov}},\
  and\ \citenamefont {{Chechetkin}}}]{Vysotsky:1978dc}%
  \BibitemOpen
  \bibfield  {author} {\bibinfo {author} {\bibfnamefont {M.~I.}\ \bibnamefont
  {{Vysotsskii}}}, \bibinfo {author} {\bibfnamefont {Y.~B.}\ \bibnamefont
  {{Zel'dovich}}}, \bibinfo {author} {\bibfnamefont {M.~Y.}\ \bibnamefont
  {{Khlopov}}},\ \bibnamefont {and}\ \bibinfo {author} {\bibfnamefont {V.~M.}\
  \bibnamefont {{Chechetkin}}},\ }\bibfield  {title} {\enquote {\bibinfo
  {title} {{Some astrophysical limitations on the axion mass}},}\ }\href
  {http://www.jetpletters.ac.ru/ps/1552/article_23764.shtml} {\bibfield
  {journal} {\bibinfo  {journal} {Soviet Journal of Experimental and
  Theoretical Physics Letters}\ }\textbf {\bibinfo {volume} {27}},\ \bibinfo
  {pages} {502} (\bibinfo {year} {1978})},\ \bibinfo {note} {also in Pis'ma v
  Zh.\ Eksp.\ Teor.\ Fiz.\ \textbf{27}, 533 (1978).}\BibitemShut {Stop}%
\bibitem [{\citenamefont {Dicus}\ \emph {et~al.}(1978)\citenamefont {Dicus},
  \citenamefont {Kolb}, \citenamefont {Teplitz},\ and\ \citenamefont
  {Wagoner}}]{Dicus:1978fp}%
  \BibitemOpen
  \bibfield  {author} {\bibinfo {author} {\bibfnamefont {D.~A.}\ \bibnamefont
  {Dicus}}, \bibinfo {author} {\bibfnamefont {E.~W.}\ \bibnamefont {Kolb}},
  \bibinfo {author} {\bibfnamefont {V.~L.}\ \bibnamefont {Teplitz}},\
  \bibnamefont {and}\ \bibinfo {author} {\bibfnamefont {R.~V.}\ \bibnamefont
  {Wagoner}},\ }\bibfield  {title} {\enquote {\bibinfo {title} {{Astrophysical
  Bounds on the Masses of Axions and Higgs Particles}},}\ }\href {\doibase
  10.1103/PhysRevD.18.1829} {\bibfield  {journal} {\bibinfo  {journal} {Phys.
  Rev. D}\ }\textbf {\bibinfo {volume} {18}},\ \bibinfo {pages} {1829}
  (\bibinfo {year} {1978})}\BibitemShut {NoStop}%
\bibitem [{\citenamefont {Dicus}\ \emph {et~al.}(1980)\citenamefont {Dicus},
  \citenamefont {Kolb}, \citenamefont {Teplitz},\ and\ \citenamefont
  {Wagoner}}]{PhysRevD.22.839}%
  \BibitemOpen
  \bibfield  {author} {\bibinfo {author} {\bibfnamefont {D.~A.}\ \bibnamefont
  {Dicus}}, \bibinfo {author} {\bibfnamefont {E.~W.}\ \bibnamefont {Kolb}},
  \bibinfo {author} {\bibfnamefont {V.~L.}\ \bibnamefont {Teplitz}},\
  \bibnamefont {and}\ \bibinfo {author} {\bibfnamefont {R.~V.}\ \bibnamefont
  {Wagoner}},\ }\bibfield  {title} {\enquote {\bibinfo {title} {Astrophysical
  bounds on very-low-mass axions},}\ }\href {\doibase 10.1103/PhysRevD.22.839}
  {\bibfield  {journal} {\bibinfo  {journal} {Phys. Rev. D}\ }\textbf {\bibinfo
  {volume} {22}},\ \bibinfo {pages} {839} (\bibinfo {year} {1980})}\BibitemShut
  {NoStop}%
\bibitem [{\citenamefont {Fukugita}\ \emph
  {et~al.}(1982{\natexlab{a}})\citenamefont {Fukugita}, \citenamefont
  {Watamura},\ and\ \citenamefont {Yoshimura}}]{PhysRevLett.48.1522}%
  \BibitemOpen
  \bibfield  {author} {\bibinfo {author} {\bibfnamefont {M.}~\bibnamefont
  {Fukugita}}, \bibinfo {author} {\bibfnamefont {S.}~\bibnamefont {Watamura}},\
  \bibnamefont {and}\ \bibinfo {author} {\bibfnamefont {M.}~\bibnamefont
  {Yoshimura}},\ }\bibfield  {title} {\enquote {\bibinfo {title} {Light
  pseudoscalar particle and stellar energy loss},}\ }\href {\doibase
  10.1103/PhysRevLett.48.1522} {\bibfield  {journal} {\bibinfo  {journal}
  {Phys. Rev. Lett.}\ }\textbf {\bibinfo {volume} {48}},\ \bibinfo {pages}
  {1522} (\bibinfo {year} {1982}{\natexlab{a}})}\BibitemShut {NoStop}%
\bibitem [{\citenamefont {Fukugita}\ \emph
  {et~al.}(1982{\natexlab{b}})\citenamefont {Fukugita}, \citenamefont
  {Watamura},\ and\ \citenamefont {Yoshimura}}]{PhysRevD.26.1840}%
  \BibitemOpen
  \bibfield  {author} {\bibinfo {author} {\bibfnamefont {M.}~\bibnamefont
  {Fukugita}}, \bibinfo {author} {\bibfnamefont {S.}~\bibnamefont {Watamura}},\
  \bibnamefont {and}\ \bibinfo {author} {\bibfnamefont {M.}~\bibnamefont
  {Yoshimura}},\ }\bibfield  {title} {\enquote {\bibinfo {title} {Astrophysical
  constraints on a new light axion and other weakly interacting particles},}\
  }\href {\doibase 10.1103/PhysRevD.26.1840} {\bibfield  {journal} {\bibinfo
  {journal} {Phys. Rev. D}\ }\textbf {\bibinfo {volume} {26}},\ \bibinfo
  {pages} {1840} (\bibinfo {year} {1982}{\natexlab{b}})}\BibitemShut {NoStop}%
\bibitem [{\citenamefont {Dearborn}\ \emph {et~al.}(1986)\citenamefont
  {Dearborn}, \citenamefont {Schramm},\ and\ \citenamefont
  {Steigman}}]{PhysRevLett.56.26}%
  \BibitemOpen
  \bibfield  {author} {\bibinfo {author} {\bibfnamefont {D.~S.~P.}\
  \bibnamefont {Dearborn}}, \bibinfo {author} {\bibfnamefont {D.~N.}\
  \bibnamefont {Schramm}},\ \bibnamefont {and}\ \bibinfo {author}
  {\bibfnamefont {G.}~\bibnamefont {Steigman}},\ }\bibfield  {title} {\enquote
  {\bibinfo {title} {Astrophysical constraints on the couplings of axions,
  majorons, and familons},}\ }\href {\doibase 10.1103/PhysRevLett.56.26}
  {\bibfield  {journal} {\bibinfo  {journal} {Phys. Rev. Lett.}\ }\textbf
  {\bibinfo {volume} {56}},\ \bibinfo {pages} {26} (\bibinfo {year}
  {1986})}\BibitemShut {NoStop}%
\bibitem [{\citenamefont {{Raffelt}}(1986)}]{Raffelt:1985nk}%
  \BibitemOpen
  \bibfield  {author} {\bibinfo {author} {\bibfnamefont {G.~G.}\ \bibnamefont
  {{Raffelt}}},\ }\bibfield  {title} {\enquote {\bibinfo {title}
  {{Astrophysical axion bounds diminished by screening effects}},}\ }\href
  {\doibase 10.1103/PhysRevD.33.897} {\bibfield  {journal} {\bibinfo  {journal}
  {\prd}\ }\textbf {\bibinfo {volume} {33}},\ \bibinfo {pages} {897} (\bibinfo
  {year} {1986})}\BibitemShut {NoStop}%
\bibitem [{\citenamefont {{Caputo}}\ \emph {et~al.}(2020)\citenamefont
  {{Caputo}}, \citenamefont {{Millar}},\ and\ \citenamefont
  {{Vitagliano}}}]{2005.00078}%
  \BibitemOpen
  \bibfield  {author} {\bibinfo {author} {\bibfnamefont {A.}~\bibnamefont
  {{Caputo}}}, \bibinfo {author} {\bibfnamefont {A.~J.}\ \bibnamefont
  {{Millar}}},\ \bibnamefont {and}\ \bibinfo {author} {\bibfnamefont
  {E.}~\bibnamefont {{Vitagliano}}},\ }\bibfield  {title} {\enquote {\bibinfo
  {title} {{Revisiting longitudinal plasmon-axion conversion in external
  magnetic fields}},}\ }\href {\doibase 10.1103/PhysRevD.101.123004} {\bibfield
   {journal} {\bibinfo  {journal} {\prd}\ }\textbf {\bibinfo {volume} {101}},\
  \bibinfo {eid} {123004} (\bibinfo {year} {2020})},\
  \Eprint{http://arxiv.org/abs/2005.00078}{arXiv:2005.00078
  [hep-ph]}\BibitemShut {NoStop}%
\bibitem [{\citenamefont {{O'Hare}}\ \emph {et~al.}(2020)\citenamefont
  {{O'Hare}}, \citenamefont {{Caputo}}, \citenamefont {{Millar}},\ and\
  \citenamefont {{Vitagliano}}}]{2006.10415}%
  \BibitemOpen
  \bibfield  {author} {\bibinfo {author} {\bibfnamefont {C.~A.~J.}\
  \bibnamefont {{O'Hare}}}, \bibinfo {author} {\bibfnamefont {A.}~\bibnamefont
  {{Caputo}}}, \bibinfo {author} {\bibfnamefont {A.~J.}\ \bibnamefont
  {{Millar}}},\ \bibnamefont {and}\ \bibinfo {author} {\bibfnamefont
  {E.}~\bibnamefont {{Vitagliano}}},\ }\bibfield  {title} {\enquote {\bibinfo
  {title} {{Axion helioscopes as solar magnetometers}},}\ }\href {\doibase
  10.1103/PhysRevD.102.043019} {\bibfield  {journal} {\bibinfo  {journal}
  {\prd}\ }\textbf {\bibinfo {volume} {102}},\ \bibinfo {eid} {043019}
  (\bibinfo {year} {2020})},\
  \Eprint{http://arxiv.org/abs/2006.10415}{arXiv:2006.10415
  [astro-ph.CO]}\BibitemShut {NoStop}%
\bibitem [{\citenamefont {{Van Tilburg}}(2020)}]{2006.12431}%
  \BibitemOpen
  \bibfield  {author} {\bibinfo {author} {\bibfnamefont {K.}~\bibnamefont {{Van
  Tilburg}}},\ }\bibfield  {title} {\enquote {\bibinfo {title} {{Stellar Basins
  of Gravitationally Bound Particles}},}\ }\href@noop {} {\bibfield  {journal}
  {\bibinfo  {journal} {arXiv e-prints}\ ,\ \bibinfo {eid} {arXiv:2006.12431}}
  (\bibinfo {year} {2020})},\
  \Eprint{http://arxiv.org/abs/2006.12431}{arXiv:2006.12431
  [hep-ph]}\BibitemShut {NoStop}%
\bibitem [{\citenamefont {{Guarini}}\ \emph {et~al.}(2020)\citenamefont
  {{Guarini}}, \citenamefont {{Carenza}}, \citenamefont {{Gal{\'a}n}},
  \citenamefont {{Giannotti}},\ and\ \citenamefont {{Mirizzi}}}]{2010.06601}%
  \BibitemOpen
  \bibfield  {author} {\bibinfo {author} {\bibfnamefont {E.}~\bibnamefont
  {{Guarini}}}, \bibinfo {author} {\bibfnamefont {P.}~\bibnamefont
  {{Carenza}}}, \bibinfo {author} {\bibfnamefont {J.}~\bibnamefont
  {{Gal{\'a}n}}}, \bibnamefont {et~al.},\ }\bibfield  {title} {\enquote
  {\bibinfo {title} {{Production of axionlike particles from photon conversions
  in large-scale solar magnetic fields}},}\ }\href {\doibase
  10.1103/PhysRevD.102.123024} {\bibfield  {journal} {\bibinfo  {journal}
  {\prd}\ }\textbf {\bibinfo {volume} {102}},\ \bibinfo {eid} {123024}
  (\bibinfo {year} {2020})},\
  \Eprint{http://arxiv.org/abs/2010.06601}{arXiv:2010.06601
  [hep-ph]}\BibitemShut {NoStop}%
\bibitem [{\citenamefont {{Aprile}}\ \emph {et~al.}(2020)\citenamefont
  {{Aprile}}, \citenamefont {{Aalbers}}, \citenamefont {{Agostini}},
  \citenamefont {{Alfonsi}}, \citenamefont {{Althueser}}, \citenamefont
  {{Amaro}}, \citenamefont {{Antochi}}, \citenamefont {{Angelino}},
  \citenamefont {{Angevaare}}, \citenamefont {{Arneodo}}, \citenamefont
  {{Barge}}, \citenamefont {{Baudis}}, \citenamefont {{Bauermeister}},
  \citenamefont {{Bellagamba}}, \citenamefont {{Benabderrahmane}},
  \citenamefont {{Berger}}, \citenamefont {{Brown}}, \citenamefont {{Brown}},
  \citenamefont {{Bruenner}}, \citenamefont {{Bruno}}, \citenamefont
  {{Budnik}}, \citenamefont {{Capelli}}, \citenamefont {{Cardoso}},
  \citenamefont {{Cichon}}, \citenamefont {{Cimmino}}, \citenamefont {{Clark}},
  \citenamefont {{Coderre}}, \citenamefont {{Colijn}}, \citenamefont
  {{Conrad}}, \citenamefont {{Cussonneau}}, \citenamefont {{Decowski}},
  \citenamefont {{Depoian}}, \citenamefont {{di Gangi}}, \citenamefont {{di
  Giovanni}}, \citenamefont {{di Stefano}}, \citenamefont {{Diglio}},
  \citenamefont {{Elykov}}, \citenamefont {{Eurin}}, \citenamefont {{Ferella}},
  \citenamefont {{Fulgione}}, \citenamefont {{Gaemers}}, \citenamefont
  {{Gaior}}, \citenamefont {{Galloway}}, \citenamefont {{Gao}}, \citenamefont
  {{Grandi}}, \citenamefont {{Hasterok}}, \citenamefont {{Hils}}, \citenamefont
  {{Hiraide}}, \citenamefont {{Hoetzsch}}, \citenamefont {{Howlett}},
  \citenamefont {{Iacovacci}}, \citenamefont {{Itow}}, \citenamefont {{Joerg}},
  \citenamefont {{Kato}}, \citenamefont {{Kazama}}, \citenamefont
  {{Kobayashi}}, \citenamefont {{Koltman}}, \citenamefont {{Kopec}},
  \citenamefont {{Landsman}}, \citenamefont {{Lang}}, \citenamefont
  {{Levinson}}, \citenamefont {{Lin}}, \citenamefont {{Lindemann}},
  \citenamefont {{Lindner}}, \citenamefont {{Lombardi}}, \citenamefont
  {{Long}}, \citenamefont {{Lopes}}, \citenamefont {{L{\'o}pez Fune}},
  \citenamefont {{Macolino}}, \citenamefont {{Mahlstedt}}, \citenamefont
  {{Mancuso}}, \citenamefont {{Manenti}}, \citenamefont {{Manfredini}},
  \citenamefont {{Marignetti}}, \citenamefont {{Marrod{\'a}n Undagoitia}},
  \citenamefont {{Martens}}, \citenamefont {{Masbou}}, \citenamefont
  {{Masson}}, \citenamefont {{Mastroianni}}, \citenamefont {{Messina}},
  \citenamefont {{Miuchi}}, \citenamefont {{Mizukoshi}}, \citenamefont
  {{Molinario}}, \citenamefont {{Mor{\^a}}}, \citenamefont {{Moriyama}},
  \citenamefont {{Mosbacher}}, \citenamefont {{Murra}}, \citenamefont
  {{Naganoma}}, \citenamefont {{Ni}}, \citenamefont {{Oberlack}}, \citenamefont
  {{Odgers}}, \citenamefont {{Palacio}}, \citenamefont {{Pelssers}},
  \citenamefont {{Peres}}, \citenamefont {{Pienaar}}, \citenamefont
  {{Pizzella}}, \citenamefont {{Plante}}, \citenamefont {{Qin}}, \citenamefont
  {{Qiu}}, \citenamefont {{Ram{\'\i}rez Garc{\'\i}a}}, \citenamefont
  {{Reichard}}, \citenamefont {{Rocchetti}}, \citenamefont {{Rupp}},
  \citenamefont {{Dos Santos}}, \citenamefont {{Sartorelli}}, \citenamefont
  {{{\v{S}}ar{\v{c}}evi{\'c}}}, \citenamefont {{Scheibelhut}}, \citenamefont
  {{Schreiner}}, \citenamefont {{Schulte}}, \citenamefont {{Schumann}},
  \citenamefont {{Scotto Lavina}}, \citenamefont {{Selvi}}, \citenamefont
  {{Semeria}}, \citenamefont {{Shagin}}, \citenamefont {{Shockley}},
  \citenamefont {{Silva}}, \citenamefont {{Simgen}}, \citenamefont {{Takeda}},
  \citenamefont {{Therreau}}, \citenamefont {{Thers}}, \citenamefont
  {{Toschi}}, \citenamefont {{Trinchero}}, \citenamefont {{Tunnell}},
  \citenamefont {{Vargas}}, \citenamefont {{Volta}}, \citenamefont {{Wang}},
  \citenamefont {{Wei}}, \citenamefont {{Weinheimer}}, \citenamefont {{Weiss}},
  \citenamefont {{Wenz}}, \citenamefont {{Wittweg}}, \citenamefont {{Xu}},
  \citenamefont {{Yamashita}}, \citenamefont {{Ye}}, \citenamefont
  {{Zavattini}}, \citenamefont {{Zhang}}, \citenamefont {{Zhu}}, \citenamefont
  {{Zopounidis}},\ and\ \citenamefont {{Xenon Collaboration}}}]{2006.09721}%
  \BibitemOpen
  \bibfield  {author} {\bibinfo {author} {\bibfnamefont {E.}~\bibnamefont
  {{Aprile}}}, \bibinfo {author} {\bibfnamefont {J.}~\bibnamefont {{Aalbers}}},
  \bibinfo {author} {\bibfnamefont {F.}~\bibnamefont {{Agostini}}},
  \bibnamefont {et~al.},\ }\bibfield  {title} {\enquote {\bibinfo {title}
  {{Excess electronic recoil events in XENON1T}},}\ }\href {\doibase
  10.1103/PhysRevD.102.072004} {\bibfield  {journal} {\bibinfo  {journal}
  {\prd}\ }\textbf {\bibinfo {volume} {102}},\ \bibinfo {eid} {072004}
  (\bibinfo {year} {2020})},\
  \Eprint{http://arxiv.org/abs/2006.09721}{arXiv:2006.09721
  [hep-ex]}\BibitemShut {NoStop}%
\bibitem [{\citenamefont {Sikivie}(1983)}]{1983_sikivie}%
  \BibitemOpen
  \bibfield  {author} {\bibinfo {author} {\bibfnamefont {P.}~\bibnamefont
  {Sikivie}},\ }\bibfield  {title} {\enquote {\bibinfo {title} {{Experimental
  Tests of the ``Invisible'' Axion}},}\ }\href {\doibase
  10.1103/PhysRevLett.51.1415} {\bibfield  {journal} {\bibinfo  {journal}
  {Physical Review Letters}\ }\textbf {\bibinfo {volume} {51}},\ \bibinfo
  {pages} {1415} (\bibinfo {year} {1983})},\ \bibinfo {note} {[\textit{Erratum
  ibid.}\ \textbf{52}, 695 (1984)]}\BibitemShut {NoStop}%
\bibitem [{\citenamefont {Sikivie}(1985)}]{1985_sikivie}%
  \BibitemOpen
  \bibfield  {author} {\bibinfo {author} {\bibfnamefont {P.}~\bibnamefont
  {Sikivie}},\ }\bibfield  {title} {\enquote {\bibinfo {title} {Detection rates
  for ``invisible''-axion searches},}\ }\href {\doibase
  10.1103/physrevd.32.2988} {\bibfield  {journal} {\bibinfo  {journal}
  {Physical Review D}\ }\textbf {\bibinfo {volume} {32}},\ \bibinfo {pages}
  {2988} (\bibinfo {year} {1985})},\ \bibinfo {note} {[\textit{Erratum ibid.}\
  \textbf{36}, 974 (1987)]}\BibitemShut {NoStop}%
\bibitem [{\citenamefont {{van Bibber}}\ \emph {et~al.}(1989)\citenamefont
  {{van Bibber}}, \citenamefont {{McIntyre}}, \citenamefont {{Morris}},\ and\
  \citenamefont {{Raffelt}}}]{1989_vanbibber}%
  \BibitemOpen
  \bibfield  {author} {\bibinfo {author} {\bibfnamefont {K.}~\bibnamefont {{van
  Bibber}}}, \bibinfo {author} {\bibfnamefont {P.~M.}\ \bibnamefont
  {{McIntyre}}}, \bibinfo {author} {\bibfnamefont {D.~E.}\ \bibnamefont
  {{Morris}}},\ \bibnamefont {and}\ \bibinfo {author} {\bibfnamefont {G.~G.}\
  \bibnamefont {{Raffelt}}},\ }\bibfield  {title} {\enquote {\bibinfo {title}
  {{Design for a practical laboratory detector for solar axions}},}\ }\href
  {\doibase 10.1103/PhysRevD.39.2089} {\bibfield  {journal} {\bibinfo
  {journal} {\prd}\ }\textbf {\bibinfo {volume} {39}},\ \bibinfo {pages} {2089}
  (\bibinfo {year} {1989})}\BibitemShut {NoStop}%
\bibitem [{\citenamefont {{Armengaud}}\ \emph {et~al.}(2014)\citenamefont
  {{Armengaud}}, \citenamefont {{Avignone}}, \citenamefont {{Betz}},
  \citenamefont {{Brax}}, \citenamefont {{Brun}}, \citenamefont {{Cantatore}},
  \citenamefont {{Carmona}}, \citenamefont {{Carosi}}, \citenamefont
  {{Caspers}}, \citenamefont {{Caspi}}, \citenamefont {{Cetin}}, \citenamefont
  {{Chelouche}}, \citenamefont {{Christensen}}, \citenamefont {{Dael}},
  \citenamefont {{Dafni}}, \citenamefont {{Davenport}}, \citenamefont
  {{Derbin}}, \citenamefont {{Desch}}, \citenamefont {{Diago}}, \citenamefont
  {{D{\"o}brich}}, \citenamefont {{Dratchnev}}, \citenamefont {{Dudarev}},
  \citenamefont {{Eleftheriadis}}, \citenamefont {{Fanourakis}}, \citenamefont
  {{Ferrer-Ribas}}, \citenamefont {{Gal{\'a}n}}, \citenamefont
  {{Garc{\'{\i}}a}}, \citenamefont {{Garza}}, \citenamefont {{Geralis}},
  \citenamefont {{Gimeno}}, \citenamefont {{Giomataris}}, \citenamefont
  {{Gninenko}}, \citenamefont {{G{\'o}mez}}, \citenamefont
  {{Gonz{\'a}lez-D{\'{\i}}az}}, \citenamefont {{Guendelman}}, \citenamefont
  {{Hailey}}, \citenamefont {{Hiramatsu}}, \citenamefont {{Hoffmann}},
  \citenamefont {{Horns}}, \citenamefont {{Iguaz}}, \citenamefont
  {{Irastorza}}, \citenamefont {{Isern}}, \citenamefont {{Imai}}, \citenamefont
  {{Jakobsen}}, \citenamefont {{Jaeckel}}, \citenamefont {{Jakov{\v c}i{\'c}}},
  \citenamefont {{Kaminski}}, \citenamefont {{Kawasaki}}, \citenamefont
  {{Karuza}}, \citenamefont {{Kr{\v c}mar}}, \citenamefont {{Kousouris}},
  \citenamefont {{Krieger}}, \citenamefont {{Laki{\'c}}}, \citenamefont
  {{Limousin}}, \citenamefont {{Lindner}}, \citenamefont {{Liolios}},
  \citenamefont {{Luz{\'o}n}}, \citenamefont {{Matsuki}}, \citenamefont
  {{Muratova}}, \citenamefont {{Nones}}, \citenamefont {{Ortega}},
  \citenamefont {{Papaevangelou}}, \citenamefont {{Pivovaroff}}, \citenamefont
  {{Raffelt}}, \citenamefont {{Redondo}}, \citenamefont {{Ringwald}},
  \citenamefont {{Russenschuck}}, \citenamefont {{Ruz}}, \citenamefont
  {{Saikawa}}, \citenamefont {{Savvidis}}, \citenamefont {{Sekiguchi}},
  \citenamefont {{Semertzidis}}, \citenamefont {{Shilon}}, \citenamefont
  {{Sikivie}}, \citenamefont {{Silva}}, \citenamefont {{ten Kate}},
  \citenamefont {{Tomas}}, \citenamefont {{Troitsky}}, \citenamefont
  {{Vafeiadis}}, \citenamefont {{van Bibber}}, \citenamefont {{Vedrine}},
  \citenamefont {{Villar}}, \citenamefont {{Vogel}}, \citenamefont
  {{Walckiers}}, \citenamefont {{Weltman}}, \citenamefont {{Wester}},
  \citenamefont {{Yildiz}},\ and\ \citenamefont {{Zioutas}}}]{1401.3233}%
  \BibitemOpen
  \bibfield  {author} {\bibinfo {author} {\bibfnamefont {E.}~\bibnamefont
  {{Armengaud}}}, \bibinfo {author} {\bibfnamefont {F.~T.}\ \bibnamefont
  {{Avignone}}}, \bibinfo {author} {\bibfnamefont {M.}~\bibnamefont {{Betz}}},
  \bibnamefont {et~al.},\ }\bibfield  {title} {\enquote {\bibinfo {title}
  {{Conceptual design of the International Axion Observatory (IAXO)}},}\ }\href
  {\doibase 10.1088/1748-0221/9/05/T05002} {\bibfield  {journal} {\bibinfo
  {journal} {Journal of Instrumentation}\ }\textbf {\bibinfo {volume} {9}},\
  \bibinfo {eid} {T05002} (\bibinfo {year} {2014})},\
  \Eprint{http://arxiv.org/abs/1401.3233}{arXiv:1401.3233
  [physics.ins-det]}\BibitemShut {NoStop}%
\bibitem [{\citenamefont {{Armengaud}}\ \emph {et~al.}(2019)\citenamefont
  {{Armengaud}}, \citenamefont {{Atti{\'e}}}, \citenamefont {{Basso}},
  \citenamefont {{Brun}}, \citenamefont {{Bykovskiy}}, \citenamefont
  {{Carmona}}, \citenamefont {{Castel}}, \citenamefont {{Cebri{\'a}n}},
  \citenamefont {{Cicoli}}, \citenamefont {{Civitani}}, \citenamefont
  {{Cogollos}}, \citenamefont {{Conlon}}, \citenamefont {{Costa}},
  \citenamefont {{Dafni}}, \citenamefont {{Daido}}, \citenamefont {{Derbin}},
  \citenamefont {{Descalle}}, \citenamefont {{Desch}}, \citenamefont
  {{Dratchnev}}, \citenamefont {{D{\"o}brich}}, \citenamefont {{Dudarev}},
  \citenamefont {{Ferrer-Ribas}}, \citenamefont {{Fleck}}, \citenamefont
  {{Gal{\'a}n}}, \citenamefont {{Galanti}}, \citenamefont {{Garrido}},
  \citenamefont {{Gascon}}, \citenamefont {{Gastaldo}}, \citenamefont
  {{Germani}}, \citenamefont {{Ghisellini}}, \citenamefont {{Giannotti}},
  \citenamefont {{Giomataris}}, \citenamefont {{Gninenko}}, \citenamefont
  {{Golubev}}, \citenamefont {{Graciani}}, \citenamefont {{Irastorza}},
  \citenamefont {{Jakov{\v{c}}i{\'c}}}, \citenamefont {{Kaminski}},
  \citenamefont {{Kr{\v{c}}mar}}, \citenamefont {{Krieger}}, \citenamefont
  {{Laki{\'c}}}, \citenamefont {{Lasserre}}, \citenamefont {{Laurent}},
  \citenamefont {{Limousin}}, \citenamefont {{Lindner}}, \citenamefont
  {{Lomskaya}}, \citenamefont {{Lubsand orzhiev}}, \citenamefont {{Luz{\'o}n}},
  \citenamefont {{Marsh}}, \citenamefont {{Margalejo}}, \citenamefont
  {{Mescia}}, \citenamefont {{Meyer}}, \citenamefont {{Miralda-Escud{\'e}}},
  \citenamefont {{Mirallas}}, \citenamefont {{Muratova}}, \citenamefont
  {{Navick}}, \citenamefont {{Nones}}, \citenamefont {{Notari}}, \citenamefont
  {{Nozik}}, \citenamefont {{Ortiz de Sol{\'o}rzano}}, \citenamefont
  {{Pantuev}}, \citenamefont {{Papaevangelou}}, \citenamefont {{Pareschi}},
  \citenamefont {{Perez}}, \citenamefont {{Picatoste}}, \citenamefont
  {{Pivovaroff}}, \citenamefont {{Redondo}}, \citenamefont {{Ringwald}},
  \citenamefont {{Roncadelli}}, \citenamefont {{Ruiz-Ch{\'o}liz}},
  \citenamefont {{Ruz}}, \citenamefont {{Saikawa}}, \citenamefont
  {{Salvad{\'o}}}, \citenamefont {{Samperiz}}, \citenamefont {{Schiffer}},
  \citenamefont {{Schmidt}}, \citenamefont {{Schneekloth}}, \citenamefont
  {{Schott}}, \citenamefont {{Silva}}, \citenamefont {{Tagliaferri}},
  \citenamefont {{Takahashi}}, \citenamefont {{Tavecchio}}, \citenamefont {{ten
  Kate}}, \citenamefont {{Tkachev}}, \citenamefont {{Troitsky}}, \citenamefont
  {{Unzhakov}}, \citenamefont {{Vedrine}}, \citenamefont {{Vogel}},
  \citenamefont {{Weinsheimer}}, \citenamefont {{Weltman}},\ and\ \citenamefont
  {{Yin}}}]{1904.09155}%
  \BibitemOpen
  \bibfield  {author} {\bibinfo {author} {\bibfnamefont {E.}~\bibnamefont
  {{Armengaud}}}, \bibinfo {author} {\bibfnamefont {D.}~\bibnamefont
  {{Atti{\'e}}}}, \bibinfo {author} {\bibfnamefont {S.}~\bibnamefont
  {{Basso}}}, \bibnamefont {et~al.},\ }\bibfield  {title} {\enquote {\bibinfo
  {title} {{Physics potential of the International Axion Observatory
  (IAXO)}},}\ }\href {\doibase 10.1088/1475-7516/2019/06/047} {\bibfield
  {journal} {\bibinfo  {journal} {\jcap}\ }\textbf {\bibinfo {volume} {2019}},\
  \bibinfo {eid} {047} (\bibinfo {year} {2019})},\
  \Eprint{http://arxiv.org/abs/1904.09155}{arXiv:1904.09155
  [hep-ph]}\BibitemShut {NoStop}%
\bibitem [{\citenamefont {{Lazarus}}\ \emph {et~al.}(1992)\citenamefont
  {{Lazarus}}, \citenamefont {{Smith}}, \citenamefont {{Cameron}},
  \citenamefont {{Melissinos}}, \citenamefont {{Ruoso}}, \citenamefont
  {{Semertzidis}},\ and\ \citenamefont {{Nezrick}}}]{Lazarus:1992ry}%
  \BibitemOpen
  \bibfield  {author} {\bibinfo {author} {\bibfnamefont {D.~M.}\ \bibnamefont
  {{Lazarus}}}, \bibinfo {author} {\bibfnamefont {G.~C.}\ \bibnamefont
  {{Smith}}}, \bibinfo {author} {\bibfnamefont {R.}~\bibnamefont {{Cameron}}},
  \bibnamefont {et~al.},\ }\bibfield  {title} {\enquote {\bibinfo {title}
  {{Search for solar axions}},}\ }\href {\doibase 10.1103/PhysRevLett.69.2333}
  {\bibfield  {journal} {\bibinfo  {journal} {\prl}\ }\textbf {\bibinfo
  {volume} {69}},\ \bibinfo {pages} {2333} (\bibinfo {year}
  {1992})}\BibitemShut {NoStop}%
\bibitem [{\citenamefont {{Moriyama}}\ \emph {et~al.}(1998)\citenamefont
  {{Moriyama}}, \citenamefont {{Minowa}}, \citenamefont {{Namba}},
  \citenamefont {{Inoue}}, \citenamefont {{Takasu}},\ and\ \citenamefont
  {{Yamamoto}}}]{hep-ex/9805026}%
  \BibitemOpen
  \bibfield  {author} {\bibinfo {author} {\bibfnamefont {S.}~\bibnamefont
  {{Moriyama}}}, \bibinfo {author} {\bibfnamefont {M.}~\bibnamefont
  {{Minowa}}}, \bibinfo {author} {\bibfnamefont {T.}~\bibnamefont {{Namba}}},
  \bibnamefont {et~al.},\ }\bibfield  {title} {\enquote {\bibinfo {title}
  {{Direct search for solar axions by using strong magnetic field and X-ray
  detectors}},}\ }\href {\doibase 10.1016/S0370-2693(98)00766-7} {\bibfield
  {journal} {\bibinfo  {journal} {Physics Letters B}\ }\textbf {\bibinfo
  {volume} {434}},\ \bibinfo {pages} {147} (\bibinfo {year} {1998})},\
  \Eprint{http://arxiv.org/abs/hep-ex/9805026}{arXiv:hep-ex/9805026
  [hep-ex]}\BibitemShut {NoStop}%
\bibitem [{\citenamefont {{Inoue}}\ \emph {et~al.}(2002)\citenamefont
  {{Inoue}}, \citenamefont {{Namba}}, \citenamefont {{Moriyama}}, \citenamefont
  {{Minowa}}, \citenamefont {{Takasu}}, \citenamefont {{Horiuchi}},\ and\
  \citenamefont {{Yamamoto}}}]{astro-ph/0204388}%
  \BibitemOpen
  \bibfield  {author} {\bibinfo {author} {\bibfnamefont {Y.}~\bibnamefont
  {{Inoue}}}, \bibinfo {author} {\bibfnamefont {T.}~\bibnamefont {{Namba}}},
  \bibinfo {author} {\bibfnamefont {S.}~\bibnamefont {{Moriyama}}},
  \bibnamefont {et~al.},\ }\bibfield  {title} {\enquote {\bibinfo {title}
  {{Search for sub-electronvolt solar axions using coherent conversion of
  axions into photons in magnetic field and gas helium}},}\ }\href {\doibase
  10.1016/S0370-2693(02)01822-1} {\bibfield  {journal} {\bibinfo  {journal}
  {Physics Letters B}\ }\textbf {\bibinfo {volume} {536}},\ \bibinfo {pages}
  {18} (\bibinfo {year} {2002})},\
  \Eprint{http://arxiv.org/abs/astro-ph/0204388}{arXiv:astro-ph/0204388
  [astro-ph]}\BibitemShut {NoStop}%
\bibitem [{\citenamefont {{Zioutas}}\ \emph {et~al.}(2005)\citenamefont
  {{Zioutas}}, \citenamefont {{Andriamonje}}, \citenamefont {{Arsov}},
  \citenamefont {{Aune}}, \citenamefont {{Autiero}}, \citenamefont
  {{Avignone}}, \citenamefont {{Barth}}, \citenamefont {{Belov}}, \citenamefont
  {{Beltr{\'a}n}}, \citenamefont {{Br{\"a}uninger}}, \citenamefont {{Carmona}},
  \citenamefont {{Cebri{\'a}n}}, \citenamefont {{Chesi}}, \citenamefont
  {{Collar}}, \citenamefont {{Creswick}}, \citenamefont {{Dafni}},
  \citenamefont {{Davenport}}, \citenamefont {{di Lella}}, \citenamefont
  {{Eleftheriadis}}, \citenamefont {{Englhauser}}, \citenamefont
  {{Fanourakis}}, \citenamefont {{Farach}}, \citenamefont {{Ferrer}},
  \citenamefont {{Fischer}}, \citenamefont {{Franz}}, \citenamefont
  {{Friedrich}}, \citenamefont {{Geralis}}, \citenamefont {{Giomataris}},
  \citenamefont {{Gninenko}}, \citenamefont {{Goloubev}}, \citenamefont
  {{Hasinoff}}, \citenamefont {{Heinsius}}, \citenamefont {{Hoffmann}},
  \citenamefont {{Irastorza}}, \citenamefont {{Jacoby}}, \citenamefont
  {{Kang}}, \citenamefont {{K{\"o}nigsmann}}, \citenamefont {{Kotthaus}},
  \citenamefont {{Kr{\v c}mar}}, \citenamefont {{Kousouris}}, \citenamefont
  {{Kuster}}, \citenamefont {{Laki{\'c}}}, \citenamefont {{Lasseur}},
  \citenamefont {{Liolios}}, \citenamefont {{Ljubi{\v c}i{\'c}}}, \citenamefont
  {{Lutz}}, \citenamefont {{Luz{\'o}n}}, \citenamefont {{Miller}},
  \citenamefont {{Morales}}, \citenamefont {{Morales}}, \citenamefont
  {{Mutterer}}, \citenamefont {{Nikolaidis}}, \citenamefont {{Ortiz}},
  \citenamefont {{Papaevangelou}}, \citenamefont {{Placci}}, \citenamefont
  {{Raffelt}}, \citenamefont {{Ruz}}, \citenamefont {{Riege}}, \citenamefont
  {{Sarsa}}, \citenamefont {{Savvidis}}, \citenamefont {{Serber}},
  \citenamefont {{Serpico}}, \citenamefont {{Semertzidis}}, \citenamefont
  {{Stewart}}, \citenamefont {{Vieira}}, \citenamefont {{Villar}},
  \citenamefont {{Walckiers}},\ and\ \citenamefont
  {{Zachariadou}}}]{hep-ex/0411033}%
  \BibitemOpen
  \bibfield  {author} {\bibinfo {author} {\bibfnamefont {K.}~\bibnamefont
  {{Zioutas}}}, \bibinfo {author} {\bibfnamefont {S.}~\bibnamefont
  {{Andriamonje}}}, \bibinfo {author} {\bibfnamefont {V.}~\bibnamefont
  {{Arsov}}}, \bibnamefont {et~al.},\ }\bibfield  {title} {\enquote {\bibinfo
  {title} {{First Results from the CERN Axion Solar Telescope}},}\ }\href
  {\doibase 10.1103/PhysRevLett.94.121301} {\bibfield  {journal} {\bibinfo
  {journal} {Physical Review Letters}\ }\textbf {\bibinfo {volume} {94}},\
  \bibinfo {eid} {121301} (\bibinfo {year} {2005})},\
  \Eprint{http://arxiv.org/abs/hep-ex/0411033}{hep-ex/0411033}\BibitemShut
  {NoStop}%
\bibitem [{\citenamefont {{Andriamonje}}\ \emph {et~al.}(2007)\citenamefont
  {{Andriamonje}}, \citenamefont {{Aune}}, \citenamefont {{Autiero}},
  \citenamefont {{Barth}}, \citenamefont {{Belov}}, \citenamefont
  {{Beltr{\'a}n}}, \citenamefont {{Br{\"a}uninger}}, \citenamefont {{Carmona}},
  \citenamefont {{Cebri{\'a}n}}, \citenamefont {{Collar}}, \citenamefont
  {{Dafni}}, \citenamefont {{Davenport}}, \citenamefont {{Di Lella}},
  \citenamefont {{Eleftheriadis}}, \citenamefont {{Englhauser}}, \citenamefont
  {{Fanourakis}}, \citenamefont {{Ferrer Ribas}}, \citenamefont {{Fischer}},
  \citenamefont {{Franz}}, \citenamefont {{Friedrich}}, \citenamefont
  {{Geralis}}, \citenamefont {{Giomataris}}, \citenamefont {{Gninenko}},
  \citenamefont {{G{\'o}mez}}, \citenamefont {{Hasinoff}}, \citenamefont
  {{Heinsius}}, \citenamefont {{Hoffmann}}, \citenamefont {{Irastorza}},
  \citenamefont {{Jacoby}}, \citenamefont {{Jakovcic}}, \citenamefont {{Kang}},
  \citenamefont {{K{\"o}nigsmann}}, \citenamefont {{Kotthaus}}, \citenamefont
  {{Krcmar}}, \citenamefont {{Kousouris}}, \citenamefont {{Kuster}},
  \citenamefont {{Lakic}}, \citenamefont {{Lasseur}}, \citenamefont
  {{Liolios}}, \citenamefont {{Ljubicic}}, \citenamefont {{Lutz}},
  \citenamefont {{Luz{\'o}n}}, \citenamefont {{Miller}}, \citenamefont
  {{Morales}}, \citenamefont {{Morales}}, \citenamefont {{Ortiz}},
  \citenamefont {{Papaevangelou}}, \citenamefont {{Placci}}, \citenamefont
  {{Raffelt}}, \citenamefont {{Riege}}, \citenamefont {{Rodr{\'\i}guez}},
  \citenamefont {{Ruz}}, \citenamefont {{Savvidis}}, \citenamefont
  {{Semertzidis}}, \citenamefont {{Serpico}}, \citenamefont {{Stewart}},
  \citenamefont {{Vieira}}, \citenamefont {{Villar}}, \citenamefont {{Vogel}},
  \citenamefont {{Walckiers}}, \citenamefont {{Zioutas}},\ and\ \citenamefont
  {{CAST Collaboration}}}]{hep-ex/0702006}%
  \BibitemOpen
  \bibfield  {author} {\bibinfo {author} {\bibfnamefont {S.}~\bibnamefont
  {{Andriamonje}}}, \bibinfo {author} {\bibfnamefont {S.}~\bibnamefont
  {{Aune}}}, \bibinfo {author} {\bibfnamefont {D.}~\bibnamefont {{Autiero}}},
  \bibnamefont {et~al.},\ }\bibfield  {title} {\enquote {\bibinfo {title} {{An
  improved limit on the axion photon coupling from the CAST experiment}},}\
  }\href {\doibase 10.1088/1475-7516/2007/04/010} {\bibfield  {journal}
  {\bibinfo  {journal} {\jcap}\ }\textbf {\bibinfo {volume} {2007}},\ \bibinfo
  {eid} {010} (\bibinfo {year} {2007})},\
  \Eprint{http://arxiv.org/abs/hep-ex/0702006}{arXiv:hep-ex/0702006
  [hep-ex]}\BibitemShut {NoStop}%
\bibitem [{\citenamefont {{Inoue}}\ \emph {et~al.}(2008)\citenamefont
  {{Inoue}}, \citenamefont {{Akimoto}}, \citenamefont {{Ohta}}, \citenamefont
  {{Mizumoto}}, \citenamefont {{Yamamoto}},\ and\ \citenamefont
  {{Minowa}}}]{0806.2230}%
  \BibitemOpen
  \bibfield  {author} {\bibinfo {author} {\bibfnamefont {Y.}~\bibnamefont
  {{Inoue}}}, \bibinfo {author} {\bibfnamefont {Y.}~\bibnamefont {{Akimoto}}},
  \bibinfo {author} {\bibfnamefont {R.}~\bibnamefont {{Ohta}}}, \bibnamefont
  {et~al.},\ }\bibfield  {title} {\enquote {\bibinfo {title} {{Search for solar
  axions with mass around 1 eV using coherent conversion of axions into
  photons}},}\ }\href {\doibase 10.1016/j.physletb.2008.08.020} {\bibfield
  {journal} {\bibinfo  {journal} {Physics Letters B}\ }\textbf {\bibinfo
  {volume} {668}},\ \bibinfo {pages} {93} (\bibinfo {year} {2008})},\
  \Eprint{http://arxiv.org/abs/0806.2230}{arXiv:0806.2230
  [astro-ph]}\BibitemShut {NoStop}%
\bibitem [{\citenamefont {{CAST Collaboration}}\ \emph
  {et~al.}(2009)\citenamefont {{CAST Collaboration}}, \citenamefont
  {{Andriamonje}}, \citenamefont {{Aune}}, \citenamefont {{Autiero}},
  \citenamefont {{Barth}}, \citenamefont {{Belov}}, \citenamefont
  {{Beltr{\'a}n}}, \citenamefont {{Br{\"a}uninger}}, \citenamefont {{Carmona}},
  \citenamefont {{Cebri{\'a}n}}, \citenamefont {{Collar}}, \citenamefont
  {{Dafni}}, \citenamefont {{Davenport}}, \citenamefont {{Di Lella}},
  \citenamefont {{Eleftheriadis}}, \citenamefont {{Englhauser}}, \citenamefont
  {{Fanourakis}}, \citenamefont {{Ferrer-Ribas}}, \citenamefont {{Fischer}},
  \citenamefont {{Franz}}, \citenamefont {{Friedrich}}, \citenamefont
  {{Geralis}}, \citenamefont {{Giomataris}}, \citenamefont {{Gninenko}},
  \citenamefont {{G{\'o}mez}}, \citenamefont {{Hasinoff}}, \citenamefont
  {{Heinsius}}, \citenamefont {{Hoffmann}}, \citenamefont {{Irastorza}},
  \citenamefont {{Jacoby}}, \citenamefont {{Jakov{\v{c}}i{\'c}}}, \citenamefont
  {{Kang}}, \citenamefont {{K{\"o}nigsmann}}, \citenamefont {{Kotthaus}},
  \citenamefont {{Krcmar}}, \citenamefont {{Kousouris}}, \citenamefont
  {{Kuster}}, \citenamefont {{Laki{\'c}}}, \citenamefont {{Lasseur}},
  \citenamefont {{Liolios}}, \citenamefont {{Ljubi{\v{c}}i{\'c}}},
  \citenamefont {{Lutz}}, \citenamefont {{Luz{\'o}n}}, \citenamefont
  {{Miller}}, \citenamefont {{Morales}}, \citenamefont {{Ortiz}}, \citenamefont
  {{Papaevangelou}}, \citenamefont {{Placci}}, \citenamefont {{Raffelt}},
  \citenamefont {{Riege}}, \citenamefont {{Rodr{\'\i}guez}}, \citenamefont
  {{Ruz}}, \citenamefont {{Savvidis}}, \citenamefont {{Semertzidis}},
  \citenamefont {{Serpico}}, \citenamefont {{Stewart}}, \citenamefont
  {{Vieira}}, \citenamefont {{Villar}}, \citenamefont {{Vogel}}, \citenamefont
  {{Walckiers}},\ and\ \citenamefont {{Zioutas}}}]{0906.4488}%
  \BibitemOpen
  \bibfield  {author} {\bibinfo {author} {\bibnamefont {{CAST Collaboration}}},
  \bibinfo {author} {\bibfnamefont {S.}~\bibnamefont {{Andriamonje}}}, \bibinfo
  {author} {\bibfnamefont {S.}~\bibnamefont {{Aune}}}, \bibnamefont {et~al.},\
  }\bibfield  {title} {\enquote {\bibinfo {title} {{Search for 14.4 keV solar
  axions emitted in the M1-transition of $^{57}$Fe nuclei with CAST}},}\ }\href
  {\doibase 10.1088/1475-7516/2009/12/002} {\bibfield  {journal} {\bibinfo
  {journal} {\jcap}\ }\textbf {\bibinfo {volume} {2009}},\ \bibinfo {eid} {002}
  (\bibinfo {year} {2009})},\
  \Eprint{http://arxiv.org/abs/0906.4488}{arXiv:0906.4488 [hep-ex]}\BibitemShut
  {NoStop}%
\bibitem [{\citenamefont {{Arik}}\ \emph {et~al.}(2011)\citenamefont {{Arik}},
  \citenamefont {{Aune}}, \citenamefont {{Barth}}, \citenamefont {{Belov}},
  \citenamefont {{Borghi}}, \citenamefont {{Br{\"a}uninger}}, \citenamefont
  {{Cantatore}}, \citenamefont {{Carmona}}, \citenamefont {{Cetin}},
  \citenamefont {{Collar}}, \citenamefont {{Dafni}}, \citenamefont
  {{Davenport}}, \citenamefont {{Eleftheriadis}}, \citenamefont {{Elias}},
  \citenamefont {{Ezer}}, \citenamefont {{Fanourakis}}, \citenamefont
  {{Ferrer-Ribas}}, \citenamefont {{Friedrich}}, \citenamefont {{Gal{\'a}n}},
  \citenamefont {{Garc{\'{\i}}a}}, \citenamefont {{Gardikiotis}}, \citenamefont
  {{Gazis}}, \citenamefont {{Geralis}}, \citenamefont {{Giomataris}},
  \citenamefont {{Gninenko}}, \citenamefont {{G{\'o}mez}}, \citenamefont
  {{Gruber}}, \citenamefont {{Guth{\"o}rl}}, \citenamefont {{Hartmann}},
  \citenamefont {{Haug}}, \citenamefont {{Hasinoff}}, \citenamefont
  {{Hoffmann}}, \citenamefont {{Iguaz}}, \citenamefont {{Irastorza}},
  \citenamefont {{Jacoby}}, \citenamefont {{Jakov{\v c}i{\'c}}}, \citenamefont
  {{Karuza}}, \citenamefont {{K{\"o}nigsmann}}, \citenamefont {{Kotthaus}},
  \citenamefont {{Kr{\v c}mar}}, \citenamefont {{Kuster}}, \citenamefont
  {{Laki{\'c}}}, \citenamefont {{Laurent}}, \citenamefont {{Liolios}},
  \citenamefont {{Ljubi{\v c}i{\'c}}}, \citenamefont {{Lozza}}, \citenamefont
  {{Lutz}}, \citenamefont {{Luz{\'o}n}}, \citenamefont {{Morales}},
  \citenamefont {{Niinikoski}}, \citenamefont {{Nordt}}, \citenamefont
  {{Papaevangelou}}, \citenamefont {{Pivovaroff}}, \citenamefont {{Raffelt}},
  \citenamefont {{Rashba}}, \citenamefont {{Riege}}, \citenamefont
  {{Rodr{\'{\i}}guez}}, \citenamefont {{Rosu}}, \citenamefont {{Ruz}},
  \citenamefont {{Savvidis}}, \citenamefont {{Silva}}, \citenamefont
  {{Solanki}}, \citenamefont {{Stewart}}, \citenamefont {{Tom{\'a}s}},
  \citenamefont {{Tsagri}}, \citenamefont {{van Bibber}}, \citenamefont
  {{Vafeiadis}}, \citenamefont {{Villar}}, \citenamefont {{Vogel}},
  \citenamefont {{Yildiz}},\ and\ \citenamefont {{Zioutas}}}]{1106.3919}%
  \BibitemOpen
  \bibfield  {author} {\bibinfo {author} {\bibfnamefont {M.}~\bibnamefont
  {{Arik}}}, \bibinfo {author} {\bibfnamefont {S.}~\bibnamefont {{Aune}}},
  \bibinfo {author} {\bibfnamefont {K.}~\bibnamefont {{Barth}}}, \bibnamefont
  {et~al.},\ }\bibfield  {title} {\enquote {\bibinfo {title} {{Search for
  Sub-eV Mass Solar Axions by the CERN Axion Solar Telescope with He3 Buffer
  Gas}},}\ }\href {\doibase 10.1103/PhysRevLett.107.261302} {\bibfield
  {journal} {\bibinfo  {journal} {Physical Review Letters}\ }\textbf {\bibinfo
  {volume} {107}},\ \bibinfo {eid} {261302} (\bibinfo {year} {2011})},\
  \Eprint{http://arxiv.org/abs/1106.3919}{arXiv:1106.3919 [hep-ex]}\BibitemShut
  {NoStop}%
\bibitem [{\citenamefont {{Barth}}\ \emph {et~al.}(2013)\citenamefont
  {{Barth}}, \citenamefont {{Belov}}, \citenamefont {{Beltran}}, \citenamefont
  {{Br{\"a}uninger}}, \citenamefont {{Carmona}}, \citenamefont {{Collar}},
  \citenamefont {{Dafni}}, \citenamefont {{Davenport}}, \citenamefont {{Di
  Lella}}, \citenamefont {{Eleftheriadis}}, \citenamefont {{Englhauser}},
  \citenamefont {{Fanourakis}}, \citenamefont {{Ferrer-Ribas}}, \citenamefont
  {{Fischer}}, \citenamefont {{Franz}}, \citenamefont {{Friedrich}},
  \citenamefont {{Gal{\'a}n}}, \citenamefont {{Garc{\'{\i}}a}}, \citenamefont
  {{Geralis}}, \citenamefont {{Giomataris}}, \citenamefont {{Gninenko}},
  \citenamefont {{G{\'o}mez}}, \citenamefont {{Hasinoff}}, \citenamefont
  {{Heinsius}}, \citenamefont {{Hoffmann}}, \citenamefont {{Irastorza}},
  \citenamefont {{Jacoby}}, \citenamefont {{Jakov{\v c}i{\'c}}}, \citenamefont
  {{Kang}}, \citenamefont {{K{\"o}nigsmann}}, \citenamefont {{Kotthaus}},
  \citenamefont {{Kousouris}}, \citenamefont {{Kr{\v c}mar}}, \citenamefont
  {{Kuster}}, \citenamefont {{Laki{\'c}}}, \citenamefont {{Liolios}},
  \citenamefont {{Ljubi{\v c}i{\'c}}}, \citenamefont {{Lutz}}, \citenamefont
  {{Luz{\'o}n}}, \citenamefont {{Miller}}, \citenamefont {{Papaevangelou}},
  \citenamefont {{Pivovaroff}}, \citenamefont {{Raffelt}}, \citenamefont
  {{Redondo}}, \citenamefont {{Riege}}, \citenamefont {{Rodr{\'{\i}}guez}},
  \citenamefont {{Ruz}}, \citenamefont {{Savvidis}}, \citenamefont
  {{Semertzidis}}, \citenamefont {{Stewart}}, \citenamefont {{Van Bibber}},
  \citenamefont {{Vieira}}, \citenamefont {{Villar}}, \citenamefont {{Vogel}},
  \citenamefont {{Walckiers}},\ and\ \citenamefont {{Zioutas}}}]{1302.6283}%
  \BibitemOpen
  \bibfield  {author} {\bibinfo {author} {\bibfnamefont {K.}~\bibnamefont
  {{Barth}}}, \bibinfo {author} {\bibfnamefont {A.}~\bibnamefont {{Belov}}},
  \bibinfo {author} {\bibfnamefont {B.}~\bibnamefont {{Beltran}}}, \bibnamefont
  {et~al.},\ }\bibfield  {title} {\enquote {\bibinfo {title} {{CAST constraints
  on the axion-electron coupling}},}\ }\href {\doibase
  10.1088/1475-7516/2013/05/010} {\bibfield  {journal} {\bibinfo  {journal}
  {\jcap}\ }\textbf {\bibinfo {volume} {5}},\ \bibinfo {eid} {010} (\bibinfo
  {year} {2013})},\ \Eprint{http://arxiv.org/abs/1302.6283}{arXiv:1302.6283
  [astro-ph.SR]}\BibitemShut {NoStop}%
\bibitem [{\citenamefont {{Arik}}\ \emph {et~al.}(2014)\citenamefont {{Arik}},
  \citenamefont {{Aune}}, \citenamefont {{Barth}}, \citenamefont {{Belov}},
  \citenamefont {{Borghi}}, \citenamefont {{Br{\"a}uninger}}, \citenamefont
  {{Cantatore}}, \citenamefont {{Carmona}}, \citenamefont {{Cetin}},
  \citenamefont {{Collar}}, \citenamefont {{Da Riva}}, \citenamefont {{Dafni}},
  \citenamefont {{Davenport}}, \citenamefont {{Eleftheriadis}}, \citenamefont
  {{Elias}}, \citenamefont {{Fanourakis}}, \citenamefont {{Ferrer-Ribas}},
  \citenamefont {{Friedrich}}, \citenamefont {{Gal{\'a}n}}, \citenamefont
  {{Garc{\'{\i}}a}}, \citenamefont {{Gardikiotis}}, \citenamefont {{Garza}},
  \citenamefont {{Gazis}}, \citenamefont {{Geralis}}, \citenamefont
  {{Georgiopoulou}}, \citenamefont {{Giomataris}}, \citenamefont {{Gninenko}},
  \citenamefont {{G{\'o}mez}}, \citenamefont {{G{\'o}mez Marzoa}},
  \citenamefont {{Gruber}}, \citenamefont {{Guth{\"o}rl}}, \citenamefont
  {{Hartmann}}, \citenamefont {{Hauf}}, \citenamefont {{Haug}}, \citenamefont
  {{Hasinoff}}, \citenamefont {{Hoffmann}}, \citenamefont {{Iguaz}},
  \citenamefont {{Irastorza}}, \citenamefont {{Jacoby}}, \citenamefont
  {{Jakov{\v c}i{\'c}}}, \citenamefont {{Karuza}}, \citenamefont
  {{K{\"o}nigsmann}}, \citenamefont {{Kotthaus}}, \citenamefont {{Kr{\v
  c}mar}}, \citenamefont {{Kuster}}, \citenamefont {{Laki{\'c}}}, \citenamefont
  {{Lang}}, \citenamefont {{Laurent}}, \citenamefont {{Liolios}}, \citenamefont
  {{Ljubi{\v c}i{\'c}}}, \citenamefont {{Luz{\'o}n}}, \citenamefont {{Neff}},
  \citenamefont {{Niinikoski}}, \citenamefont {{Nordt}}, \citenamefont
  {{Papaevangelou}}, \citenamefont {{Pivovaroff}}, \citenamefont {{Raffelt}},
  \citenamefont {{Riege}}, \citenamefont {{Rodr{\'{\i}}guez}}, \citenamefont
  {{Rosu}}, \citenamefont {{Ruz}}, \citenamefont {{Savvidis}}, \citenamefont
  {{Shilon}}, \citenamefont {{Silva}}, \citenamefont {{Solanki}}, \citenamefont
  {{Stewart}}, \citenamefont {{Tom{\'a}s}}, \citenamefont {{Tsagri}},
  \citenamefont {{van Bibber}}, \citenamefont {{Vafeiadis}}, \citenamefont
  {{Villar}}, \citenamefont {{Vogel}}, \citenamefont {{Yildiz}}, \citenamefont
  {{Zioutas}},\ and\ \citenamefont {{CAST Collaboration}}}]{1307.1985}%
  \BibitemOpen
  \bibfield  {author} {\bibinfo {author} {\bibfnamefont {M.}~\bibnamefont
  {{Arik}}}, \bibinfo {author} {\bibfnamefont {S.}~\bibnamefont {{Aune}}},
  \bibinfo {author} {\bibfnamefont {K.}~\bibnamefont {{Barth}}}, \bibnamefont
  {et~al.},\ }\bibfield  {title} {\enquote {\bibinfo {title} {{Search for Solar
  Axions by the CERN Axion Solar Telescope with He3 Buffer Gas: Closing the Hot
  Dark Matter Gap}},}\ }\href {\doibase 10.1103/PhysRevLett.112.091302}
  {\bibfield  {journal} {\bibinfo  {journal} {Physical Review Letters}\
  }\textbf {\bibinfo {volume} {112}},\ \bibinfo {eid} {091302} (\bibinfo {year}
  {2014})},\ \Eprint{http://arxiv.org/abs/1307.1985}{arXiv:1307.1985
  [hep-ex]}\BibitemShut {NoStop}%
\bibitem [{\citenamefont {{Arik}}\ \emph {et~al.}(2015)\citenamefont {{Arik}},
  \citenamefont {{Aune}}, \citenamefont {{Barth}}, \citenamefont {{Belov}},
  \citenamefont {{Br{\"a}uninger}}, \citenamefont {{Bremer}}, \citenamefont
  {{Burwitz}}, \citenamefont {{Cantatore}}, \citenamefont {{Carmona}},
  \citenamefont {{Cetin}}, \citenamefont {{Collar}}, \citenamefont {{Da Riva}},
  \citenamefont {{Dafni}}, \citenamefont {{Davenport}}, \citenamefont
  {{Dermenev}}, \citenamefont {{Eleftheriadis}}, \citenamefont {{Elias}},
  \citenamefont {{Fanourakis}}, \citenamefont {{Ferrer-Ribas}}, \citenamefont
  {{Gal{\'a}n}}, \citenamefont {{Garc{\'{\i}}a}}, \citenamefont
  {{Gardikiotis}}, \citenamefont {{Garza}}, \citenamefont {{Gazis}},
  \citenamefont {{Geralis}}, \citenamefont {{Georgiopoulou}}, \citenamefont
  {{Giomataris}}, \citenamefont {{Gninenko}}, \citenamefont {{G{\'o}mez
  Marzoa}}, \citenamefont {{Hasinoff}}, \citenamefont {{Hoffmann}},
  \citenamefont {{Iguaz}}, \citenamefont {{Irastorza}}, \citenamefont
  {{Jacoby}}, \citenamefont {{Jakov{\v c}i{\'c}}}, \citenamefont {{Karuza}},
  \citenamefont {{Kavuk}}, \citenamefont {{Kr{\v c}mar}}, \citenamefont
  {{Kuster}}, \citenamefont {{Laki{\'c}}}, \citenamefont {{Laurent}},
  \citenamefont {{Liolios}}, \citenamefont {{Ljubi{\v c}i{\'c}}}, \citenamefont
  {{Luz{\'o}n}}, \citenamefont {{Neff}}, \citenamefont {{Niinikoski}},
  \citenamefont {{Nordt}}, \citenamefont {{Ortega}}, \citenamefont
  {{Papaevangelou}}, \citenamefont {{Pivovaroff}}, \citenamefont {{Raffelt}},
  \citenamefont {{Rodr{\'{\i}}guez}}, \citenamefont {{Rosu}}, \citenamefont
  {{Ruz}}, \citenamefont {{Savvidis}}, \citenamefont {{Shilon}}, \citenamefont
  {{Solanki}}, \citenamefont {{Stewart}}, \citenamefont {{Tom{\'a}s}},
  \citenamefont {{Vafeiadis}}, \citenamefont {{Villar}}, \citenamefont
  {{Vogel}}, \citenamefont {{Yildiz}}, \citenamefont {{Zioutas}},\ and\
  \citenamefont {{CAST Collaboration}}}]{1503.00610}%
  \BibitemOpen
  \bibfield  {author} {\bibinfo {author} {\bibfnamefont {M.}~\bibnamefont
  {{Arik}}}, \bibinfo {author} {\bibfnamefont {S.}~\bibnamefont {{Aune}}},
  \bibinfo {author} {\bibfnamefont {K.}~\bibnamefont {{Barth}}}, \bibnamefont
  {et~al.},\ }\bibfield  {title} {\enquote {\bibinfo {title} {{New solar axion
  search using the CERN Axion Solar Telescope with $^{4}$He filling}},}\ }\href
  {\doibase 10.1103/PhysRevD.92.021101} {\bibfield  {journal} {\bibinfo
  {journal} {\prd}\ }\textbf {\bibinfo {volume} {92}},\ \bibinfo {eid} {021101}
  (\bibinfo {year} {2015})},\
  \Eprint{http://arxiv.org/abs/1503.00610}{arXiv:1503.00610
  [hep-ex]}\BibitemShut {NoStop}%
\bibitem [{\citenamefont {{Anastassopoulos}}\ \emph {et~al.}(2017)\citenamefont
  {{Anastassopoulos}}, \citenamefont {{Aune}}, \citenamefont {{Barth}},
  \citenamefont {{Belov}}, \citenamefont {{Br{\"a}uninger}}, \citenamefont
  {{Cantatore}}, \citenamefont {{Carmona}}, \citenamefont {{Castel}},
  \citenamefont {{Cetin}}, \citenamefont {{Christensen}}, \citenamefont
  {{Collar}}, \citenamefont {{Dafni}}, \citenamefont {{Davenport}},
  \citenamefont {{Decker}}, \citenamefont {{Dermenev}}, \citenamefont
  {{Desch}}, \citenamefont {{Eleftheriadis}}, \citenamefont {{Fanourakis}},
  \citenamefont {{Ferrer-Ribas}}, \citenamefont {{Fischer}}, \citenamefont
  {{Garc{\'{\i}}a}}, \citenamefont {{Gardikiotis}}, \citenamefont {{Garza}},
  \citenamefont {{Gazis}}, \citenamefont {{Geralis}}, \citenamefont
  {{Giomataris}}, \citenamefont {{Gninenko}}, \citenamefont {{Hailey}},
  \citenamefont {{Hasinoff}}, \citenamefont {{Hoffmann}}, \citenamefont
  {{Iguaz}}, \citenamefont {{Irastorza}}, \citenamefont {{Jakobsen}},
  \citenamefont {{Jacoby}}, \citenamefont {{Jakov{\v c}i{\'c}}}, \citenamefont
  {{Kaminski}}, \citenamefont {{Karuza}}, \citenamefont {{Kralj}},
  \citenamefont {{Kr{\v c}mar}}, \citenamefont {{Kostoglou}}, \citenamefont
  {{Krieger}}, \citenamefont {{Laki{\'c}}}, \citenamefont {{Laurent}},
  \citenamefont {{Liolios}}, \citenamefont {{Ljubi{\v c}i{\'c}}}, \citenamefont
  {{Luz{\'o}n}}, \citenamefont {{Maroudas}}, \citenamefont {{Miceli}},
  \citenamefont {{Neff}}, \citenamefont {{Ortega}}, \citenamefont
  {{Papaevangelou}}, \citenamefont {{Paraschou}}, \citenamefont {{Pivovaroff}},
  \citenamefont {{Raffelt}}, \citenamefont {{Rosu}}, \citenamefont {{Ruz}},
  \citenamefont {{Ch{\'o}liz}}, \citenamefont {{Savvidis}}, \citenamefont
  {{Schmidt}}, \citenamefont {{Semertzidis}}, \citenamefont {{Solanki}},
  \citenamefont {{Stewart}}, \citenamefont {{Vafeiadis}}, \citenamefont
  {{Vogel}}, \citenamefont {{Yildiz}},\ and\ \citenamefont
  {{Zioutas}}}]{1705.02290}%
  \BibitemOpen
  \bibfield  {author} {\bibinfo {author} {\bibfnamefont {V.}~\bibnamefont
  {{Anastassopoulos}}}, \bibinfo {author} {\bibfnamefont {S.}~\bibnamefont
  {{Aune}}}, \bibinfo {author} {\bibfnamefont {K.}~\bibnamefont {{Barth}}},
  \bibnamefont {et~al.},\ }\bibfield  {title} {\enquote {\bibinfo {title} {{New
  CAST limit on the axion-photon interaction}},}\ }\href {\doibase
  10.1038/nphys4109} {\bibfield  {journal} {\bibinfo  {journal} {Nature
  Physics}\ }\textbf {\bibinfo {volume} {13}},\ \bibinfo {pages} {584}
  (\bibinfo {year} {2017})},\
  \Eprint{http://arxiv.org/abs/1705.02290}{arXiv:1705.02290
  [hep-ex]}\BibitemShut {NoStop}%
\bibitem [{\citenamefont {{Jaeckel}}\ and\ \citenamefont
  {{Thormaehlen}}(2019{\natexlab{a}})}]{1811.09278}%
  \BibitemOpen
  \bibfield  {author} {\bibinfo {author} {\bibfnamefont {J.}~\bibnamefont
  {{Jaeckel}}}\ \bibnamefont {and}\ \bibinfo {author} {\bibfnamefont {L.~J.}\
  \bibnamefont {{Thormaehlen}}},\ }\bibfield  {title} {\enquote {\bibinfo
  {title} {{Distinguishing axion models with IAXO}},}\ }\href {\doibase
  10.1088/1475-7516/2019/03/039} {\bibfield  {journal} {\bibinfo  {journal}
  {\jcap}\ }\textbf {\bibinfo {volume} {2019}},\ \bibinfo {eid} {039} (\bibinfo
  {year} {2019}{\natexlab{a}})},\
  \Eprint{http://arxiv.org/abs/1811.09278}{arXiv:1811.09278
  [hep-ph]}\BibitemShut {NoStop}%
\bibitem [{\citenamefont {{Dafni}}\ \emph {et~al.}(2019)\citenamefont
  {{Dafni}}, \citenamefont {{O'Hare}}, \citenamefont {{Laki{\'c}}},
  \citenamefont {{Gal{\'a}n}}, \citenamefont {{Iguaz}}, \citenamefont
  {{Irastorza}}, \citenamefont {{Jakov{\v{c}}i{\'c}}}, \citenamefont
  {{Luz{\'o}n}}, \citenamefont {{Redondo}},\ and\ \citenamefont {{Ruiz
  Ch{\'o}liz}}}]{1811.09290}%
  \BibitemOpen
  \bibfield  {author} {\bibinfo {author} {\bibfnamefont {T.}~\bibnamefont
  {{Dafni}}}, \bibinfo {author} {\bibfnamefont {C.~A.~J.}\ \bibnamefont
  {{O'Hare}}}, \bibinfo {author} {\bibfnamefont {B.}~\bibnamefont
  {{Laki{\'c}}}}, \bibnamefont {et~al.},\ }\bibfield  {title} {\enquote
  {\bibinfo {title} {{Weighing the solar axion}},}\ }\href {\doibase
  10.1103/PhysRevD.99.035037} {\bibfield  {journal} {\bibinfo  {journal}
  {\prd}\ }\textbf {\bibinfo {volume} {99}},\ \bibinfo {eid} {035037} (\bibinfo
  {year} {2019})},\ \Eprint{http://arxiv.org/abs/1811.09290}{arXiv:1811.09290
  [hep-ph]}\BibitemShut {NoStop}%
\bibitem [{\citenamefont {{Jaeckel}}\ and\ \citenamefont
  {{Thormaehlen}}(2019{\natexlab{b}})}]{1908.10878}%
  \BibitemOpen
  \bibfield  {author} {\bibinfo {author} {\bibfnamefont {J.}~\bibnamefont
  {{Jaeckel}}}\ \bibnamefont {and}\ \bibinfo {author} {\bibfnamefont {L.~J.}\
  \bibnamefont {{Thormaehlen}}},\ }\bibfield  {title} {\enquote {\bibinfo
  {title} {{Axions as a probe of solar metals}},}\ }\href {\doibase
  10.1103/PhysRevD.100.123020} {\bibfield  {journal} {\bibinfo  {journal}
  {\prd}\ }\textbf {\bibinfo {volume} {100}},\ \bibinfo {eid} {123020}
  (\bibinfo {year} {2019}{\natexlab{b}})},\
  \Eprint{http://arxiv.org/abs/1908.10878}{arXiv:1908.10878
  [astro-ph.SR]}\BibitemShut {NoStop}%
\bibitem [{\citenamefont {{Raffelt}}(1988)}]{Raffelt:1987np}%
  \BibitemOpen
  \bibfield  {author} {\bibinfo {author} {\bibfnamefont {G.~G.}\ \bibnamefont
  {{Raffelt}}},\ }\bibfield  {title} {\enquote {\bibinfo {title} {{Plasmon
  decay into low-mass bosons in stars}},}\ }\href {\doibase
  10.1103/PhysRevD.37.1356} {\bibfield  {journal} {\bibinfo  {journal} {\prd}\
  }\textbf {\bibinfo {volume} {37}},\ \bibinfo {pages} {1356} (\bibinfo {year}
  {1988})}\BibitemShut {NoStop}%
\bibitem [{\citenamefont {{Redondo}}(2013)}]{1310.0823}%
  \BibitemOpen
  \bibfield  {author} {\bibinfo {author} {\bibfnamefont {J.}~\bibnamefont
  {{Redondo}}},\ }\bibfield  {title} {\enquote {\bibinfo {title} {{Solar axion
  flux from the axion-electron coupling}},}\ }\href {\doibase
  10.1088/1475-7516/2013/12/008} {\bibfield  {journal} {\bibinfo  {journal}
  {\jcap}\ }\textbf {\bibinfo {volume} {12}},\ \bibinfo {eid} {008} (\bibinfo
  {year} {2013})},\ \Eprint{http://arxiv.org/abs/1310.0823}{arXiv:1310.0823
  [hep-ph]}\BibitemShut {NoStop}%
\bibitem [{\citenamefont {{Hoof}}\ \emph {et~al.}(2019)\citenamefont {{Hoof}},
  \citenamefont {{Kahlhoefer}}, \citenamefont {{Scott}}, \citenamefont
  {{Weniger}},\ and\ \citenamefont {{White}}}]{1810.07192}%
  \BibitemOpen
  \bibfield  {author} {\bibinfo {author} {\bibfnamefont {S.}~\bibnamefont
  {{Hoof}}}, \bibinfo {author} {\bibfnamefont {F.}~\bibnamefont
  {{Kahlhoefer}}}, \bibinfo {author} {\bibfnamefont {P.}~\bibnamefont
  {{Scott}}}, \bibnamefont {et~al.},\ }\bibfield  {title} {\enquote {\bibinfo
  {title} {{Axion global fits with Peccei-Quinn symmetry breaking before
  inflation using GAMBIT}},}\ }\href {\doibase 10.1007/JHEP03(2019)191}
  {\bibfield  {journal} {\bibinfo  {journal} {Journal of High Energy Physics}\
  }\textbf {\bibinfo {volume} {3}},\ \bibinfo {eid} {191} (\bibinfo {year}
  {2019})},\ \Eprint{http://arxiv.org/abs/1810.07192}{arXiv:1810.07192
  [hep-ph]}\BibitemShut {NoStop}%
\bibitem [{\citenamefont {{Bahcall}}\ \emph {et~al.}(2006)\citenamefont
  {{Bahcall}}, \citenamefont {{Serenelli}},\ and\ \citenamefont
  {{Basu}}}]{astro-ph/0511337}%
  \BibitemOpen
  \bibfield  {author} {\bibinfo {author} {\bibfnamefont {J.~N.}\ \bibnamefont
  {{Bahcall}}}, \bibinfo {author} {\bibfnamefont {A.~M.}\ \bibnamefont
  {{Serenelli}}},\ \bibnamefont {and}\ \bibinfo {author} {\bibfnamefont
  {S.}~\bibnamefont {{Basu}}},\ }\bibfield  {title} {\enquote {\bibinfo {title}
  {{10,000 Standard Solar Models: A Monte Carlo Simulation}},}\ }\href
  {\doibase 10.1086/504043} {\bibfield  {journal} {\bibinfo  {journal} {\apjs}\
  }\textbf {\bibinfo {volume} {165}},\ \bibinfo {pages} {400} (\bibinfo {year}
  {2006})},\
  \Eprint{http://arxiv.org/abs/astro-ph/0511337}{arXiv:astro-ph/0511337
  [astro-ph]}\BibitemShut {NoStop}%
\bibitem [{\citenamefont {{Serenelli}}\ \emph {et~al.}(2009)\citenamefont
  {{Serenelli}}, \citenamefont {{Basu}}, \citenamefont {{Ferguson}},\ and\
  \citenamefont {{Asplund}}}]{Serenelli:2009yc}%
  \BibitemOpen
  \bibfield  {author} {\bibinfo {author} {\bibfnamefont {A.~M.}\ \bibnamefont
  {{Serenelli}}}, \bibinfo {author} {\bibfnamefont {S.}~\bibnamefont {{Basu}}},
  \bibinfo {author} {\bibfnamefont {J.~W.}\ \bibnamefont {{Ferguson}}},\
  \bibnamefont {and}\ \bibinfo {author} {\bibfnamefont {M.}~\bibnamefont
  {{Asplund}}},\ }\bibfield  {title} {\enquote {\bibinfo {title} {{New Solar
  Composition: The Problem with Solar Models Revisited}},}\ }\href {\doibase
  10.1088/0004-637X/705/2/L123} {\bibfield  {journal} {\bibinfo  {journal}
  {\apjl}\ }\textbf {\bibinfo {volume} {705}},\ \bibinfo {pages} {L123}
  (\bibinfo {year} {2009})},\
  \Eprint{http://arxiv.org/abs/0909.2668}{arXiv:0909.2668
  [astro-ph.SR]}\BibitemShut {NoStop}%
\bibitem [{\citenamefont {{Vinyoles}}\ \emph {et~al.}(2017)\citenamefont
  {{Vinyoles}}, \citenamefont {{Serenelli}}, \citenamefont {{Villante}},
  \citenamefont {{Basu}}, \citenamefont {{Bergstr{\"o}m}}, \citenamefont
  {{Gonzalez-Garcia}}, \citenamefont {{Maltoni}}, \citenamefont
  {{Pe{\~n}a-Garay}},\ and\ \citenamefont {{Song}}}]{1611.09867}%
  \BibitemOpen
  \bibfield  {author} {\bibinfo {author} {\bibfnamefont {N.}~\bibnamefont
  {{Vinyoles}}}, \bibinfo {author} {\bibfnamefont {A.~M.}\ \bibnamefont
  {{Serenelli}}}, \bibinfo {author} {\bibfnamefont {F.~L.}\ \bibnamefont
  {{Villante}}}, \bibnamefont {et~al.},\ }\bibfield  {title} {\enquote
  {\bibinfo {title} {{A New Generation of Standard Solar Models}},}\ }\href
  {\doibase 10.3847/1538-4357/835/2/202} {\bibfield  {journal} {\bibinfo
  {journal} {\apj}\ }\textbf {\bibinfo {volume} {835}},\ \bibinfo {eid} {202}
  (\bibinfo {year} {2017})},\
  \Eprint{http://arxiv.org/abs/1611.09867}{arXiv:1611.09867
  [astro-ph.SR]}\BibitemShut {NoStop}%
\bibitem [{\citenamefont {{Quevillon}}\ and\ \citenamefont
  {{Smith}}(2019)}]{Quevillon:2019zrd}%
  \BibitemOpen
  \bibfield  {author} {\bibinfo {author} {\bibfnamefont {J.}~\bibnamefont
  {{Quevillon}}}\ \bibnamefont {and}\ \bibinfo {author} {\bibfnamefont
  {C.}~\bibnamefont {{Smith}}},\ }\bibfield  {title} {\enquote {\bibinfo
  {title} {{Axions are blind to anomalies}},}\ }\href {\doibase
  10.1140/epjc/s10052-019-7304-4} {\bibfield  {journal} {\bibinfo  {journal}
  {European Physical Journal C}\ }\textbf {\bibinfo {volume} {79}},\ \bibinfo
  {eid} {822} (\bibinfo {year} {2019})},\
  \Eprint{http://arxiv.org/abs/1903.12559}{arXiv:1903.12559
  [hep-ph]}\BibitemShut {NoStop}%
\bibitem [{\citenamefont {{Raffelt}}\ and\ \citenamefont
  {{Stodolsky}}(1982)}]{1982PhLB..119..323R}%
  \BibitemOpen
  \bibfield  {author} {\bibinfo {author} {\bibfnamefont {G.}~\bibnamefont
  {{Raffelt}}}\ \bibnamefont {and}\ \bibinfo {author} {\bibfnamefont
  {L.}~\bibnamefont {{Stodolsky}}},\ }\bibfield  {title} {\enquote {\bibinfo
  {title} {{New particles from nuclear reactions in the sun}},}\ }\href
  {\doibase 10.1016/0370-2693(82)90680-3} {\bibfield  {journal} {\bibinfo
  {journal} {Physics Letters B}\ }\textbf {\bibinfo {volume} {119}},\ \bibinfo
  {pages} {323} (\bibinfo {year} {1982})}\BibitemShut {NoStop}%
\bibitem [{\citenamefont {{CUORE Collaboration}}(2013)}]{1209.2800}%
  \BibitemOpen
  \bibfield  {author} {\bibinfo {author} {\bibnamefont {{CUORE
  Collaboration}}},\ }\bibfield  {title} {\enquote {\bibinfo {title} {{Search
  for 14.4 keV solar axions from M1 transition of $^{57}$Fe with CUORE
  crystals}},}\ }\href {\doibase 10.1088/1475-7516/2013/05/007} {\bibfield
  {journal} {\bibinfo  {journal} {\jcap}\ }\textbf {\bibinfo {volume} {2013}},\
  \bibinfo {eid} {007} (\bibinfo {year} {2013})},\
  \Eprint{http://arxiv.org/abs/1209.2800}{arXiv:1209.2800 [hep-ex]}\BibitemShut
  {NoStop}%
\bibitem [{\citenamefont {Moriyama}(1995)}]{Moriyama:1995bz}%
  \BibitemOpen
  \bibfield  {author} {\bibinfo {author} {\bibfnamefont {S.}~\bibnamefont
  {Moriyama}},\ }\bibfield  {title} {\enquote {\bibinfo {title} {{A Proposal to
  search for a monochromatic component of solar axions using Fe-57}},}\ }\href
  {\doibase 10.1103/PhysRevLett.75.3222} {\bibfield  {journal} {\bibinfo
  {journal} {Phys. Rev. Lett.}\ }\textbf {\bibinfo {volume} {75}},\ \bibinfo
  {pages} {3222} (\bibinfo {year} {1995})},\
  \Eprint{http://arxiv.org/abs/hep-ph/9504318}{arXiv:hep-ph/9504318}\BibitemShut
  {NoStop}%
\bibitem [{\citenamefont {Krcmar}\ \emph {et~al.}(1998)\citenamefont {Krcmar},
  \citenamefont {Krecak}, \citenamefont {Stipcevic}, \citenamefont {Ljubicic},\
  and\ \citenamefont {Bradley}}]{Krcmar:1998xn}%
  \BibitemOpen
  \bibfield  {author} {\bibinfo {author} {\bibfnamefont {M.}~\bibnamefont
  {Krcmar}}, \bibinfo {author} {\bibfnamefont {Z.}~\bibnamefont {Krecak}},
  \bibinfo {author} {\bibfnamefont {M.}~\bibnamefont {Stipcevic}}, \bibnamefont
  {et~al.},\ }\bibfield  {title} {\enquote {\bibinfo {title} {{Search for
  invisible axions using Fe-57}},}\ }\href {\doibase
  10.1016/S0370-2693(98)01231-3} {\bibfield  {journal} {\bibinfo  {journal}
  {Phys. Lett. B}\ }\textbf {\bibinfo {volume} {442}},\ \bibinfo {pages} {38}
  (\bibinfo {year} {1998})},\
  \Eprint{http://arxiv.org/abs/nucl-ex/9801005}{arXiv:nucl-ex/9801005}\BibitemShut
  {NoStop}%
\bibitem [{\citenamefont {Gavrilyuk}\ \emph {et~al.}(2015)\citenamefont
  {Gavrilyuk} \emph {et~al.}}]{Gavrilyuk:2014mch}%
  \BibitemOpen
  \bibfield  {author} {\bibinfo {author} {\bibfnamefont {Y.~M.}\ \bibnamefont
  {Gavrilyuk}} \bibnamefont {et~al.},\ }\bibfield  {title} {\enquote {\bibinfo
  {title} {{First result of the experimental search for the 9.4 keV solar axion
  reactions with $^{83}$Kr in the copper proportional counter}},}\ }\href
  {\doibase 10.1134/S1063779615020094} {\bibfield  {journal} {\bibinfo
  {journal} {Phys. Part. Nucl.}\ }\textbf {\bibinfo {volume} {46}},\ \bibinfo
  {pages} {152} (\bibinfo {year} {2015})},\
  \Eprint{http://arxiv.org/abs/1405.1271}{arXiv:1405.1271
  [nucl-ex]}\BibitemShut {NoStop}%
\bibitem [{\citenamefont {Kramida}\ \emph {et~al.}(2020)\citenamefont
  {Kramida}, \citenamefont {{Yu.~Ralchenko}}, \citenamefont {Reader},\ and\
  \citenamefont {{and NIST ASD Team}}}]{NIST_ASD}%
  \BibitemOpen
  \bibfield  {author} {\bibinfo {author} {\bibfnamefont {A.}~\bibnamefont
  {Kramida}}, \bibinfo {author} {\bibnamefont {{Yu.~Ralchenko}}}, \bibinfo
  {author} {\bibfnamefont {J.}~\bibnamefont {Reader}},\ \bibnamefont {and}\
  \bibinfo {author} {\bibnamefont {{and NIST ASD Team}}},\ }\href@noop {}
  {}\bibinfo {howpublished} {{NIST Atomic Spectra Database (ver. 5.8),
  [Online]. Available: {\tt{https://physics.nist.gov/asd}} [2021, January 19].
  National Institute of Standards and Technology, Gaithersburg, MD.}} (\bibinfo
  {year} {2020})\BibitemShut {NoStop}%
\bibitem [{\citenamefont {Raffelt}(1996)}]{Raffelt:1996wa}%
  \BibitemOpen
  \bibfield  {author} {\bibinfo {author} {\bibfnamefont {G.~G.}\ \bibnamefont
  {Raffelt}},\ }\href
  {https://wwwth.mpp.mpg.de/members/raffelt/mypapers/Stars.pdf} {\emph
  {\bibinfo {title} {Stars as Laboratories for Fundamental Physics: The
  Astrophysics of Neutrinos, Axions, and Other Weakly Interacting Particles}}}\
  (\bibinfo  {publisher} {University Of Chicago Press},\ \bibinfo {year}
  {1996})\BibitemShut {NoStop}%
\bibitem [{\citenamefont {Jaeckel}\ \emph {et~al.}(2007)\citenamefont
  {Jaeckel}, \citenamefont {Masso}, \citenamefont {Redondo}, \citenamefont
  {Ringwald},\ and\ \citenamefont {Takahashi}}]{Jaeckel:2006xm}%
  \BibitemOpen
  \bibfield  {author} {\bibinfo {author} {\bibfnamefont {J.}~\bibnamefont
  {Jaeckel}}, \bibinfo {author} {\bibfnamefont {E.}~\bibnamefont {Masso}},
  \bibinfo {author} {\bibfnamefont {J.}~\bibnamefont {Redondo}}, \bibnamefont
  {et~al.},\ }\bibfield  {title} {\enquote {\bibinfo {title} {{The Need for
  purely laboratory-based axion-like particle searches}},}\ }\href {\doibase
  10.1103/PhysRevD.75.013004} {\bibfield  {journal} {\bibinfo  {journal} {Phys.
  Rev. D}\ }\textbf {\bibinfo {volume} {75}},\ \bibinfo {pages} {013004}
  (\bibinfo {year} {2007})},\
  \Eprint{http://arxiv.org/abs/hep-ph/0610203}{arXiv:hep-ph/0610203}\BibitemShut
  {NoStop}%
\bibitem [{\citenamefont {{Bludman}}\ and\ \citenamefont {{van
  Riper}}(1977)}]{FermiGas}%
  \BibitemOpen
  \bibfield  {author} {\bibinfo {author} {\bibfnamefont {S.~A.}\ \bibnamefont
  {{Bludman}}}\ \bibnamefont {and}\ \bibinfo {author} {\bibfnamefont {K.~A.}\
  \bibnamefont {{van Riper}}},\ }\bibfield  {title} {\enquote {\bibinfo {title}
  {{Equation of state of an ideal Fermi gas.}}}\ }\href {\doibase
  10.1086/155110} {\bibfield  {journal} {\bibinfo  {journal} {\apj}\ }\textbf
  {\bibinfo {volume} {212}},\ \bibinfo {pages} {859} (\bibinfo {year}
  {1977})}\BibitemShut {NoStop}%
\bibitem [{\citenamefont {Landau}(1946)}]{Landau:1946jc}%
  \BibitemOpen
  \bibfield  {author} {\bibinfo {author} {\bibfnamefont {L.~D.}\ \bibnamefont
  {Landau}},\ }\bibfield  {title} {\enquote {\bibinfo {title} {{On the
  vibrations of the electronic plasma}},}\ }\href@noop {} {\bibfield  {journal}
  {\bibinfo  {journal} {J. Phys. (USSR)}\ }\textbf {\bibinfo {volume} {10}},\
  \bibinfo {pages} {25} (\bibinfo {year} {1946})}\BibitemShut {NoStop}%
\bibitem [{\citenamefont {Krief}\ \emph {et~al.}(2016)\citenamefont {Krief},
  \citenamefont {Feigel},\ and\ \citenamefont {Gazit}}]{Krief:2016znd}%
  \BibitemOpen
  \bibfield  {author} {\bibinfo {author} {\bibfnamefont {M.}~\bibnamefont
  {Krief}}, \bibinfo {author} {\bibfnamefont {A.}~\bibnamefont {Feigel}},\
  \bibnamefont {and}\ \bibinfo {author} {\bibfnamefont {D.}~\bibnamefont
  {Gazit}},\ }\bibfield  {title} {\enquote {\bibinfo {title} {{Solar opacity
  calculations using the super-transition-array method}},}\ }\href {\doibase
  10.3847/0004-637X/821/1/45} {\bibfield  {journal} {\bibinfo  {journal}
  {Astrophys. J.}\ }\textbf {\bibinfo {volume} {821}},\ \bibinfo {pages} {45}
  (\bibinfo {year} {2016})},\
  \Eprint{http://arxiv.org/abs/1601.01930}{arXiv:1601.01930
  [astro-ph.SR]}\BibitemShut {NoStop}%
\bibitem [{\citenamefont {Huebner}\ and\ \citenamefont
  {Barfield}(2014)}]{2014_opacity}%
  \BibitemOpen
  \bibfield  {author} {\bibinfo {author} {\bibfnamefont {W.~F.}\ \bibnamefont
  {Huebner}}\ \bibnamefont {and}\ \bibinfo {author} {\bibfnamefont {W.~D.}\
  \bibnamefont {Barfield}},\ }\href {\doibase 10.1007/978-1-4614-8797-5} {\emph
  {\bibinfo {title} {Opacity}}},\ \bibinfo {series} {Astrophysics and Space
  Science Library}, Vol.\ \bibinfo {volume} {402}\ (\bibinfo  {publisher}
  {Springer-Verlag New York},\ \bibinfo {year} {2014})\BibitemShut {NoStop}%
\bibitem [{\citenamefont {Vitagliano}\ \emph {et~al.}(2017)\citenamefont
  {Vitagliano}, \citenamefont {Redondo},\ and\ \citenamefont
  {Raffelt}}]{Vitagliano:2017odj}%
  \BibitemOpen
  \bibfield  {author} {\bibinfo {author} {\bibfnamefont {E.}~\bibnamefont
  {Vitagliano}}, \bibinfo {author} {\bibfnamefont {J.}~\bibnamefont
  {Redondo}},\ \bibnamefont {and}\ \bibinfo {author} {\bibfnamefont
  {G.}~\bibnamefont {Raffelt}},\ }\bibfield  {title} {\enquote {\bibinfo
  {title} {{Solar neutrino flux at keV energies}},}\ }\href {\doibase
  10.1088/1475-7516/2017/12/010} {\bibfield  {journal} {\bibinfo  {journal}
  {JCAP}\ }\textbf {\bibinfo {volume} {12}},\ \bibinfo {pages} {010} (\bibinfo
  {year} {2017})},\ \Eprint{http://arxiv.org/abs/1708.02248}{arXiv:1708.02248
  [hep-ph]}\BibitemShut {NoStop}%
\bibitem [{\citenamefont {{Carenza}}\ and\ \citenamefont
  {{Lucente}}(2021)}]{Carenza:2021osu}%
  \BibitemOpen
  \bibfield  {author} {\bibinfo {author} {\bibfnamefont {P.}~\bibnamefont
  {{Carenza}}}\ \bibnamefont {and}\ \bibinfo {author} {\bibfnamefont
  {G.}~\bibnamefont {{Lucente}}},\ }\bibfield  {title} {\enquote {\bibinfo
  {title} {{Revisiting axion-electron bremsstrahlung emission rates in
  astrophysical environments}},}\ }\href@noop {} {\bibfield  {journal}
  {\bibinfo  {journal} {arXiv e-prints}\ ,\ \bibinfo {eid} {arXiv:2104.09524}}
  (\bibinfo {year} {2021})},\
  \Eprint{http://arxiv.org/abs/2104.09524}{arXiv:2104.09524
  [hep-ph]}\BibitemShut {NoStop}%
\bibitem [{\citenamefont {{Cox}}\ and\ \citenamefont
  {{Stewart}}(1965)}]{1965ApJS...11...22C}%
  \BibitemOpen
  \bibfield  {author} {\bibinfo {author} {\bibfnamefont {A.~N.}\ \bibnamefont
  {{Cox}}}\ \bibnamefont {and}\ \bibinfo {author} {\bibfnamefont {J.~N.}\
  \bibnamefont {{Stewart}}},\ }\bibfield  {title} {\enquote {\bibinfo {title}
  {{Radiative and Conductive Opacities for Eleven Astrophysical Mixtures.}}}\
  }\href {\doibase 10.1086/190108} {\bibfield  {journal} {\bibinfo  {journal}
  {\apjs}\ }\textbf {\bibinfo {volume} {11}},\ \bibinfo {pages} {22} (\bibinfo
  {year} {1965})}\BibitemShut {NoStop}%
\bibitem [{\citenamefont {{Rogers}}\ and\ \citenamefont
  {{Iglesias}}(1992)}]{Rogers:1992ud}%
  \BibitemOpen
  \bibfield  {author} {\bibinfo {author} {\bibfnamefont {F.~J.}\ \bibnamefont
  {{Rogers}}}\ \bibnamefont {and}\ \bibinfo {author} {\bibfnamefont {C.~A.}\
  \bibnamefont {{Iglesias}}},\ }\bibfield  {title} {\enquote {\bibinfo {title}
  {{Radiative Atomic Rosseland Mean Opacity Tables}},}\ }\href {\doibase
  10.1086/191659} {\bibfield  {journal} {\bibinfo  {journal} {\apjs}\ }\textbf
  {\bibinfo {volume} {79}},\ \bibinfo {pages} {507} (\bibinfo {year}
  {1992})}\BibitemShut {NoStop}%
\bibitem [{\citenamefont {{Badnell}}\ \emph {et~al.}(2005)\citenamefont
  {{Badnell}}, \citenamefont {{Bautista}}, \citenamefont {{Butler}},
  \citenamefont {{Delahaye}}, \citenamefont {{Mendoza}}, \citenamefont
  {{Palmeri}}, \citenamefont {{Zeippen}},\ and\ \citenamefont
  {{Seaton}}}]{astro-ph/0410744}%
  \BibitemOpen
  \bibfield  {author} {\bibinfo {author} {\bibfnamefont {N.~R.}\ \bibnamefont
  {{Badnell}}}, \bibinfo {author} {\bibfnamefont {M.~A.}\ \bibnamefont
  {{Bautista}}}, \bibinfo {author} {\bibfnamefont {K.}~\bibnamefont
  {{Butler}}}, \bibnamefont {et~al.},\ }\bibfield  {title} {\enquote {\bibinfo
  {title} {{Updated opacities from the Opacity Project}},}\ }\href {\doibase
  10.1111/j.1365-2966.2005.08991.x} {\bibfield  {journal} {\bibinfo  {journal}
  {\mnras}\ }\textbf {\bibinfo {volume} {360}},\ \bibinfo {pages} {458}
  (\bibinfo {year} {2005})},\
  \Eprint{http://arxiv.org/abs/astro-ph/0410744}{arXiv:astro-ph/0410744
  [astro-ph]}\BibitemShut {NoStop}%
\bibitem [{\citenamefont {{Magee}}\ \emph {et~al.}(1995)\citenamefont
  {{Magee}}, \citenamefont {{Abdallah}}, \citenamefont {{Clark}}, \citenamefont
  {{Cohen}}, \citenamefont {{Collins}}, \citenamefont {{Csanak}}, \citenamefont
  {{Fontes}}, \citenamefont {{Gauger}}, \citenamefont {{Keady}}, \citenamefont
  {{Kilcrease}},\ and\ \citenamefont {{Merts}}}]{1995_LEDCOP}%
  \BibitemOpen
  \bibfield  {author} {\bibinfo {author} {\bibfnamefont {N.~H.}\ \bibnamefont
  {{Magee}}}, \bibinfo {author} {\bibfnamefont {J.}~\bibnamefont {{Abdallah}},
  \bibfnamefont {J.}}, \bibinfo {author} {\bibfnamefont {R.~E.~H.}\
  \bibnamefont {{Clark}}}, \bibnamefont {et~al.},\ }\bibfield  {title}
  {\enquote {\bibinfo {title} {{Atomic Structure Calculations and New LOS
  Alamos Astrophysical Opacities}},}\ }in\ \href@noop {} {\emph {\bibinfo
  {booktitle} {Astrophysical Applications of Powerful New Databases}}},\
  \bibinfo {series} {Astronomical Society of the Pacific Conference Series},
  Vol.~\bibinfo {volume} {78},\ \bibinfo {editor} {edited by\ \bibinfo {editor}
  {\bibfnamefont {S.~J.}\ \bibnamefont {{Adelman}}}\ \bibnamefont {and}\
  \bibinfo {editor} {\bibfnamefont {W.~L.}\ \bibnamefont {{Wiese}}}}\ (\bibinfo
  {year} {1995})\ p.~\bibinfo {pages} {51}\BibitemShut {NoStop}%
\bibitem [{\citenamefont {{Blancard}}\ \emph {et~al.}(2012)\citenamefont
  {{Blancard}}, \citenamefont {{Coss{\'e}}},\ and\ \citenamefont
  {{Faussurier}}}]{2012_OPAS}%
  \BibitemOpen
  \bibfield  {author} {\bibinfo {author} {\bibfnamefont {C.}~\bibnamefont
  {{Blancard}}}, \bibinfo {author} {\bibfnamefont {P.}~\bibnamefont
  {{Coss{\'e}}}},\ \bibnamefont {and}\ \bibinfo {author} {\bibfnamefont
  {G.}~\bibnamefont {{Faussurier}}},\ }\bibfield  {title} {\enquote {\bibinfo
  {title} {{Solar Mixture Opacity Calculations Using Detailed Configuration and
  Level Accounting Treatments}},}\ }\href {\doibase 10.1088/0004-637X/745/1/10}
  {\bibfield  {journal} {\bibinfo  {journal} {\apj}\ }\textbf {\bibinfo
  {volume} {745}},\ \bibinfo {eid} {10} (\bibinfo {year} {2012})}\BibitemShut
  {NoStop}%
\bibitem [{\citenamefont {{Seaton}}(2005)}]{astro-ph/0411010}%
  \BibitemOpen
  \bibfield  {author} {\bibinfo {author} {\bibfnamefont {M.~J.}\ \bibnamefont
  {{Seaton}}},\ }\bibfield  {title} {\enquote {\bibinfo {title} {{Opacity
  Project data on CD for mean opacities and radiative accelerations}},}\ }\href
  {\doibase 10.1111/j.1365-2966.2005.00019.x} {\bibfield  {journal} {\bibinfo
  {journal} {\mnras}\ }\textbf {\bibinfo {volume} {362}},\ \bibinfo {pages}
  {L1} (\bibinfo {year} {2005})},\
  \Eprint{http://arxiv.org/abs/astro-ph/0411010}{arXiv:astro-ph/0411010
  [astro-ph]}\BibitemShut {NoStop}%
\bibitem [{\citenamefont {{Asplund}}\ \emph {et~al.}(2009)\citenamefont
  {{Asplund}}, \citenamefont {{Grevesse}}, \citenamefont {{Sauval}},\ and\
  \citenamefont {{Scott}}}]{Asplund:2009fu}%
  \BibitemOpen
  \bibfield  {author} {\bibinfo {author} {\bibfnamefont {M.}~\bibnamefont
  {{Asplund}}}, \bibinfo {author} {\bibfnamefont {N.}~\bibnamefont
  {{Grevesse}}}, \bibinfo {author} {\bibfnamefont {A.~J.}\ \bibnamefont
  {{Sauval}}},\ \bibnamefont {and}\ \bibinfo {author} {\bibfnamefont
  {P.}~\bibnamefont {{Scott}}},\ }\bibfield  {title} {\enquote {\bibinfo
  {title} {{The Chemical Composition of the Sun}},}\ }\href {\doibase
  10.1146/annurev.astro.46.060407.145222} {\bibfield  {journal} {\bibinfo
  {journal} {\araa}\ }\textbf {\bibinfo {volume} {47}},\ \bibinfo {pages} {481}
  (\bibinfo {year} {2009})},\
  \Eprint{http://arxiv.org/abs/0909.0948}{arXiv:0909.0948
  [astro-ph.SR]}\BibitemShut {NoStop}%
\bibitem [{\citenamefont {Williams}(2020)}]{NASA}%
  \BibitemOpen
  \bibfield  {author} {\bibinfo {author} {\bibfnamefont {D.~R.}\ \bibnamefont
  {Williams}},\ }\href@noop {} {\enquote {\bibinfo {title} {Earth fact
  sheet},}\ }\bibinfo {howpublished}
  {\url{https://nssdc.gsfc.nasa.gov/planetary/factsheet/earthfact.html}}
  (\bibinfo {year} {2020}),\ \bibinfo {note} {accessed: 19.01.2021}\BibitemShut
  {NoStop}%
\bibitem [{\citenamefont {{Serenelli}}(2010)}]{0910.3690}%
  \BibitemOpen
  \bibfield  {author} {\bibinfo {author} {\bibfnamefont {A.~M.}\ \bibnamefont
  {{Serenelli}}},\ }\bibfield  {title} {\enquote {\bibinfo {title} {{New
  results on standard solar models}},}\ }\href {\doibase
  10.1007/s10509-009-0174-8} {\bibfield  {journal} {\bibinfo  {journal}
  {\apss}\ }\textbf {\bibinfo {volume} {328}},\ \bibinfo {pages} {13} (\bibinfo
  {year} {2010})},\ \Eprint{http://arxiv.org/abs/0910.3690}{arXiv:0910.3690
  [astro-ph.SR]}\BibitemShut {NoStop}%
\bibitem [{\citenamefont {{Bahcall}}\ \emph {et~al.}(1998)\citenamefont
  {{Bahcall}}, \citenamefont {{Basu}},\ and\ \citenamefont
  {{Pinsonneault}}}]{astro-ph/9805135}%
  \BibitemOpen
  \bibfield  {author} {\bibinfo {author} {\bibfnamefont {J.~N.}\ \bibnamefont
  {{Bahcall}}}, \bibinfo {author} {\bibfnamefont {S.}~\bibnamefont {{Basu}}},\
  \bibnamefont {and}\ \bibinfo {author} {\bibfnamefont {M.~H.}\ \bibnamefont
  {{Pinsonneault}}},\ }\bibfield  {title} {\enquote {\bibinfo {title} {{How
  uncertain are solar neutrino predictions?}}}\ }\href {\doibase
  10.1016/S0370-2693(98)00657-1} {\bibfield  {journal} {\bibinfo  {journal}
  {Physics Letters B}\ }\textbf {\bibinfo {volume} {433}},\ \bibinfo {pages}
  {1} (\bibinfo {year} {1998})},\
  \Eprint{http://arxiv.org/abs/astro-ph/9805135}{arXiv:astro-ph/9805135
  [astro-ph]}\BibitemShut {NoStop}%
\bibitem [{\citenamefont {{Bahcall}}\ \emph {et~al.}(2001)\citenamefont
  {{Bahcall}}, \citenamefont {{Pinsonneault}},\ and\ \citenamefont
  {{Basu}}}]{astro-ph/0010346}%
  \BibitemOpen
  \bibfield  {author} {\bibinfo {author} {\bibfnamefont {J.~N.}\ \bibnamefont
  {{Bahcall}}}, \bibinfo {author} {\bibfnamefont {M.~H.}\ \bibnamefont
  {{Pinsonneault}}},\ \bibnamefont {and}\ \bibinfo {author} {\bibfnamefont
  {S.}~\bibnamefont {{Basu}}},\ }\bibfield  {title} {\enquote {\bibinfo {title}
  {{Solar Models: Current Epoch and Time Dependences, Neutrinos, and
  Helioseismological Properties}},}\ }\href {\doibase 10.1086/321493}
  {\bibfield  {journal} {\bibinfo  {journal} {\apj}\ }\textbf {\bibinfo
  {volume} {555}},\ \bibinfo {pages} {990} (\bibinfo {year} {2001})},\
  \Eprint{http://arxiv.org/abs/astro-ph/0010346}{arXiv:astro-ph/0010346
  [astro-ph]}\BibitemShut {NoStop}%
\bibitem [{\citenamefont {{Bahcall}}\ and\ \citenamefont
  {{Pinsonneault}}(2004)}]{astro-ph/0402114}%
  \BibitemOpen
  \bibfield  {author} {\bibinfo {author} {\bibfnamefont {J.~N.}\ \bibnamefont
  {{Bahcall}}}\ \bibnamefont {and}\ \bibinfo {author} {\bibfnamefont {M.~H.}\
  \bibnamefont {{Pinsonneault}}},\ }\bibfield  {title} {\enquote {\bibinfo
  {title} {{What Do We (Not) Know Theoretically about Solar Neutrino
  Fluxes?}}}\ }\href {\doibase 10.1103/PhysRevLett.92.121301} {\bibfield
  {journal} {\bibinfo  {journal} {Physical Review Letters}\ }\textbf {\bibinfo
  {volume} {92}},\ \bibinfo {eid} {121301} (\bibinfo {year} {2004})},\
  \Eprint{http://arxiv.org/abs/astro-ph/0402114}{astro-ph/0402114}\BibitemShut
  {NoStop}%
\bibitem [{\citenamefont {{Bahcall}}\ \emph {et~al.}(2005)\citenamefont
  {{Bahcall}}, \citenamefont {{Serenelli}},\ and\ \citenamefont
  {{Basu}}}]{astro-ph/0412440}%
  \BibitemOpen
  \bibfield  {author} {\bibinfo {author} {\bibfnamefont {J.~N.}\ \bibnamefont
  {{Bahcall}}}, \bibinfo {author} {\bibfnamefont {A.~M.}\ \bibnamefont
  {{Serenelli}}},\ \bibnamefont {and}\ \bibinfo {author} {\bibfnamefont
  {S.}~\bibnamefont {{Basu}}},\ }\bibfield  {title} {\enquote {\bibinfo {title}
  {{New Solar Opacities, Abundances, Helioseismology, and Neutrino Fluxes}},}\
  }\href {\doibase 10.1086/428929} {\bibfield  {journal} {\bibinfo  {journal}
  {\apjl}\ }\textbf {\bibinfo {volume} {621}},\ \bibinfo {pages} {L85}
  (\bibinfo {year} {2005})},\
  \Eprint{http://arxiv.org/abs/astro-ph/0412440}{arXiv:astro-ph/0412440
  [astro-ph]}\BibitemShut {NoStop}%
\bibitem [{\citenamefont {Bahcall}\ \emph {et~al.}(2005)\citenamefont
  {Bahcall}, \citenamefont {Basu}, \citenamefont {Pinsonneault},\ and\
  \citenamefont {Serenelli}}]{Bahcall:2004yr}%
  \BibitemOpen
  \bibfield  {author} {\bibinfo {author} {\bibfnamefont {J.~N.}\ \bibnamefont
  {Bahcall}}, \bibinfo {author} {\bibfnamefont {S.}~\bibnamefont {Basu}},
  \bibinfo {author} {\bibfnamefont {M.}~\bibnamefont {Pinsonneault}},\
  \bibnamefont {and}\ \bibinfo {author} {\bibfnamefont {A.~M.}\ \bibnamefont
  {Serenelli}},\ }\bibfield  {title} {\enquote {\bibinfo {title}
  {{Helioseismological implications of recent solar abundance
  determinations}},}\ }\href {\doibase 10.1086/426070} {\bibfield  {journal}
  {\bibinfo  {journal} {Astrophys. J.}\ }\textbf {\bibinfo {volume} {618}},\
  \bibinfo {pages} {1049} (\bibinfo {year} {2005})},\
  \Eprint{http://arxiv.org/abs/astro-ph/0407060}{arXiv:astro-ph/0407060}\BibitemShut
  {NoStop}%
\bibitem [{\citenamefont {Antia}\ and\ \citenamefont
  {Basu}(2005)}]{Antia:2005mg}%
  \BibitemOpen
  \bibfield  {author} {\bibinfo {author} {\bibfnamefont {H.}~\bibnamefont
  {Antia}}\ \bibnamefont {and}\ \bibinfo {author} {\bibfnamefont
  {S.}~\bibnamefont {Basu}},\ }\bibfield  {title} {\enquote {\bibinfo {title}
  {{The Discrepancy between solar abundances and helioseismology}},}\ }\href
  {\doibase 10.1086/428652} {\bibfield  {journal} {\bibinfo  {journal}
  {Astrophys. J. Lett.}\ }\textbf {\bibinfo {volume} {620}},\ \bibinfo {pages}
  {L129} (\bibinfo {year} {2005})},\
  \Eprint{http://arxiv.org/abs/astro-ph/0501129}{arXiv:astro-ph/0501129}\BibitemShut
  {NoStop}%
\bibitem [{\citenamefont {{Pena-Garay}}\ and\ \citenamefont
  {{Serenelli}}(2008)}]{PenaGaray:2008qe}%
  \BibitemOpen
  \bibfield  {author} {\bibinfo {author} {\bibfnamefont {C.}~\bibnamefont
  {{Pena-Garay}}}\ \bibnamefont {and}\ \bibinfo {author} {\bibfnamefont
  {A.}~\bibnamefont {{Serenelli}}},\ }\bibfield  {title} {\enquote {\bibinfo
  {title} {{Solar neutrinos and the solar composition problem}},}\ }\href@noop
  {} {\bibfield  {journal} {\bibinfo  {journal} {arXiv e-prints}\ ,\ \bibinfo
  {eid} {arXiv:0811.2424}} (\bibinfo {year} {2008})},\
  \Eprint{http://arxiv.org/abs/0811.2424}{arXiv:0811.2424
  [astro-ph]}\BibitemShut {NoStop}%
\bibitem [{\citenamefont {Mendoza}(2018)}]{Mendoza:2018iyx}%
  \BibitemOpen
  \bibfield  {author} {\bibinfo {author} {\bibfnamefont {C.}~\bibnamefont
  {Mendoza}},\ }\bibfield  {title} {\enquote {\bibinfo {title} {{Computation of
  Atomic Astrophysical Opacities}},}\ }\href {\doibase 10.3390/atoms6020028}
  {\bibfield  {journal} {\bibinfo  {journal} {Atoms}\ }\textbf {\bibinfo
  {volume} {6}},\ \bibinfo {pages} {28} (\bibinfo {year} {2018})},\
  \Eprint{http://arxiv.org/abs/1704.03528}{arXiv:1704.03528
  [astro-ph.SR]}\BibitemShut {NoStop}%
\bibitem [{\citenamefont {{Colgan}}\ \emph {et~al.}(2016)\citenamefont
  {{Colgan}}, \citenamefont {{Kilcrease}}, \citenamefont {{Magee}},
  \citenamefont {{Sherrill}}, \citenamefont {{Abdallah}}, \citenamefont
  {{Hakel}}, \citenamefont {{Fontes}}, \citenamefont {{Guzik}},\ and\
  \citenamefont {{Mussack}}}]{1601.01005}%
  \BibitemOpen
  \bibfield  {author} {\bibinfo {author} {\bibfnamefont {J.}~\bibnamefont
  {{Colgan}}}, \bibinfo {author} {\bibfnamefont {D.~P.}\ \bibnamefont
  {{Kilcrease}}}, \bibinfo {author} {\bibfnamefont {N.~H.}\ \bibnamefont
  {{Magee}}}, \bibnamefont {et~al.},\ }\bibfield  {title} {\enquote {\bibinfo
  {title} {{A New Generation of Los Alamos Opacity Tables}},}\ }\href {\doibase
  10.3847/0004-637X/817/2/116} {\bibfield  {journal} {\bibinfo  {journal}
  {\apj}\ }\textbf {\bibinfo {volume} {817}},\ \bibinfo {eid} {116} (\bibinfo
  {year} {2016})},\ \Eprint{http://arxiv.org/abs/1601.01005}{arXiv:1601.01005
  [astro-ph.SR]}\BibitemShut {NoStop}%
\bibitem [{\citenamefont {{Mondet}}\ \emph {et~al.}(2015)\citenamefont
  {{Mondet}}, \citenamefont {{Blancard}}, \citenamefont {{Coss{\'e}}},\ and\
  \citenamefont {{Faussurier}}}]{2015_OPAS}%
  \BibitemOpen
  \bibfield  {author} {\bibinfo {author} {\bibfnamefont {G.}~\bibnamefont
  {{Mondet}}}, \bibinfo {author} {\bibfnamefont {C.}~\bibnamefont
  {{Blancard}}}, \bibinfo {author} {\bibfnamefont {P.}~\bibnamefont
  {{Coss{\'e}}}},\ \bibnamefont {and}\ \bibinfo {author} {\bibfnamefont
  {G.}~\bibnamefont {{Faussurier}}},\ }\bibfield  {title} {\enquote {\bibinfo
  {title} {{Opacity Calculations for Solar Mixtures}},}\ }\href {\doibase
  10.1088/0067-0049/220/1/2} {\bibfield  {journal} {\bibinfo  {journal}
  {\apjs}\ }\textbf {\bibinfo {volume} {220}},\ \bibinfo {eid} {2} (\bibinfo
  {year} {2015})}\BibitemShut {NoStop}%
\bibitem [{\citenamefont {Weiss}\ and\ \citenamefont
  {Schlattl}(2008)}]{Weiss2008}%
  \BibitemOpen
  \bibfield  {author} {\bibinfo {author} {\bibfnamefont {A.}~\bibnamefont
  {Weiss}}\ \bibnamefont {and}\ \bibinfo {author} {\bibfnamefont
  {H.}~\bibnamefont {Schlattl}},\ }\bibfield  {title} {\enquote {\bibinfo
  {title} {{GARSTEC—the Garching Stellar Evolution Code}},}\ }\href {\doibase
  10.1007/s10509-007-9606-5} {\bibfield  {journal} {\bibinfo  {journal}
  {Astrophysics and Space Science}\ }\textbf {\bibinfo {volume} {316}},\
  \bibinfo {pages} {99} (\bibinfo {year} {2008})}\BibitemShut {NoStop}%
\bibitem [{\citenamefont {Abe}\ \emph {et~al.}(2021)\citenamefont {Abe},
  \citenamefont {Hamaguchi},\ and\ \citenamefont {Nagata}}]{Abe:2021ocf}%
  \BibitemOpen
  \bibfield  {author} {\bibinfo {author} {\bibfnamefont {T.}~\bibnamefont
  {Abe}}, \bibinfo {author} {\bibfnamefont {K.}~\bibnamefont {Hamaguchi}},\
  \bibnamefont {and}\ \bibinfo {author} {\bibfnamefont {N.}~\bibnamefont
  {Nagata}},\ }\bibfield  {title} {\enquote {\bibinfo {title} {{Atomic Form
  Factors and Inverse Primakoff Scattering of Axion}},}\ }\href {\doibase
  10.1016/j.physletb.2021.136174} {\bibfield  {journal} {\bibinfo  {journal}
  {Phys. Lett. B}\ }\textbf {\bibinfo {volume} {815}},\ \bibinfo {pages}
  {136174} (\bibinfo {year} {2021})},\
  \Eprint{http://arxiv.org/abs/2012.02508}{arXiv:2012.02508
  [hep-ph]}\BibitemShut {NoStop}%
\bibitem [{\citenamefont {Hubbell}\ \emph {et~al.}(1975)\citenamefont
  {Hubbell}, \citenamefont {Veigele}, \citenamefont {Briggs}, \citenamefont
  {Brown}, \citenamefont {Cromer},\ and\ \citenamefont
  {Howerton}}]{doi:10.1063/1.555523}%
  \BibitemOpen
  \bibfield  {author} {\bibinfo {author} {\bibfnamefont {J.~H.}\ \bibnamefont
  {Hubbell}}, \bibinfo {author} {\bibfnamefont {W.~J.}\ \bibnamefont
  {Veigele}}, \bibinfo {author} {\bibfnamefont {E.~A.}\ \bibnamefont {Briggs}},
  \bibnamefont {et~al.},\ }\bibfield  {title} {\enquote {\bibinfo {title}
  {Atomic form factors, incoherent scattering functions, and photon scattering
  cross sections},}\ }\href {\doibase 10.1063/1.555523} {\bibfield  {journal}
  {\bibinfo  {journal} {Journal of Physical and Chemical Reference Data}\
  }\textbf {\bibinfo {volume} {4}},\ \bibinfo {pages} {471} (\bibinfo {year}
  {1975})},\
  \Eprint{http://arxiv.org/abs/https://doi.org/10.1063/1.555523}{https://doi.org/10.1063/1.555523}\BibitemShut
  {NoStop}%
\bibitem [{\citenamefont {{Elwert}}(1939)}]{Elwert1939}%
  \BibitemOpen
  \bibfield  {author} {\bibinfo {author} {\bibfnamefont {G.}~\bibnamefont
  {{Elwert}}},\ }\bibfield  {title} {\enquote {\bibinfo {title}
  {{Versch{\"a}rfte Berechnung von Intensit{\"a}t und Polarisation im
  kontinuierlichen R{\"o}ntgenspektrum}},}\ }\href {\doibase
  10.1002/andp.19394260206} {\bibfield  {journal} {\bibinfo  {journal} {Annalen
  der Physik}\ }\textbf {\bibinfo {volume} {426}},\ \bibinfo {pages} {178}
  (\bibinfo {year} {1939})}\BibitemShut {NoStop}%
\bibitem [{\citenamefont {{Chen}}\ and\ \citenamefont
  {{Dawson}}(2013)}]{1301.0309}%
  \BibitemOpen
  \bibfield  {author} {\bibinfo {author} {\bibfnamefont {C.-Y.}\ \bibnamefont
  {{Chen}}}\ \bibnamefont {and}\ \bibinfo {author} {\bibfnamefont
  {S.}~\bibnamefont {{Dawson}}},\ }\bibfield  {title} {\enquote {\bibinfo
  {title} {{Exploring two Higgs doublet models through Higgs production}},}\
  }\href {\doibase 10.1103/PhysRevD.87.055016} {\bibfield  {journal} {\bibinfo
  {journal} {\prd}\ }\textbf {\bibinfo {volume} {87}},\ \bibinfo {eid} {055016}
  (\bibinfo {year} {2013})},\
  \Eprint{http://arxiv.org/abs/1301.0309}{arXiv:1301.0309 [hep-ph]}\BibitemShut
  {NoStop}%
\bibitem [{\citenamefont {Kim}(1979)}]{1979_kim_ksvz}%
  \BibitemOpen
  \bibfield  {author} {\bibinfo {author} {\bibfnamefont {J.~E.}\ \bibnamefont
  {Kim}},\ }\bibfield  {title} {\enquote {\bibinfo {title} {{Weak-Interaction
  Singlet and Strong {CP} Invariance}},}\ }\href
  {http://dx.doi.org/10.1103/PhysRevLett.43.103} {\bibfield  {journal}
  {\bibinfo  {journal} {Physical Review Letters}\ }\textbf {\bibinfo {volume}
  {43}},\ \bibinfo {pages} {103} (\bibinfo {year} {1979})}\BibitemShut
  {NoStop}%
\bibitem [{\citenamefont {Shifman}\ \emph {et~al.}(1980)\citenamefont
  {Shifman}, \citenamefont {Vainshtein},\ and\ \citenamefont
  {Zakharov}}]{1980_shifman_ksvz}%
  \BibitemOpen
  \bibfield  {author} {\bibinfo {author} {\bibfnamefont {M.}~\bibnamefont
  {Shifman}}, \bibinfo {author} {\bibfnamefont {A.}~\bibnamefont
  {Vainshtein}},\ \bibnamefont {and}\ \bibinfo {author} {\bibfnamefont
  {V.}~\bibnamefont {Zakharov}},\ }\bibfield  {title} {\enquote {\bibinfo
  {title} {{Can confinement ensure natural {CP} invariance of strong
  interactions?}}}\ }\href {http://dx.doi.org/10.1016/0550-3213(80)90209-6}
  {\bibfield  {journal} {\bibinfo  {journal} {Nuclear Physics B}\ }\textbf
  {\bibinfo {volume} {166}},\ \bibinfo {pages} {493} (\bibinfo {year}
  {1980})}\BibitemShut {NoStop}%
\bibitem [{\citenamefont {{Di Luzio}}\ \emph
  {et~al.}(2017{\natexlab{a}})\citenamefont {{Di Luzio}}, \citenamefont
  {{Mescia}},\ and\ \citenamefont {{Nardi}}}]{1610.07593}%
  \BibitemOpen
  \bibfield  {author} {\bibinfo {author} {\bibfnamefont {L.}~\bibnamefont {{Di
  Luzio}}}, \bibinfo {author} {\bibfnamefont {F.}~\bibnamefont {{Mescia}}},\
  \bibnamefont {and}\ \bibinfo {author} {\bibfnamefont {E.}~\bibnamefont
  {{Nardi}}},\ }\bibfield  {title} {\enquote {\bibinfo {title} {{Redefining the
  Axion Window}},}\ }\href {\doibase 10.1103/PhysRevLett.118.031801} {\bibfield
   {journal} {\bibinfo  {journal} {Physical Review Letters}\ }\textbf {\bibinfo
  {volume} {118}},\ \bibinfo {eid} {031801} (\bibinfo {year}
  {2017}{\natexlab{a}})},\
  \Eprint{http://arxiv.org/abs/1610.07593}{arXiv:1610.07593
  [hep-ph]}\BibitemShut {NoStop}%
\bibitem [{\citenamefont {{Di Luzio}}\ \emph
  {et~al.}(2017{\natexlab{b}})\citenamefont {{Di Luzio}}, \citenamefont
  {{Mescia}},\ and\ \citenamefont {{Nardi}}}]{1705.05370}%
  \BibitemOpen
  \bibfield  {author} {\bibinfo {author} {\bibfnamefont {L.}~\bibnamefont {{Di
  Luzio}}}, \bibinfo {author} {\bibfnamefont {F.}~\bibnamefont {{Mescia}}},\
  \bibnamefont {and}\ \bibinfo {author} {\bibfnamefont {E.}~\bibnamefont
  {{Nardi}}},\ }\bibfield  {title} {\enquote {\bibinfo {title} {{Window for
  preferred axion models}},}\ }\href {\doibase 10.1103/PhysRevD.96.075003}
  {\bibfield  {journal} {\bibinfo  {journal} {\prd}\ }\textbf {\bibinfo
  {volume} {96}},\ \bibinfo {eid} {075003} (\bibinfo {year}
  {2017}{\natexlab{b}})},\
  \Eprint{http://arxiv.org/abs/1705.05370}{arXiv:1705.05370
  [hep-ph]}\BibitemShut {NoStop}%
\bibitem [{\citenamefont {{Di Luzio}}\ \emph {et~al.}(2020)\citenamefont {{Di
  Luzio}}, \citenamefont {{Giannotti}}, \citenamefont {{Nardi}},\ and\
  \citenamefont {{Visinelli}}}]{2003.01100}%
  \BibitemOpen
  \bibfield  {author} {\bibinfo {author} {\bibfnamefont {L.}~\bibnamefont {{Di
  Luzio}}}, \bibinfo {author} {\bibfnamefont {M.}~\bibnamefont {{Giannotti}}},
  \bibinfo {author} {\bibfnamefont {E.}~\bibnamefont {{Nardi}}},\ \bibnamefont
  {and}\ \bibinfo {author} {\bibfnamefont {L.}~\bibnamefont {{Visinelli}}},\
  }\bibfield  {title} {\enquote {\bibinfo {title} {{The landscape of QCD axion
  models}},}\ }\href {\doibase 10.1016/j.physrep.2020.06.002} {\bibfield
  {journal} {\bibinfo  {journal} {\physrep}\ }\textbf {\bibinfo {volume}
  {870}},\ \bibinfo {pages} {1} (\bibinfo {year} {2020})},\
  \Eprint{http://arxiv.org/abs/2003.01100}{arXiv:2003.01100
  [hep-ph]}\BibitemShut {NoStop}%
\bibitem [{\citenamefont {{di Cortona}}\ \emph {et~al.}(2016)\citenamefont {{di
  Cortona}}, \citenamefont {{Hardy}}, \citenamefont {{Vega}},\ and\
  \citenamefont {{Villadoro}}}]{1511.02867}%
  \BibitemOpen
  \bibfield  {author} {\bibinfo {author} {\bibfnamefont {G.~G.}\ \bibnamefont
  {{di Cortona}}}, \bibinfo {author} {\bibfnamefont {E.}~\bibnamefont
  {{Hardy}}}, \bibinfo {author} {\bibfnamefont {J.~P.}\ \bibnamefont
  {{Vega}}},\ \bibnamefont {and}\ \bibinfo {author} {\bibfnamefont
  {G.}~\bibnamefont {{Villadoro}}},\ }\bibfield  {title} {\enquote {\bibinfo
  {title} {{The QCD axion, precisely}},}\ }\href {\doibase
  10.1007/JHEP01(2016)034} {\bibfield  {journal} {\bibinfo  {journal} {Journal
  of High Energy Physics}\ }\textbf {\bibinfo {volume} {1}},\ \bibinfo {eid}
  {34} (\bibinfo {year} {2016})},\
  \Eprint{http://arxiv.org/abs/1511.02867}{arXiv:1511.02867
  [hep-ph]}\BibitemShut {NoStop}%
\bibitem [{\citenamefont {{Gorghetto}}\ and\ \citenamefont
  {{Villadoro}}(2019)}]{1812.01008}%
  \BibitemOpen
  \bibfield  {author} {\bibinfo {author} {\bibfnamefont {M.}~\bibnamefont
  {{Gorghetto}}}\ \bibnamefont {and}\ \bibinfo {author} {\bibfnamefont
  {G.}~\bibnamefont {{Villadoro}}},\ }\bibfield  {title} {\enquote {\bibinfo
  {title} {{Topological susceptibility and QCD axion mass: QED and NNLO
  corrections}},}\ }\href {\doibase 10.1007/JHEP03(2019)033} {\bibfield
  {journal} {\bibinfo  {journal} {Journal of High Energy Physics}\ }\textbf
  {\bibinfo {volume} {2019}},\ \bibinfo {eid} {33} (\bibinfo {year} {2019})},\
  \Eprint{http://arxiv.org/abs/1812.01008}{arXiv:1812.01008
  [hep-ph]}\BibitemShut {NoStop}%
\bibitem [{\citenamefont {{Abeln}}\ \emph {et~al.}(2020)\citenamefont
  {{Abeln}}, \citenamefont {{Altenm{\"u}ller}}, \citenamefont {{Arguedas
  Cuendis}}, \citenamefont {{Armengaud}}, \citenamefont {{Atti{\'e}}},
  \citenamefont {{Aune}}, \citenamefont {{Basso}}, \citenamefont {{Berg{\'e}}},
  \citenamefont {{Biasuzzi}}, \citenamefont {{Borges De Sousa}}, \citenamefont
  {{Brun}}, \citenamefont {{Bykovskiy}}, \citenamefont {{Calvet}},
  \citenamefont {{Carmona}}, \citenamefont {{Castel}}, \citenamefont
  {{Cebri{\'a}n}}, \citenamefont {{Chernov}}, \citenamefont {{Christensen}},
  \citenamefont {{Civitani}}, \citenamefont {{Cogollos}}, \citenamefont
  {{Dafn{\'\i}}}, \citenamefont {{Derbin}}, \citenamefont {{Desch}},
  \citenamefont {{D{\'\i}ez}}, \citenamefont {{Dinter}}, \citenamefont
  {{D{\"o}brich}}, \citenamefont {{Drachnev}}, \citenamefont {{Dudarev}},
  \citenamefont {{Dumoulin}}, \citenamefont {{Ferreira}}, \citenamefont
  {{Ferrer-Ribas}}, \citenamefont {{Fleck}}, \citenamefont {{Gal{\'a}n}},
  \citenamefont {{Gasc{\'o}n}}, \citenamefont {{Gastaldo}}, \citenamefont
  {{Giannotti}}, \citenamefont {{Giomataris}}, \citenamefont {{Giuliani}},
  \citenamefont {{Gninenko}}, \citenamefont {{Golm}}, \citenamefont
  {{Golubev}}, \citenamefont {{Hagge}}, \citenamefont {{Hahn}}, \citenamefont
  {{Hailey}}, \citenamefont {{Hengstler}}, \citenamefont {{Henriksen}},
  \citenamefont {{Houdy}}, \citenamefont {{Iglesias-Marzoa}}, \citenamefont
  {{Iguaz-Gutierrez}}, \citenamefont {{Irastorza}}, \citenamefont
  {{I{\~n}iguez}}, \citenamefont {{Jakovcic}}, \citenamefont {{Kaminski}},
  \citenamefont {{Kanoute}}, \citenamefont {{Karstensen}}, \citenamefont
  {{Kravchuk}}, \citenamefont {{Lakic}}, \citenamefont {{Lasserre}},
  \citenamefont {{Laurent}}, \citenamefont {{Limousin}}, \citenamefont
  {{Lindner}}, \citenamefont {{Loidl}}, \citenamefont {{Lomskaya}},
  \citenamefont {{L{\'o}pez-Alegre}}, \citenamefont {{Lubsandorzhiev}},
  \citenamefont {{Ludwig}}, \citenamefont {{Luz{\'o}n}}, \citenamefont
  {{Malbrunot}}, \citenamefont {{Margalejo}}, \citenamefont {{Marin-Franch}},
  \citenamefont {{Marnieros}}, \citenamefont {{Marutzky}}, \citenamefont
  {{Mauricio}}, \citenamefont {{Menesguen}}, \citenamefont {{Mentink}},
  \citenamefont {{Mertens}}, \citenamefont {{Mescia}}, \citenamefont
  {{Miralda-Escud{\'e}}}, \citenamefont {{Mirallas}}, \citenamefont {{Mols}},
  \citenamefont {{Muratova}}, \citenamefont {{Navick}}, \citenamefont
  {{Nones}}, \citenamefont {{Notari}}, \citenamefont {{Nozik}}, \citenamefont
  {{Obis}}, \citenamefont {{Oriol}}, \citenamefont {{Orsini}}, \citenamefont
  {{Ortiz de Sol{\'o}rzano}}, \citenamefont {{Oster}}, \citenamefont {{Pais Da
  Silva}}, \citenamefont {{Pantuev}}, \citenamefont {{Papaevangelou}},
  \citenamefont {{Pareschi}}, \citenamefont {{Perez}}, \citenamefont
  {{P{\'e}rez}}, \citenamefont {{Picatoste}}, \citenamefont {{Pivovaroff}},
  \citenamefont {{Poda}}, \citenamefont {{Redondo}}, \citenamefont
  {{Ringwald}}, \citenamefont {{Rodrigues}}, \citenamefont {{Rueda-Teruel}},
  \citenamefont {{Rueda-Teruel}}, \citenamefont {{Ruiz-Choliz}}, \citenamefont
  {{Ruz}}, \citenamefont {{Saemann}}, \citenamefont {{Salvado}}, \citenamefont
  {{Schiffer}}, \citenamefont {{Schmidt}}, \citenamefont {{Schneekloth}},
  \citenamefont {{Schott}}, \citenamefont {{Segui}}, \citenamefont
  {{Tavecchio}}, \citenamefont {{ten Kate}}, \citenamefont {{Tkachev}},
  \citenamefont {{Troitsky}}, \citenamefont {{Unger}}, \citenamefont
  {{Unzhakov}}, \citenamefont {{Ushakov}}, \citenamefont {{Vogel}},
  \citenamefont {{Voronin}}, \citenamefont {{Weltman}}, \citenamefont
  {{Werthenbach}}, \citenamefont {{Wuensch}},\ and\ \citenamefont
  {{Yanes-D{\'\i}az}}}]{2010.12076}%
  \BibitemOpen
  \bibfield  {author} {\bibinfo {author} {\bibfnamefont {A.}~\bibnamefont
  {{Abeln}}}, \bibinfo {author} {\bibfnamefont {K.}~\bibnamefont
  {{Altenm{\"u}ller}}}, \bibinfo {author} {\bibfnamefont {S.}~\bibnamefont
  {{Arguedas Cuendis}}}, \bibnamefont {et~al.},\ }\bibfield  {title} {\enquote
  {\bibinfo {title} {{Conceptual Design of BabyIAXO, the intermediate stage
  towards the International Axion Observatory}},}\ }\href@noop {} {\bibfield
  {journal} {\bibinfo  {journal} {arXiv e-prints}\ ,\ \bibinfo {eid}
  {arXiv:2010.12076}} (\bibinfo {year} {2020})},\
  \Eprint{http://arxiv.org/abs/2010.12076}{arXiv:2010.12076
  [physics.ins-det]}\BibitemShut {NoStop}%
\bibitem [{\citenamefont {Landau}\ and\ \citenamefont
  {Lifshitz}(1968)}]{Landau:1968}%
  \BibitemOpen
  \bibfield  {author} {\bibinfo {author} {\bibfnamefont {L.~D.}\ \bibnamefont
  {Landau}}\ \bibnamefont {and}\ \bibinfo {author} {\bibfnamefont {E.~M.}\
  \bibnamefont {Lifshitz}},\ }\href@noop {} {\emph {\bibinfo {title}
  {{Statistical Physics}}}},\ \bibinfo {edition} {2nd}\ ed.\ (\bibinfo
  {publisher} {Pergamon Press},\ \bibinfo {address} {Oxford},\ \bibinfo {year}
  {1968})\BibitemShut {NoStop}%
\end{thebibliography}%

\end{document}